\documentclass[]{aa}  

\usepackage{pdfpages}

\usepackage{booktabs}
\usepackage{graphicx}
\usepackage{multirow}
\usepackage{amsmath}
\usepackage{multicol}
\usepackage{txfonts}
\usepackage{natbib}

\bibpunct{(}{)}{;}{a}{}{,} 
\def\ltsim{\raise 2pt \hbox {$<$} \kern-1.1em \lower 4pt \hbox {$\sim$}}
\def\gtsim{\raise 2pt \hbox {$>$} \kern-1.1em \lower 4pt \hbox {$\sim$}}

\begin{document} 

\title{Radio and X-ray connection in radio mini-halos:\\implications for hadronic models}

\author{A. Ignesti\inst{1,2}, G. Brunetti\inst{2}, M. Gitti\inst{1,2}, S. Giacintucci\inst{3}}

\institute{
  Dipartimento di Fisica e Astronomia, Universit\`a di Bologna, via Gobetti 93/2, 40129 Bologna, Italy \\
  \email{ alessandro.ignesti2@unibo.it} 
\and 
INAF, Istituto di Radioastronomia di Bologna, via Gobetti 101, 40129 Bologna, Italy
\and
Naval Research Laboratory, 4555 Overlook Avenue SW, Code 7213, Washington, DC 20375, USA
}
\authorrunning{Ignesti et al.}
\titlerunning{Radio and X-ray connection in radio mini-halos: implications for hadronic models}
\date{Accepted for publication on A$\&$A.}

\abstract 
{
  A large fraction of cool-core clusters are known to host diffuse, steep-spectrum radio sources, called radio mini halos, in their cores. Mini-halos probe the presence of relativistic particles on scales of hundreds of kiloparsecs, beyond the scales directly influenced by the central AGN, but the nature of the mechanism that produces such a population of radio-emitting, relativistic electrons is still debated.  It is also unclear to what extent the AGN plays a role in the formation of mini-halos by providing the seeds of the relativistic population. }
{
 In this work we explore the connection between thermal and non-thermal components of the intra-cluster medium in a sample of radio mini-halos and we study the implications in the framework of a hadronic model for the origin of the emitting electrons. }
{
  For the first time, we studied the thermal and non-thermal connection by carrying out a point-to-point comparison of the radio and the X-ray surface brightness in a sample of radio mini-halos. We extended the method generally applied to giant radio halos by considering the effects of a grid randomly generated through a Monte Carlo chain. Then we used the radio and X-ray correlation to constrain the physical parameters of a hadronic model and we compared the model predictions with current observations.}
{
Contrary to what is generally reported in the literature for giant radio halos, we find that the mini-halos in our sample have super-linear scaling between radio and X-rays, which suggests a peaked distribution of relativistic electrons and magnetic field. We explore the consequences of our findings on models of mini-halos. We use the four mini-halos in the sample that have roundish brightness distribution to constrain model parameters in the case of a hadronic origin of the mini-halos. Specifically, we focus on a model where cosmic rays are injected by the central AGN and they generate secondaries in the intra-cluster medium, and we assume that the role of turbulent re-acceleration is negligible. This simple model allows us to constrain the AGN cosmic ray luminosity in the range $\sim10^{44-46}$ erg s$^{-1}$ and the central magnetic field in the range 10-40 $\mu$G. The resulting $\gamma$-ray fluxes calculated assuming these model parameters do not violate the upper limits on $\gamma$-ray diffuse emission set by the Fermi-LAT telescope. Further studies are now required to explore the consistency of these large magnetic fields with Faraday rotation studies and to study the interplay between the secondary electrons and the intra-cluster medium turbulence. 
 } {}

\maketitle

\section{Introduction} 

Cluster scale radio emission probes magnetic field and relativistic particles in the intra-cluster medium (ICM) on hundred-kpc to Mpc scales, thus posing fundamental questions on the ICM micro-physics  \citep[e.g.,][for review]{Brunetti-Jones_2014}. 

In particular, in the past decades radio observations have revealed the presence of diffuse radio emission with steep spectrum ($\alpha>1$ with synchrotron flux at the frequency $\nu$ $S\propto \nu^{-\alpha}$) at the center of massive, relaxed clusters, the so-called radio mini-halos (MHs). The radio emission, whose emissivity is generally higher than that of giant radio halos \citep[e.g.,][]{Cassano-Gitti_2008,Murgia_2009}, has been observed surrounding the central radio galaxy and extending up to 300 kpc \citep[e.g,][and the references therein]{VanWeeren_2019}. The radio emission is often confined in the cool cores of the clusters, thus suggesting a connection between the non-thermal ICM components and the thermal plasma \citep[e.g.,][]{Mazzotta-Giacintucci_2008,Giacintucci_2014b}. This intrinsic connection is supported by the correlations observed between global radio and X-ray luminosities \citep[][]{Bravi_2016,Gitti_2015,Gitti_2018,Giacintucci_2019}. At the present time, we know 23 MHs, observed in almost all the massive cool-core clusters \citep[the incidence is $\sim 80 \%$ for $M_\text{500}>6\cdot10^{14}$ $M_{\odot}$,][]{Giacintucci_2017}, but current and future facilities, as LOFAR and SKA, may have the potential to discover up to $10^4$ new MHs \citep[][]{Gitti_2018}. \\

The origin of radio-emitting, cosmic-ray electrons (CRe) in the MH volume is still debated. Two possible scenarios have been proposed. One is the leptonic scenario, where the CRe, possibly injected by the active galactic nucleus (AGN) of the central radio galaxy, are re-accelerated by ICM turbulence. In this scenario the turbulence in the cool cores may be injected by the cooling flow of the ICM on the central galaxy \citep[e.g.,][]{Gitti_2002}, by the AGN itself during the so-called "radio-mode" AGN feedback \citep[e.g.,][for a review]{McNamara-Nulsen_2012, Gitti_2012}, or by the gas dynamics driven by the cold fronts \citep[e.g.,][]{ZuHone_2013}. The other one is the hadronic scenario, where CRe are produced by collisions between cosmic-ray protons (CRp) and the thermal protons of the ICM \citep[e.g.,][]{Pfrommer-Ensslin_2004}.
Once the CRp are released in the cluster, most likely by the central AGN, they can diffuse on the scale of the observed radio emission due to their longer radiative times \citep[$\tau_\text{CRp}\simeq10^{10}$ yrs $>>\tau_\text{CRe}\simeq10^{8}$ years, e.g.][]{Brunetti-Jones_2014}. \\

The two scenarios share two common aspects: the possible role played by the AGN as the source of relativistic particles and the physical connection between the CRe and thermal plasma (as background medium for the turbulence or targets for CRp collisions). 
This connection would induce a spatial correlation between the radio, $I_\text{R}$, and X-ray, $I_\text{X}$, surface brightness. The importance of this correlation has been discussed in the case of giant radio halos \citep[e.g.,][]{Govoni_2001, Brunetti_2004, Pfrommer_2008,Donnert_2010, Brunetti-Jones_2014}. In re-acceleration models the $I_\text{R}$-$I_\text{X}$ correlation is sensitive to the way turbulence is generated in the thermal background plasma and relativistic particles are accelerated and transported in that turbulence. The correlation is particularly straightforward in the case of secondary models, where the thermal plasma, which generates the X-ray emission, provides also the targets for the inelastic collisions with the CRp that produce the secondary electrons emitting in the radio band. In this latter case, it is generally expected a super-linear scaling between radio and X-ray brightness.\\

Similar considerations apply to the case of MH, thus following this idea, in this work we study the connection between $I_\text{R}$ and $I_\text{X}$ surface brightness for the first time for a sample of MHs. Our results allow constraints on the dynamics of CR and magnetic fields in the emitting region. In particular, we explore the case of a hadronic model assuming that CRp are generated only by the central AGN and we show that is possible to obtain constraints on the magnetic field in the MH and CRp luminosity of the AGN.\\

The paper is structured as follows. In Sec. \ref{data_an} we present the sample of MHs, the Chandra data reduction and we introduce a new tool to evaluate the $I_\text{R}$-$I_\text{X}$ connection. The results are presented in Sec. \ref{results}. In Sec. \ref{model} we present a pure hadronic model, based on the diffusion of CRp from a central source, and we derive the physical conditions that allow the model to reproduce the observed radio emission for a sub-sample of MHs in the pure hadronic framework. The results are discussed and summarized in Sec. \ref{discussion}. In Appendix \ref{descr} we report a brief, morphological description of each cluster analyzed in this work, whereas the radio and X-ray maps that we used are presented in Appendix \ref{images} and the role of central sources is briefly discussed in Appendix \ref{subtr}. In Appendix \ref{D0} we present our considerations on the diffusion coefficient that we adopted in Sec. \ref{model}. We adopted $\Lambda$CDM cosmology, with H$_\text{0}=73$ km s$^{-1}$ Mpc$^{-1}$, $\Omega _\text{m}=1-\Omega_{\Lambda}=0.27$ .

\section{Data analysis}
\label{data_an}
Radio and X-ray correlations can be studied through the point-to-point connection between radio and X-ray surface brightness. For giant radio halos, these studies generally found a sub-linear scaling between the two quantities, as $I_\text{R}\propto I_\text{X}^k$, with k $\leq$ 1  \citep[][]{Govoni_2001, Feretti_2001,Giacintucci_2005, Vacca_2010,Hoang_2019}. \\

Here we extend the analysis of $I_\text{R}$-$I_\text{X}$ connection to the case of radio MHs. Exploring these correlations potentially provides important information on the origin of these sources and their connection with the central AGN. However, these studies are difficult for MHs because of their limited extent, which may affect the spatial sampling, and the presence of bright radio galaxies at their center, that may contaminate the diffuse emission. 
 For these reasons, we selected a sample of targets with deep and well-resolved radio images in literature. Our sample consists of seven MHs (Tab. \ref{phys.tab}). We then produced the X-ray images of each cluster from archival Chandra observations. We report in Appendix \ref{descr} a brief, morphological description of the clusters of our sample.
\subsection{Data preparation}
The radio images used in this paper have been presented in previous works (see list of references in Tab. \ref{obs.tab}). The images were obtained from high-sensitivity, pointed radio observations with the Very Large Array (VLA) at 1.4 and 5.5 GHz and Giant Metrewave Radio Telescope (GMRT) at 0.3 and 0.6 GHz. To enhance the diffuse emission, a weighting scheme close to natural weighting was typically adopted during the data imaging. Higher-resolution images, showing the smaller-scale emission associated with the central galaxy, are also presented in the previous works. All observations used to produce our MH images have a good sampling of the $uv$ plane at short antenna spacings, that ensures the detection of large-scale emission (above the image sensitivity) on scales significantly larger than the measured extent of the MH \citep[for details see][Tab. 10 and Fig. 12]{Giacintucci_2017}. Furthermore, the MH sizes do not appear to correlate with the signal-to-noise ratio of the radio images \citep[][Appendix B]{Giacintucci_2017}, thus ensuring that the measured extent is not biased by the image sensitivity. Besides a good $uv$ coverage at short spacings, the observations have also a sufficiently high resolution to disentangle the central radio galaxy from the surrounding diffuse emission. Nevertheless, to avoid any possible contamination of the radio galaxy emission into the diffuse MH, we masked the central region of each MH using an appropriate mask with a size larger than the radio beam. We report the mask of each cluster in Appendix \ref{images}, whereas in Appendix \ref{subtr} we briefly discuss the comparison between masked and source-subtracted images and we show that the two approaches are equivalent for the purpose of the study of the $I_\text{R}$-$I_\text{X}$ spatial correlation.\\

Concerning the X-ray images, we retrieved the Chandra observations of the clusters from the archive\footnote{{\ttfamily http://cxc.harvard.edu/cda/}} to produce the X-ray images and to derive the physical quantities of the thermal ICM. When it was possible, we collected multiple observations to improve the sensitivity of our analysis.
The datasets were reprocessed with CIAO v.4.9 and corrected for known time-dependent gain and charge transfer inefficiency problems following techniques similar to those described in the Chandra analysis threads\footnote{{\ttfamily http://cxc.harvard.edu/ciao/threads/index.html}}. To filter out strong background flares, we also applied screening of the event files. We used CALDB v.4.7.8 blank-sky background files normalized to the count rate of the source image in the 9-12 keV band to produce the appropriate background image for each observation. We produced the exposure-corrected, background-subtracted brightness maps in the energy range 0.5-2.0 keV. We used this energy band because it is where the thermal ICM emission and Chandra sensitivity are at their maximum, thus it assures an optimal count statistic for the analysis with our data. We checked for the presence of X-ray point sources embedded in the cluster emission and, if any, we masked them.\\

We report the details of the radio images and archival X-ray observations in Tab. \ref{obs.tab}, while the X-ray images with the radio contours are presented in the Appendix \ref{images}.  

\begin{table*}
\begin{center}
  \caption{ Physical properties of the clusters analyzed in this work. }

\begin{tabular}{lccccccc}
\hline
\hline
  Cluster name&RA$_\text{J2000}$&DEC$_\text{J2000}$& $z$ & M$_\text{500}^{\dag}$ & R$_\text{500}^{\dag}$&R$_\text{MH}^{\ddag}$&$L_\text{X,$R_{500}$}$ \\
  &[h, m, s]&[deg, ', "]&&[$10^{14}$ $M_{\odot}$]&[Mpc]&[kpc]&[$10^{44}$ erg s$^{-1}$]\\
\hline
~&~&~\\
2A0335+096&03 38 44.4&+09 56 34&0.035&2.3$^{+0.2}_\text{-0.3}$&0.92&70&$4.4\pm0.5$\\
RBS 797&09 47 00.2&+76 23 44&0.345&6.3$^{+0.6}_\text{-0.7}$&1.16&120&$41.9\pm5.4$ \\
Abell 3444&10 23 54.8&-27 17 09&0.254&7.6$^{+0.5}_\text{-0.6}$&1.27&120&$28.3\pm4.0$\\
MS 1455.0+2232&14 57 15.1&+22 20 34&0.258&3.5$^{+0.4}_\text{-0.4}$&0.98&120&$21.1\pm2.2$ \\

RXC J1504.1-0248&15 04 05.4&-02 47 54&0.215&7.0$^{+0.6}_\text{-0.6}$&0.98&140&$68.4\pm7.0$\\
RX J1532.9+3021&15 32 53.8&+30 20 58&0.345&4.7$^{+0.6}_\text{-0.6}$&1.04&100&$41.6\pm4.5$\\
RX J1720.1+2637&17 20 12.6&+26 37 23&0.164&6.3$^{+0.4}_\text{-0.4}$&1.24&140&$17.2\pm1.7$ \\
&&&&&\\
\hline
\label{phys.tab}
\end{tabular}
 \tablefoot{$^{\dag}$ Radius and total mass at a mean over-density of 500 with respect to the cosmological critical density at redshift $z$ ; $^{\ddag}$ Average radius of the diffuse emission defined as R$_\text{MH}=\sqrt{R_\text{max}\times R_\text{min}}$ where $R_\text{max}$ and $R_\text{min}$ are the maximum and minimum radius as derived from the $3\sigma$ iso-contour emission; Bolometric X-ray luminosity measured within $R_{500}$. The values are taken from \citet[][]{Giacintucci_2017} and references therein.}
\end{center}
\end{table*}
\begin{table*}
\begin{center}
  \caption{ Archival radio and X-ray observations used in this work. }
\begin{tabular}{lcccccc}
\hline
\hline
& \multicolumn{4}{c}{Radio}   &   \multicolumn{2}{c}{X-ray} \\
\cmidrule(rl){2-5}
\cmidrule(rl){6-7}
Cluster name & Reference & Frequency & Beam & RMS &Chandra Obs ID & Total exposure time \\
&&[GHz]&[arcsec$\times$arcsec]&[$\mu$Jy beam$^{-1}$]&&[ks]\\
\hline
~&~&~\\
\vspace{0.2cm}
2A0335+096&1&\begin{tabular}{c} 1.4  \\ 5.5 \end{tabular}&\begin{tabular}{c} 23.0$\times$22.0  \\ 18.5$\times$16.0 \end{tabular}&\begin{tabular}{c} 56  \\ 16 \end{tabular}&919, 7939, 9792&106 \\
\vspace{0.2cm}
RBS 797&2&1.4&3.0$\times$3.0&10&7902&40 \\
\vspace{0.2cm}
Abell 3444&1&\begin{tabular}{c} 0.6  \\ 1.4 \end{tabular}&\begin{tabular}{c} 8.0$\times$8.0  \\ 8.0$\times$8.0 \end{tabular}&\begin{tabular}{c} 58  \\ 35 \end{tabular}&9400&37 \\

MS 1455.0+2232&3&0.6&6.0$\times$5.0&50&4192&92 \\
RXC J1504.1-0248&4&0.3&11.3$\times$10.4&75&17197, 17669, 17670 & 109 \\
RX J1532.9+3021&5&1.4&3.4$\times$2.9&15&14009&88 \\

RX J1720.1+2637&6&0.6&7.8$\times$6.1&30&3224, 4361&50 \\
\hline
\label{obs.tab}
\end{tabular}
\tablefoot{References of the radio maps: (1) \citet[][]{Giacintucci_2019} (2) \citet[][]{Doria_2012,Gitti_2013a} (3) \citet[][]{Mazzotta-Giacintucci_2008} (4) \citet[][]{Giacintucci_2011a} (5) \citet[][]{Giacintucci_2014a} (6) \citet[][]{Giacintucci_2014b}  }

\end{center}
\vspace{-0.2in}
\end{table*}

\subsection{Monte Carlo point-to-point analysis}
\label{ptp.sec}
\begin{figure*}
    
    \centering
    \includegraphics[width=.6\linewidth]{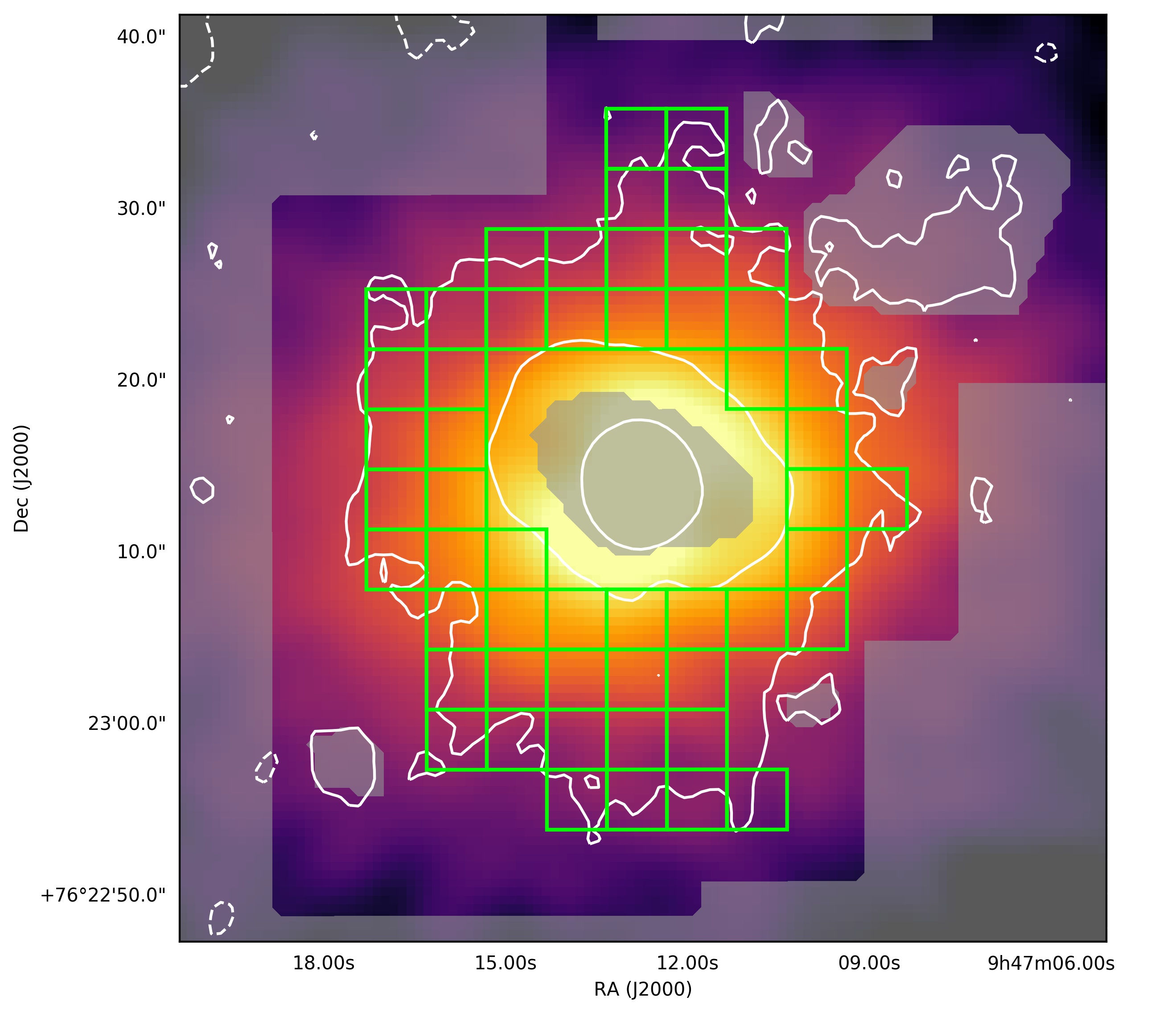}
\begin{multicols}{2}
    
    \includegraphics[width=\linewidth]{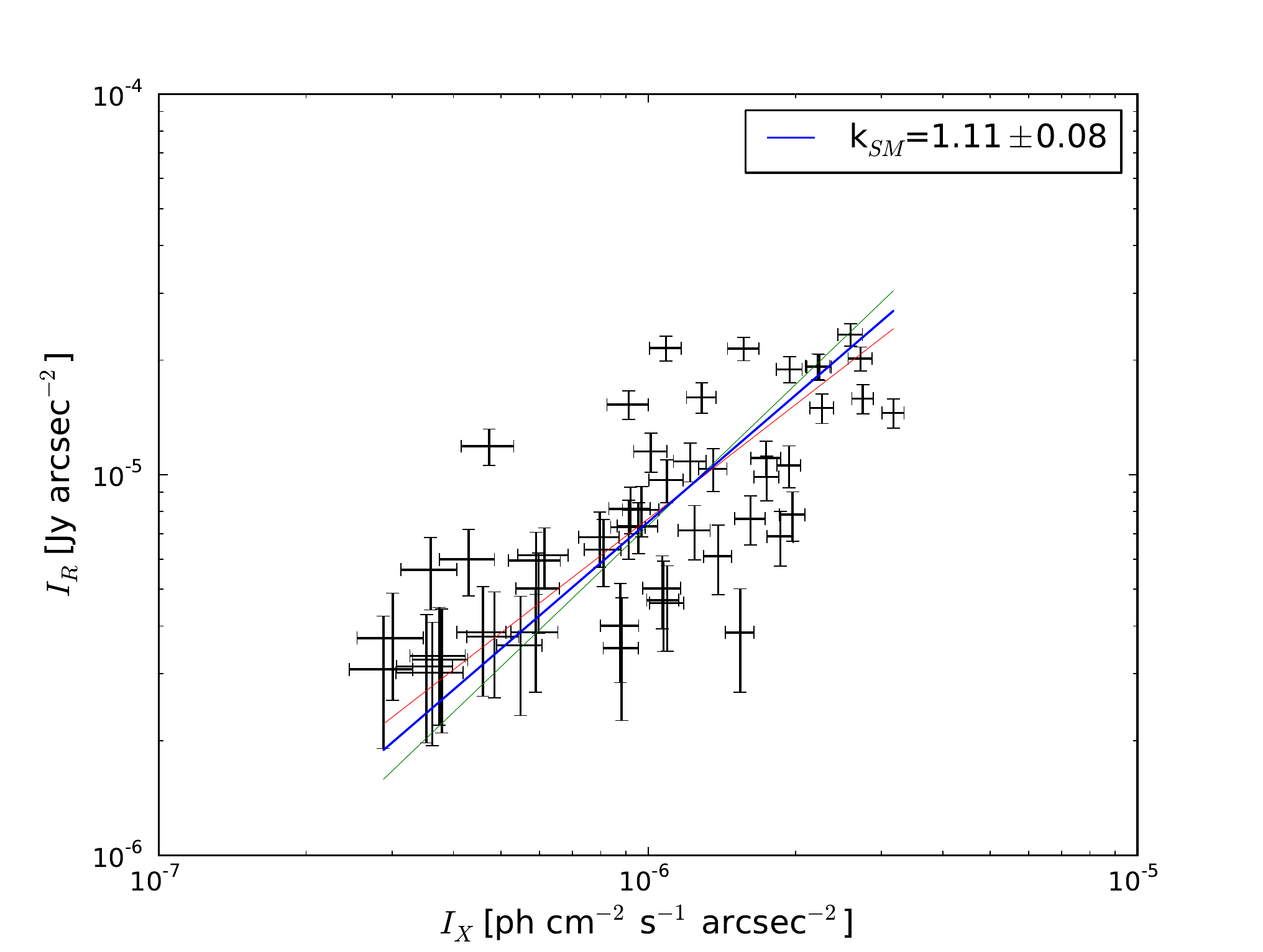}\par 
    \includegraphics[width=\linewidth]{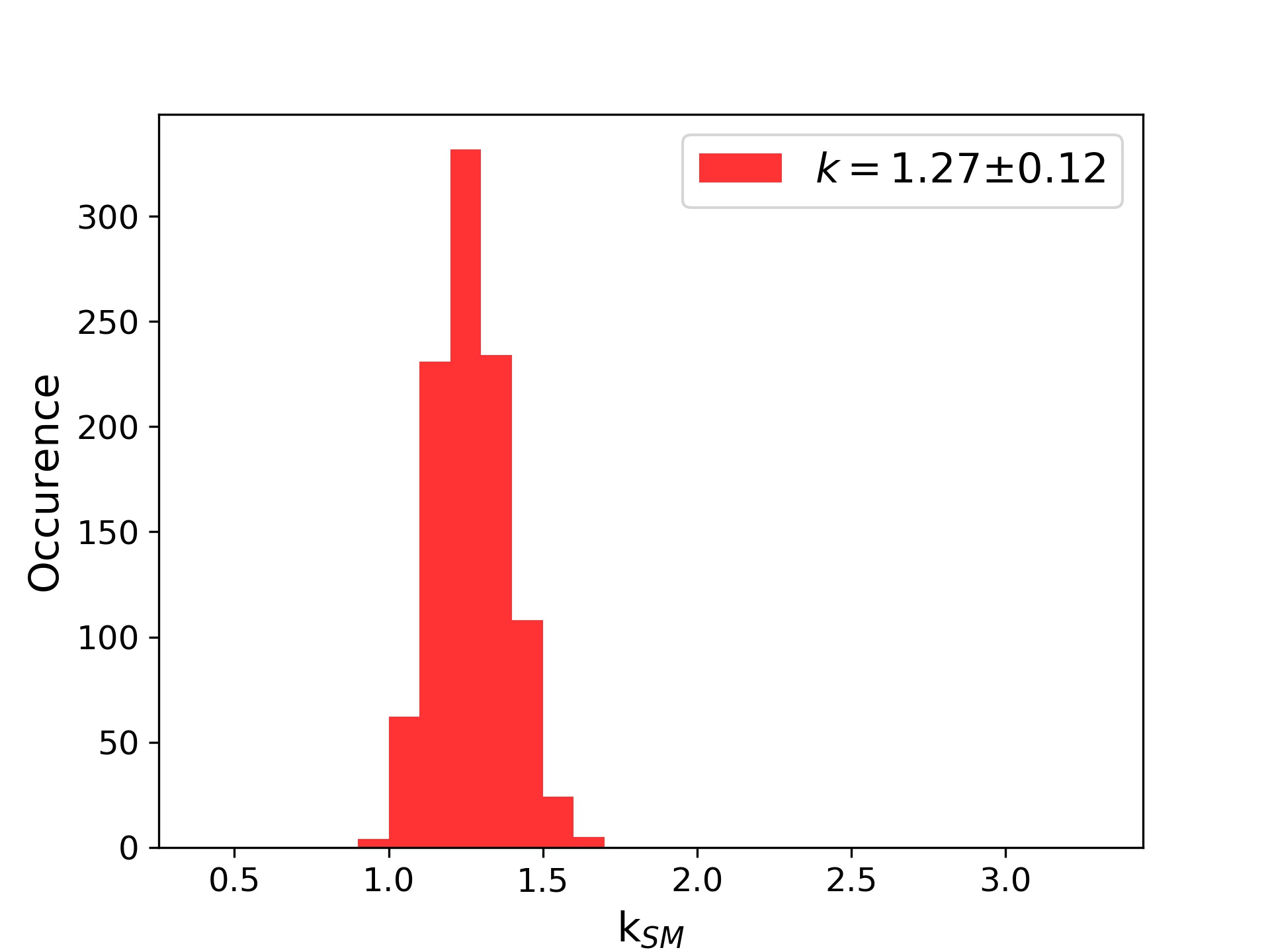}\par 
    \end{multicols}

\caption{\label{rbs797.fig} {\it Top:} Chandra image of the RBS 797 cluster with the contours at the -3, 3, 24, 96$\sigma$ levels of the radio emission at 1.4 GHz \citep[][]{Doria_2012}. The resolution of the radio maps is 3$''$x3$''$ and 1$\sigma=10$ $\mu$Jy beam$^{-1}$. Shown in green there is the final mesh that samples the emission above the 3$\sigma$ level with cells of 4$''$x4$''$ size. The central cavities and the external sources were masked (grey) and, therefore, they were excluded from the sampling ;} {\it Bottom-left:} $I_\text{R}$ vs $I_\text{X}$ plot where each point corresponds to a cell of the sampling mesh. The red and green lines are, respectively, the best-fit power-law estimated for ($I_\text{R}\mid I_\text{X}$) and ($I_\text{X}\mid I_\text{R}$). The blue line is their bisector power-law. The value of $k_\text{SM}$ is reported in the legend ; {\it Bottom-right:} Histogram of the distribution of values of $k_\text{SM}$ produced by the MCptp analysis with 1000 cycles. The best estimate of $k$ is reported in the legend with associated the 1$\sigma$ error.
\end{figure*}

 \citet[][]{Govoni_2001} performed a point-to-point analysis of the radio and X-ray emission for a sample of clusters. In their work they adopted a single grid of cells to sample the diffuse radio emission. Here we extend the single-mesh point-to-point (SMptp) analysis introduced in \citet[][]{Govoni_2001}. The case of MH is more complex than that of giant, well resolved, radio halos, because of the small number of independent beams sampling the surface brightness. Furthermore, the sampling scale that allows the maximum number of cells is the angular resolution of the image itself. However, using grids with cells as large as the beam of the image could generate biases in the analysis of the spatial correlations, because contiguous cells are not statistically independent.
For these reasons, we carried out a Monte Carlo point-to-point (MCptp) analysis. We perform several cycles of SMptp analysis with a randomly-generated mesh tailored to the diffuse radio emission for each one. Each cycle produces a different estimate of the $k$ index ($k_\text{SM}$) that we combine to obtain a more reliable estimate of the real scaling.\\
Basically, we have developed the MCptp analysis as follow:
\begin{enumerate}
\item Each cycle begins with the generation of the sampling mesh (Fig. \ref{rbs797.fig}, top panel). The grid is generated starting from a random point close to the center of the MH within a radius of $\sim1/4$ the radius of the MH. The size of the cells of the grid matches the resolution of the radio image to maximize the number of sampling points. The grid starts as a rectangular covering the whole MH, then its final shape is tailored by a given lower threshold on the radio surface brightness level and a mask provided by the user. The mask is produced by identifying the regions that are meant to be excluded from the analysis of the diffuse emission, i.e. emission related to the central galaxy or field sources. The $I_\text{R}$ is obtained from the total flux measured in every cell and, then, all those cells that cover a region of the sky previously masked or with a $I_\text{R}$ below the threshold are rejected;

\item $I_\text{R}$ and $I_\text{X}$ are measured in the cells of the final sampling grid obtained in step 1. When several Chandra observations of the same cluster are involved, we compute the total $I_\text{X}$ of a cell as:
 \begin{equation}
  I_\text{X}=\frac{\sum N_\text{cnt,i} - \sum N_\text{bkg,i} }{\sum q_\text{exp,i}}\frac{1}{\Omega_\text{c}}=\frac{\sum S_\text{X,i}\cdot q_\text{exp,i}}{\sum q_\text{exp,i}}\frac{1}{\Omega_\text{c}}
  \label{Itot.math}
 \end{equation}
where
 \begin{equation}
  S_\text{X,i}=\frac{N_\text{cnt,i}-N_\text{bkg,i}}{q_\text{exp,i}}
 \end{equation}
 is the flux measured for the $i$-th Chandra observation, $\Omega_\text{c}$ is the angular area of the cell in units of arcsec$^2$
and $N_\text{cnt,i}$, $N_\text{bkg,i}$ (in units of counts) and $q_\text{exp,i}$ (in units of counts cm$^{2}$ s photons$^{-1}$) are, respectively, the values measured on the counts, background and exposure map of the $i$-th Chandra observation. We also computed the associated error on each measure. For the $I_\text{R}$, we obtain the value of the uncertainties as the root mean square of the contribute from the noise of the map and the calibration error. We assume a calibration error of 5$\%$ of the amplitude, that is a value acceptable for both VLA and GMRT observations \citep[e.g.,][]{Chandra_2004}. On the other hand, we derive the associated errors on $S_\text{X,i}$ by assuming a Poisson error for $N_\text{cnt,i}$ and $N_\text{bkg,i}$ and computing the error propagation of Eq. \ref{Itot.math}. The cells measuring upper limits for $I_\text{R}$ or $I_\text{X}$ are excluded from the following steps;

\item We estimate the value of $k_\text{SM}$ with the BCES (Bivariate Correlated Errors and intrinsic Scatter) estimator proposed by \citet[][]{Akritas_1996}. We adopted this tool over the ordinary least squares estimator adopted in \citet[][]{Govoni_2001} because it takes into account the possible intrinsic scatter of the data respect to the power-law fit and the errors associated to each measure. 
BCES estimates both the ($I_\text{R}\mid I_\text{X}$) and ($I_\text{X}\mid I_\text{R}$) slopes and then it estimates $k_\text{SM}$, with the associated error $\sigma_\text{SM}$, as index of the bisector slope between them (Fig. \ref{rbs797.fig}, bottom-left panel).

\item  At the end of each cycle, we bootstrap a value of $k$ from a normal distribution centered at $k_\text{SM}$ with a dispersion equal to $\sigma_\text{SM}$. This procedure permits us to transpose the error of the fit in the following step;
\item Finally, the result of the MCptp analysis, $k$, is:
\begin{equation}
k=\overline{k}_\text{SM}\pm\sigma_{k_\text{SM}} 
\end{equation}
where $\overline{k}_\text{SM}$ and $\sigma_{k_\text{SM}}$ are the mean and the standard deviation of the distribution of bootstrapped $k$ obtained at the end of each cycle (Fig. \ref{rbs797.fig}, bottom-right panel).  
\end{enumerate}
We developed a Python script to perform the steps of the MCptp analysis. The code handles the radio and X-ray maps trough the NRAO radio analysis package CASA (Common Astronomy Software Applications). The results presented here were obtained by using CASA v4.7 .
\section{Results}
\label{results}
We performed the MCptp analysis with 1000 cycles on each MH of the sample. For each cluster, we set the size of the cells to match the resolution of the radio map and we set the $I_\text{R}$ minimum threshold for the brightness measured in each cell (flux/cell area) to 3$\sigma$. We excluded the region of the radio-filled cavities from the analysis for 2A0335+096, RBS 797 and RX J1532.9+3021. In Appendix \ref{images} we report an example of a random SMptp analysis for each cluster (with exception of RBS 797, that is shown in Fig. \ref{rbs797.fig}). 

We found clear evidence of a spatial correlation between radio and X-ray emission. To further corroborate this result, we run the Spearman test for each cluster, finding statistical dependence $\rho_s > 0.6$ and two-sided significance levels of deviation from zero $P_c<2\cdot10^{-2}$.
We also tested if the sampling size may affect the results of the MCptp. We tried to vary the size of the cells from the 1$\times$ beam size to 1.5$\times$beam size, finding that the increment of the cell size does not produce significant differences in the results. We limit our analysis to the band 0.5-2 keV (Sect. 2.1). We could not test the presence of the correlation at higher energies because the low count statistics limits the quality of the brightness maps.

\begin{table}
\begin{center}
  \caption{ Results of the MCptp analysis.}

\begin{tabular}{lc}
\hline
\hline
  Cluster name & $k$\\
\hline
~&\\
2A035+096 [1.4 GHz]&$1.33\pm0.23$\\
2A035+096 [5.5 GHz]&$1.01\pm0.15$\\
RBS 797&$1.27\pm0.12$\\
Abell 3444 [0.6 GHz]&$1.29\pm0.14$\\
Abell 3444 [1.4 GHz]&$1.23\pm0.14$\\
MS 1455.0+2232&$1.00\pm0.12$\\
RXC J1504.1-0248&$2.09\pm0.33$\\
RX J1532.9+3021& $1.12\pm0.17$\\
RX J1720.1+2637&$1.73\pm0.14$\\
\hline
\label{res.tab}
\end{tabular}

\end{center}
\vspace{-0.2in}
\end{table}

For the whole sample we estimate k$\geq$ 1 and for the 2 cases with radio observations at two frequencies we do not find a significant variation of $k$ with frequency (Tab. \ref{res.tab}). This is different from the sub-linear or linear scalings that are reported in the literature for giant radio halos \citep[][]{Govoni_2001,Feretti_2001,Giacintucci_2005, Vacca_2010,Hoang_2019,Xie_2020}\footnote{The only exception is 1RXS J0603.3+4214 \citep[][]{Rajpurohit_2018}. 
This difference may suggest an intrinsic difference of the origin and dynamics of the CRs in MHs and giant radio halos, although a re-analysis of the case of giant radio halos adopting our procedure is required for a more quantitative statement.}

\section{Implications for hadronic models}
\label{model}
The study of radio and X-ray brightness distribution provides important information on the origin of diffuse radio sources and on the model parameters \citep[e.g.,][]{Govoni_2001, Brunetti_2004, Pfrommer_2008, Brunetti-Jones_2014}. In this paper we focus on the hadronic model. The super-linear scaling between $I_\text{R}$ and $I_\text{X}$ for MHs suggest that the number density of emitting electrons rapidly declines from the center to the external regions.
One possibility is that CRp propagate from the central AGN and generate secondary particles, from inelastic collisions with thermal protons in the ICM, that emit the observed radio emission.
As we will show in the following, in this scenario the observed radio and X-ray spatial correlation can constrain the model parameters, including the CRp luminosity of the AGN and the magnetic fields in the MH volume.
We note, however, that steep $I_\text{R}$ profiles can be explained also by pure leptonic models \citep[e.g.,][for the MH in the Perseus cluster]{Gitti_2002}. 

\subsection{Model}
\label{eq_model}
In the context of a pure hadronic scenario, we assume the central AGN as primary source of CRp, that are injected with a rate:
\begin{equation}
 Q(p)=Q_0 p^{-\delta}
 \label{inj}
\end{equation}
where $p$ is the proton momentum for which we assumed a power-law distribution in momentum as typically assumed for CR sources in the ICM \citep[][for a theoretical discussion]{Brunetti-Jones_2014}. We assume a diffusive propagation of CR on scales much larger than the coherent scale of the magnetic field in the ICM, \citep[$>>$10 kpc, e.g.,][]{Brunetti-Jones_2014}. In this case, the time required to a particle to diffuse up to the observed MH radius, $R_\text{MH}$, is $\tau=R_\text{MH}^2/4D_0$, where $D_0$ is the spatial diffusion coefficient. In this work we assumed a diffusion coefficient $D_0$ that does to not depend on CRp energy. 
The resulting spectrum of CRp as a function of momentum, distance and time is:
\begin{equation}
N_p(p,r,t)=\frac{1}{2\pi^{3/2} r}\frac{Q(p)}{D_0}\int_{r/r_\text{max}}^{\infty}e^{-y^2}dy
\label{dens}
\end{equation}
where $r$ is the distance form the source and $r_\text{max}=\sqrt{4D_0t}$ is the distance reached by CRp in an interval of time $t$ \citep[e.g.,][]{Blasi-Colafrancesco_1999}. 
In this paper we assume the simplified case where stationary conditions for CRp distribution are established. These conditions are generated by the interplay of diffusion and injection from the central AGN and are valid under the following assumptions: 
\begin{itemize}
 \item CRp diffuse on a MH scale on a time-scale that is considerably smaller than the time-scale of the energy losses of CRp with energy $\sim$ 100 GeV \citep[$\sim10^{10}$ yrs, e.g.,][]{Brunetti-Jones_2014}; this condition selects a minimum value of the spatial diffusion coefficient (see Appendix. \ref{D0} for details);
 \item The CRp injection rate from the AGN, $L_\text{CRp}$, is fairly constant when averaged and sampled on a sufficiently long time scale (smaller than the diffusion time) of about 100 Myr or longer. We note that this includes also the possibility of a AGN duty-cycle, provided that its time-scale is considerably smaller than the diffusion time.
\end{itemize}
Assuming stationary conditions, Eq. \ref{dens} gives the stationary solution:
\begin{equation}
\label{stat}
 N_p(p,r)=\frac{1}{4 \pi r}\frac{Q(p)}{D_0}
\end{equation}
While they are diffusing over the cluster volume, CRp with kinetic energy above 300 MeV \citep[e.g.,][]{Brunetti_2017} interact with the ICM thermal protons in the cool core, for which we assumed a $\beta$-model distribution:
\begin{equation}
 n_\text{th}(r)=n_\text{0}\left[1+\left(\frac{r}{r_\text{c}}\right)^{2}\right]^{-\frac{3}{2}\beta}
 \label{therm}
\end{equation}
where $n_\text{0}$ is the central proton density, $r_\text{c}$ is the core radius and $\beta$ describes the ratio between thermal and gravitational energy of the plasma \citep[e.g.,][]{Cavaliere-Fusco_1976}.
As results of these interactions, secondary particles, i.e. $\pi^0$, positrons and electrons, are continuously injected in the cluster volume \citep[e.g.,][]{Blasi-Colafrancesco_1999,Pfrommer-Ensslin_2004, Brunetti-Blasi_2005}.

We follow the procedures in \citet[][(Sec. 3.4)]{Brunetti_2017} to calculate the injection spectrum of secondary electrons and positrons, $Q_e^{\pm}(p,t)$, and calculate the spectrum of electrons/positrons, $N_e^{\pm}(p,t)$, assuming stationary conditions:
\begin{equation}
 N_e^{\pm}(p,t)=\frac{1}{\sum_\text{rad, i}\left|\frac{\text{d}p}{\text{d}t}\right|}\int_p Q_e^{\pm}(p,t)pdt 
\end{equation}
where $\left|\text{d}p/\text{d}t\right|_\text{rad, i}$ are the radiative and Coulomb losses and:
\begin{equation}
 \begin{aligned}
Q_{e}^{\pm}(p, t)=& \frac{8 \beta_{\mu}^{\prime} m_{\pi}^{2} n_\text{th} c^{2}}{m_{\pi}^{2}-m_{\mu}^{2}} \int_{E_{\min }} \int_{p_{*}} \frac{\text{d} E_{\pi} \text{d} p}{E_{\pi} \bar{\beta}_{\mu}} \beta_{p} N_{p}(p, t) \\
& \times \frac{\text{d} \sigma^{\pm,0}}{\text{d} E}\left(E_{\pi}, E_{p}\right) F_{e}\left(E_{e}, E_{\pi}\right)
\end{aligned}
\end{equation}
where $F_{e}\left(E_{e}, E_{\pi}\right)$ is given in \citet[][]{Brunetti-Blasi_2005} (Eq. 36-37), $\bar{\beta}_{\mu}=\sqrt{1-m_\mu^2/\bar{E_\mu^2}}$, $\bar{E_\mu}=1/2E_\pi(m^2_\pi-m^2_\mu)/(\beta_{\mu}^{\prime}m_\pi^2)$, $\beta_{\mu}^{\prime}=0.2714.$ and $d\sigma^{\pm,0}/dE$ is the differential inclusive cross-section for the production of charged and neutral pions \citep[we assume the cross-section as in][]{Brunetti_2017}.
The secondary CRe injected in the ICM magnetic field, $B(r)$, can generate, in turn, synchrotron radio emission with an emissivity:
\begin{equation}
\begin{split}
 j_\text{R}(\nu,r)&=\sqrt{3}\frac{e^3}{m_\text{e}c^2}\int_0^{\pi/2}\text{sin}^2\theta\text{d}\theta\int N_e^{\pm}(p)F\left(\frac{\nu}{\nu_c}\right)\text{d}p\\
 &\propto N_p(p,r)n_\text{th}(r)\frac{B(r)^{1+\alpha}}{B(r)^2+B_\text{CMB}^2}\nu^{-\alpha}\\
 &\propto \frac{1}{4 \pi r}\frac{Q(p)}{D_0}n_\text{th}(r)\frac{B(r)^{1+\alpha}}{B(r)^2+B_\text{CMB}^2}\nu^{-\alpha}
\end{split}
\label{prof}
\end{equation}
 
where $F\left(\frac{\nu}{\nu_c}\right)$ is the synchrotron kernel \citep[e.g.][]{Rybicki-Lightman_1979}, $B_\text{CMB}=3.25(1+z)^2$ $\mu$G is the CMB equivalent magnetic field and the spectral index is $\alpha\simeq\delta/2$ \citep[][and references therein]{Brunetti_2017}. We assumed the ICM magnetic field radial profile to scale with the gas density profile, $n_\text{th}(r)$, as:
\begin{equation}
 B(r)=B_\text{0}\left[\frac{n_\text{th}(r)}{n_0}\right]^{\eta}
 \label{mag}
\end{equation}
where $\eta$ is the parameter that describes the scaling and $n_0$ is the central gas density (Eq. \ref{therm}.)\\ 

We follow the procedures in \citet[][]{Brunetti_2017} to calculate the injection spectrum of neutral pions:
\begin{equation} 
Q_{\pi}^0(E_\pi,t)=n_\text{th}c\int_{p_*}\text{d}pN_p (p,t)\beta_p \frac{d\sigma^{\pm,0}}{dE}(E_p,E_\pi)
\end{equation}
Then the $\gamma$-ray emissivity produced by the $\pi_0$ decay is:
\begin{equation}
\begin{split}
 j_{\gamma}(r)&=2\int_{E_\text{min}}^{E_{max}}\frac{Q_{\pi}^0(E_\pi,t)}{\sqrt{E_{\pi}^2-{m_\pi}^2c^4}}\text{d}E_\pi \\
 &\propto N_p(p,r)n_\text{th}(r)\\
 &\propto \frac{1}{4 \pi r}\frac{Q(p)}{D_0}n_\text{th}(r)
 \end{split}
 \label{jgamma}
\end{equation}
that produces a large-scale $\gamma$-ray halo surrounding the AGN.\\
Finally, due to the spherical symmetry of our model, radio and $\gamma$-ray emissivities can be straightforwardly converted in a surface brightness profile with the Abel transformation:
\begin{equation}
 I_\text{R,$\gamma$}(b)=\int_{b}^{+\infty}\frac{2rj_\text{R,$\gamma$}(r)}{\sqrt{r^2-b^2}}dr
 \label{abel}
\end{equation}
where $I_\text{R,$\gamma$}(b)$ is the surface brightness at the projected distance $b$ obtained by integrating the emissivity $j_\text{R,$\gamma$}(r)$ along the line of sight. In this pure, hadronic framework, from the ratio of Eq. \ref{prof} and Eq. \ref{jgamma} we can derive a relation between radio and $\gamma$-ray emission:
\begin{equation}
\frac{L_\text{$\gamma$}}{L_\text{R}}\simeq A(\alpha) < \frac{B^2+B_\text{CMB}^2}{B^{\alpha+1}} > 
\label{gamma-b}
\end{equation}
where $A(\alpha)$ is function of the spectral index and the quantities are averaged in the emitting volume. Eq. \ref{gamma-b} shows that for a source with an observed $L_\text{R}$, which is assumed to be generated only by secondary electrons, a larger (smaller) $\gamma$-ray luminosity is predicted for weaker (stronger) magnetic fields. This is simply because for weaker (stronger) magnetic fields a larger (smaller) number of secondary electrons is necessary to explain the observed radio luminosity which also implies a larger (smaller) number of CR that generate the neutral pions and the $\gamma$-rays.

\subsection{Application to a sample of MH}
\label{application}
The model presented in Sec. \ref{eq_model} is based on spherical symmetry. For this reason we select only the most roundish MHs of our sample, namely RBS 797, RXC J1504.1-0248, RX J1532.9+3021 and Abell 3444, for which we could extend our assumption of spherical symmetry. Specifically, our model depends on a set of physical parameters:
\begin{itemize}
\item CRp injection spectrum;
\item Number density of thermal targets;
\item ICM magnetic field;
\item AGN CRp luminosity.
\end{itemize}
In the hadronic framework the spectrum of CRp can be constrained from the radio spectrum of MHs as $\delta\simeq\alpha/2$. We infer the parameters that describe the distribution of thermal plasma inside the MHs ($n_0$, $\beta$ and $r_c$, see Tab. \ref{i0}) from the observed $I_\text{X}$ profile as \citep[e.g.,][]{Sarazin_1986}:
\begin{equation}
 I_\text{x}(r)=\sqrt{\pi}n_\text{0}^{2}r_\text{c}\Lambda(T)\frac{\Gamma(3\beta-0.5)}{\Gamma(3\beta)}\left[1+\left(\frac{r}{r_\text{c}}\right)^{2}\right]^{\frac{1}{2}-3\beta}
 \label{ix}
\end{equation}
where $\Lambda(T)$ is the cooling function that describes the emissivity of a plasma with a temperature $T$ that we measured from the X-ray spectra \citep{Sutherland-Dopita_1993}. 
\begin{table}
\centering
  \caption{Parameters of the $n_\text{th}$ profile estimated within $R_\text{MH}$}
\begin{tabular}{lccc}
\hline
\hline
  Cluster name & $n_0$ &$r_\text{c}$& $\beta$  \\
  &[10$^{-3}$ cm$^{-3}$]&[kpc]&\\
\hline
~&~&~\\

RBS 797&21.0&26.9&0.6\\
Abell 3444 &14.5&24.1&0.5 \\
RXC J1504.1-0248&15.9&22.2&0.5 \\
RX J1532.9+3021&21.9&23.0&0.6\\
\hline
\label{i0}
\end{tabular}
\tablefoot{From left to right: Cluster name; Central proton density; Core radius; $\beta$ index (Eq. \ref{therm}).}

\end{table}
The remaining model parameters to constrain are, thus, the AGN CRp luminosity and the magnetic field in the ICM.
In the following we describe in details the steps of our analysis. We report the results of this analysis in Tab. \ref{res_gamma}.
\subsubsection{ICM magnetic field implied by the $I_\text{R}$-$I_\text{X}$ correlation}
\label{magnetic}
\begin{figure*}
\begin{multicols}{2}
    \includegraphics[width=\linewidth]{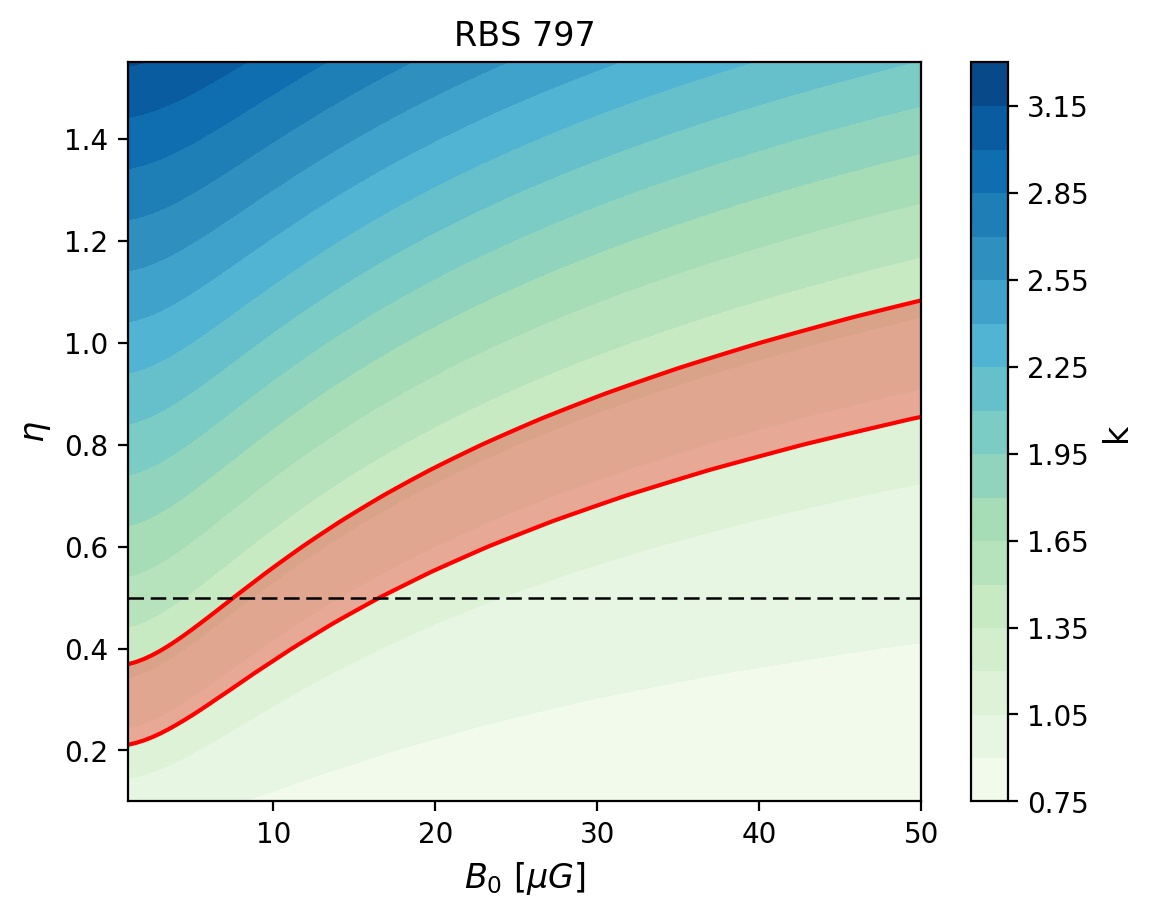}\par 
    \includegraphics[width=\linewidth]{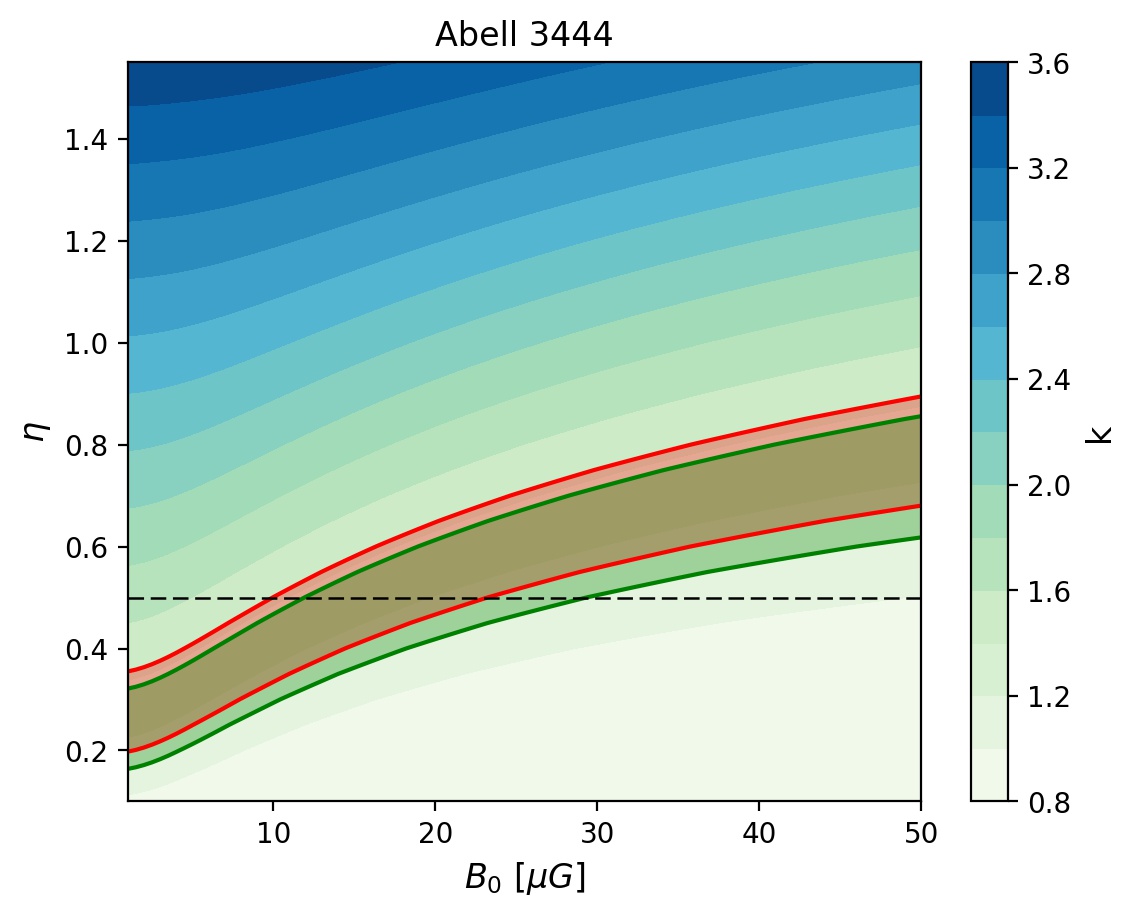}\par 
    \end{multicols}
    
    \begin{multicols}{2}
    \includegraphics[width=\linewidth]{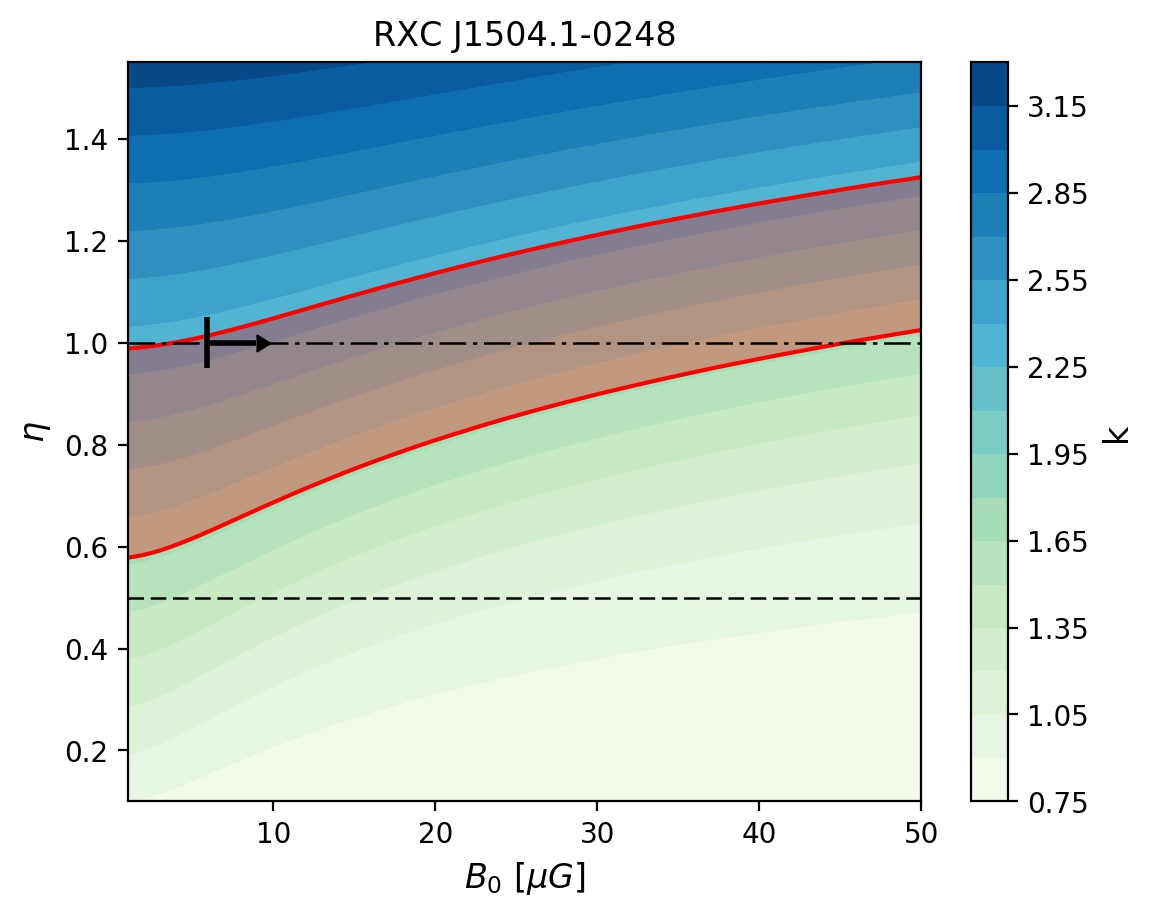}\par
    \includegraphics[width=\linewidth]{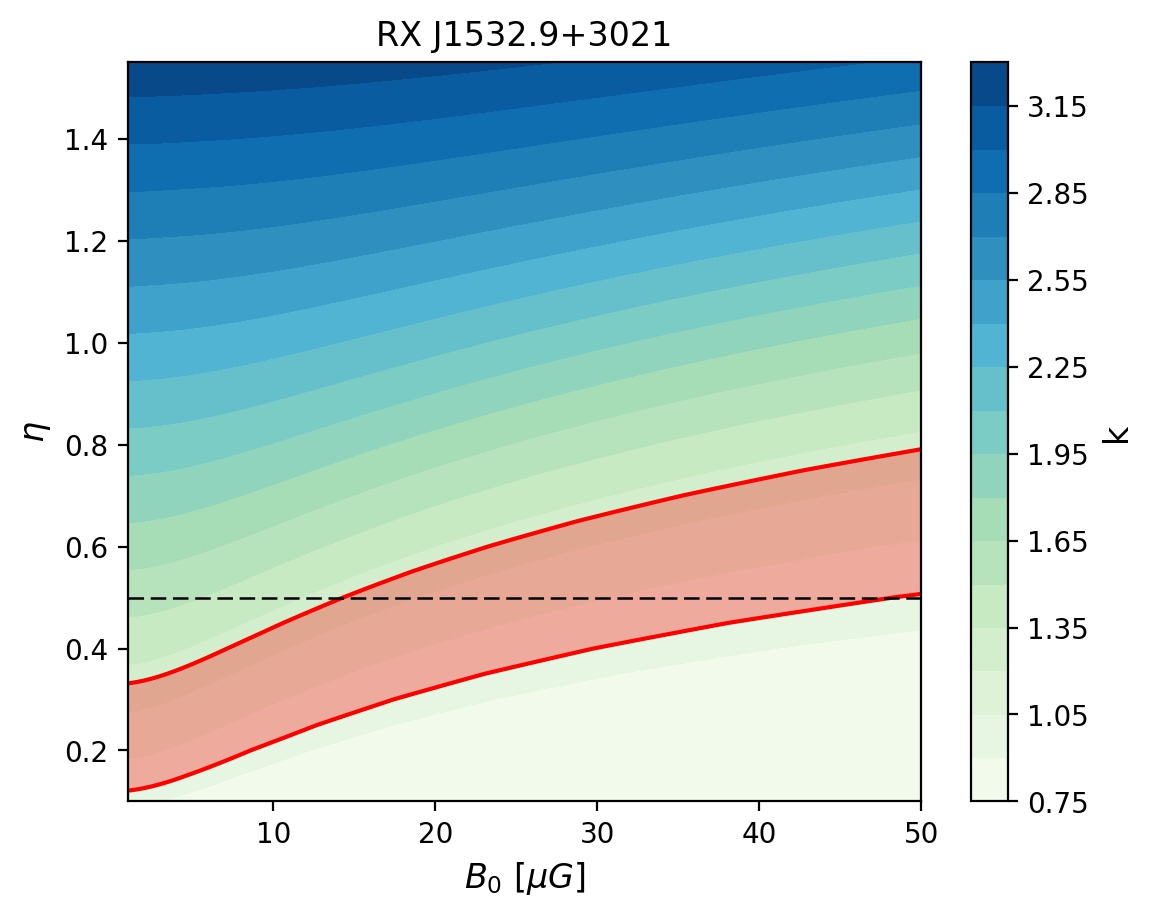}\par
\end{multicols}

\caption{\label{mag.fig} Parameter spaces of $k$-index for the spherical MHs. The red lines locate the 1-$\sigma$ confidence interval for $k$ measured for each MH. For RBS 797 we report the result obtained with $\alpha=1.1$. For Abell 3444 we report in red the result at 610 MHz and in green the result at 1.4 GHz. The horizontal, dashed line points out the level $\eta=0.5$ that reproduces the equilibrium between thermal and non-thermal energy. For RXC J1504.1-0248 we report the $\eta=1$ level with the black dash-dotted line and the lower limit derived from Fermi-LAT observation presented in \citet[][]{Dutson_2013}. The lower-limits for $B_0$ for the other clusters are below 1 $\mu$G and they are reported in Tab. \ref{B0}. }
\end{figure*}
In the case of spherical symmetry, the radial profile of the ratio $I_\text{R}/I_\text{X}$ depends on the magnetic field model (Eq. \ref{prof}, \ref{mag}, \ref{abel} and \ref{gamma-b}). 
Specifically, the values of the index $k$ in Sect. \ref{results} constrain a range of values for the couples $B_0$-$\eta$.  \\

Therefore, for each cluster, we calculated numerically the $I_\text{R}$ within the MH radius (Tab. \ref{phys.tab}) by testing a wide range of combinations $B_0$-$\eta$, then we compared them with the observed $I_\text{X}$ to estimate the corresponding $k$-index. In Fig. \ref{mag.fig} we report the numerical estimates compared with the observed $k$ for each cluster. For RBS 797 the spectral index of the diffuse emission was not measured unambiguously \citep[][]{Doria_2012}, therefore we tested two extreme possibilities, $\alpha=1.1$ ($\delta=2.2$) and $\alpha=1.5$ ($\delta=3.0$).\\
We found that for a given value of $k$, larger values of $B_0$ are obtained for larger values of $\eta$.
As a reference value, we assumed  $\eta=0.5$, which is the case where magnetic field energy scales linearly with thermal energy for isothermal ICM.
Under this assumption, we constrain central values of the magnetic field of 11.8$\pm$4.8 and 18.5$\pm$5.5 $\mu$G for RBS 797 assuming $\alpha$=1.1 and $\alpha=1.5$, 18.8$\pm$5.5 $\mu$G for Abell 3444, and we derive a lower limit of 14.5 $\mu$G for RX J1532.9+3021. \\
A value $\eta=0.5$ is inconsistent with the case of RXC J1504.1-0248, whose steep scaling ($k\simeq2$) is reproduced only by a peaked spatial distribution of the magnetic field ($0.6<\eta<1.3$). On the one hand, by assuming a steeper profile for the magnetic field ($\eta=1$), we estimate a central magnetic field $B_0=20.0\pm18.5$ $\mu$G. On the other hand, assuming $\eta$=0.5 would require a CRp density radial profile steeper than the $\propto 1/r$ profile of the stationary solution (Eq. \ref{stat}) to produce a final $I_\text{R}$ as peaked in the center as the observed one. This case would imply a more complicated situation, including (1) a non-constant CRp luminosity of the central AGN showing a significant enhancement across the duty cycle or (2) that the diffusion time that is necessary to CRp to cover the MH scale is longer than (i) AGN activity time scale, and/or (ii) energy losses of CRp.

\subsubsection{AGN CRp luminosity}
\label{dc}
Once the scaling between the magnetic field and the thermal density is constrained by the observed scaling between $I_\text{R}$ and $I_\text{X}$, we can derive the CRp luminosity of the central AGN that is required to substain the observed radio luminosity of MHs.  
The AGN luminosity, $L_\text{CRp}$, is:
\begin{equation}
 L_\text{CRp}=\int_{p_\text{0.2 GeV}} Q_0 p^{-\delta}\sqrt{c^2p^2+m_\text{p}^2c^4}dp %
 \label{lum}
\end{equation}
where $m_\text{p}$ is the proton mass and $p_\text{0.2 GeV}$ is the momentum for which the kinetic energy, $pc$, is 0.2 GeV. We note that for $\delta<3$ the exact choice of the minimum energy is not relevant for the final result.\\

In order to obtain the value of $Q_0$ to compute $L_\text{CRp}$ (Eq. \ref{lum}) we matched the observations with the $I_\text{R}$ profiles predicted by our model. We estimated the synchrotron emissivity numericcaly with Eq. \ref{prof} by following the formalism presented in Sec. \ref{eq_model} and by assuming the $B(r)$ configurations that we constrained in Sec. \ref{magnetic}. The radio emission depends on the ratio $Q_0/D_0$ (Eq. \ref{prof}), therefore we estimated $Q_0$, and thus $L_\text{CRp}$, by assuming a diffusion coefficient $D_0^{1 Gyr}$ for which the diffusion time of CRp in the MH is $\tau=R_\text{MH}^2/4D_0$=1 Gyr. This implies an optimistic diffusion coefficient and consequently a upper bound to the $L_\text{CRp}$ that is required by the model. In Appendix \ref{D0} we discuss the consequences of different assumptions, including the scenario of total dissipation of CRp within $R_\text{MH}$, that entails the lower bound for $L_\text{CRp}$ in our model.
\begin{figure*}
 \includegraphics[width=\textwidth]{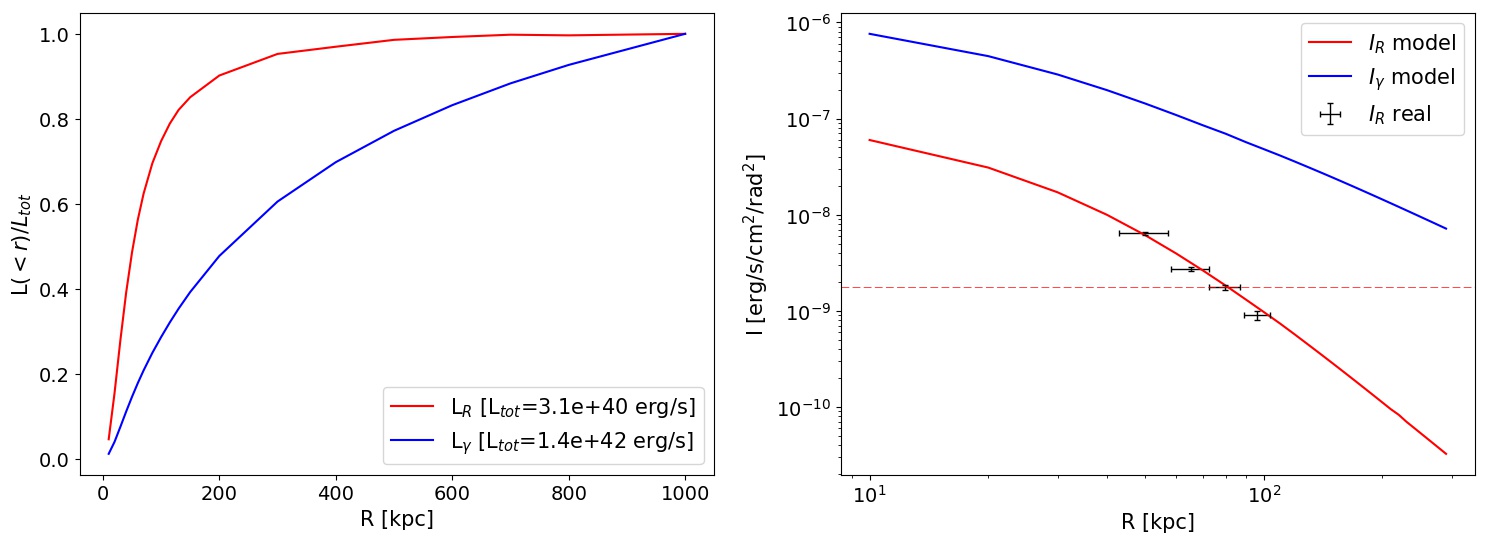}
 \includegraphics[width=\textwidth]{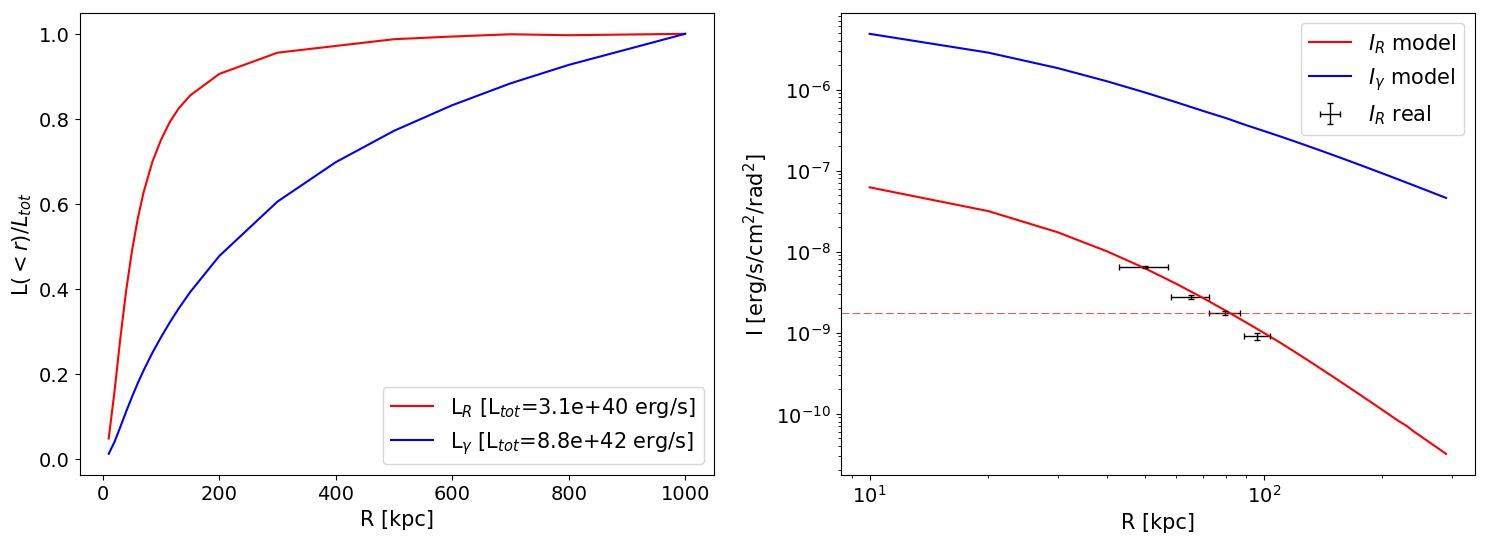}
 \caption{\label{prof_RBS}Results for RBS 797 for $\delta$=2.2 (top) and $\delta$=3.0 (bottom). {\it Left:} Integrated radio and $\gamma$-ray luminosity; {\it Right: } radio and $\gamma$-ray surface brightness. We report the observed $I_\text{R}$ profile and the 3$\sigma$ level of the observation (dashed red line).} 
\end{figure*}
\begin{figure*}
 \includegraphics[width=\textwidth]{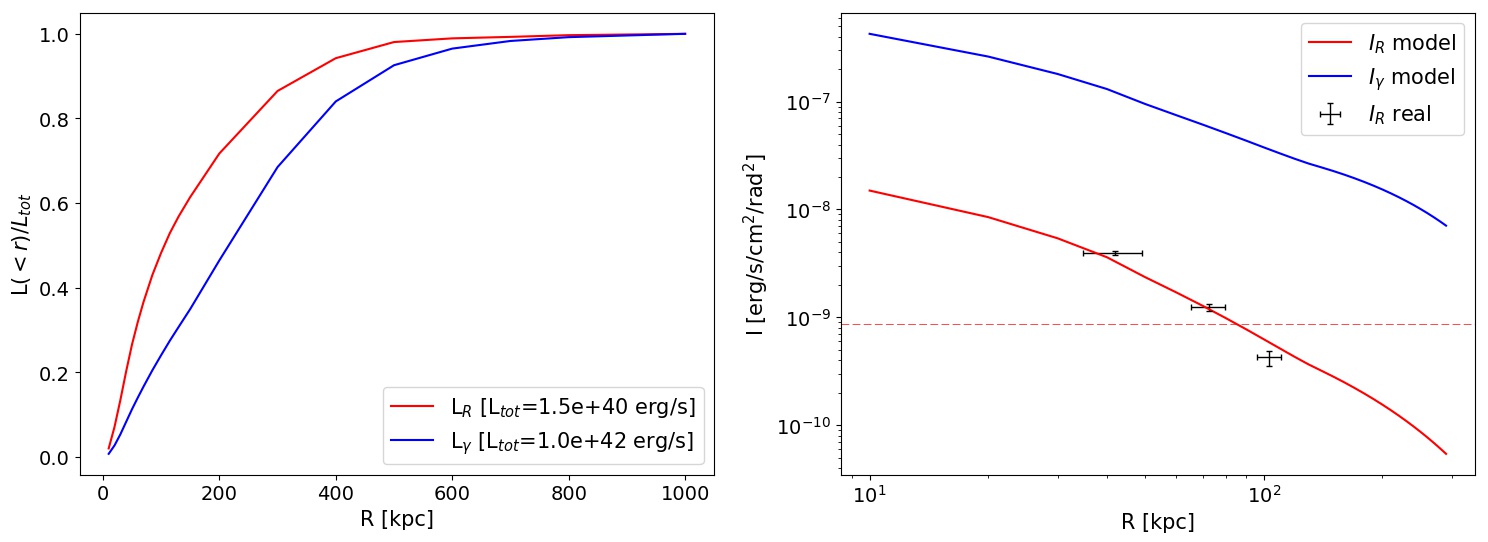}
 \includegraphics[width=\textwidth]{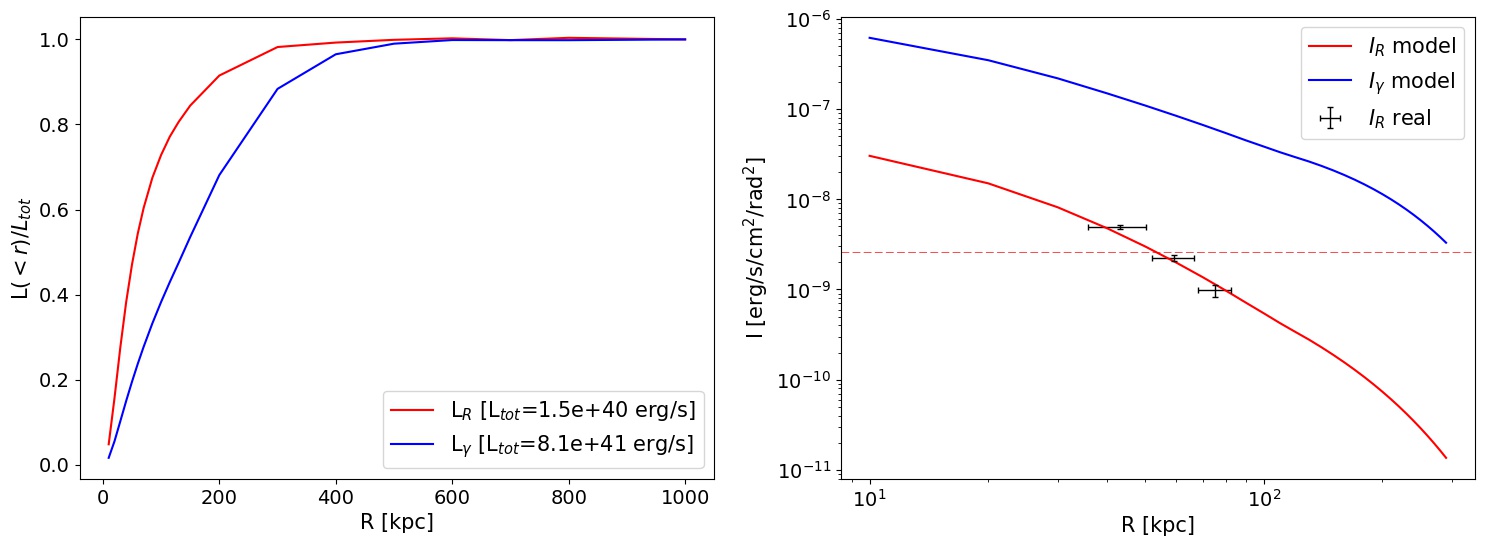}
 \caption{\label{prof_altri}Results for Abell 3444 at 1.4 GHz (top) and RX J1532.9+3021 (bottom). For the latter, we report reference values of $L_\text{R}$ and $L_{\gamma}$ derived from the upper limit of the magnetic field (Tab. \ref{res_gamma}). {\it Left:} Integrated radio and $\gamma$-ray luminosity; {\it Right: } radio and $\gamma$-ray surface brightness. We report the observed $I_\text{R}$ profile and the 3$\sigma$ level of the observation (dashed red line).} 
\end{figure*}
\begin{figure*}
 \includegraphics[width=\textwidth]{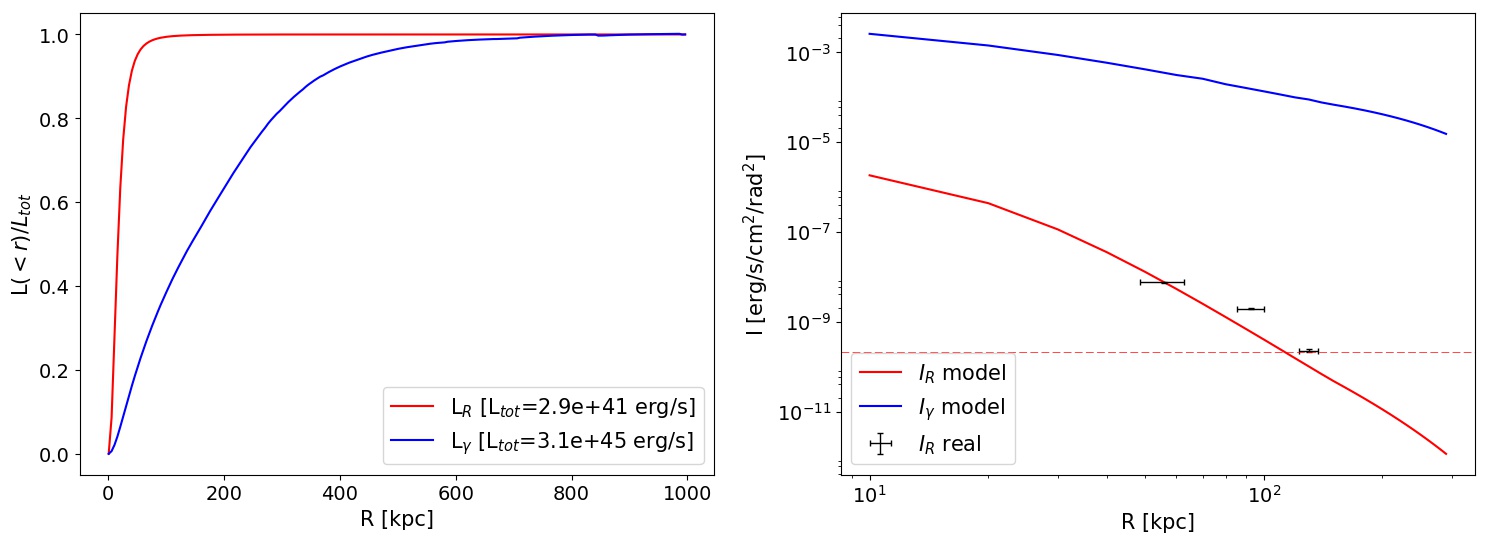}
 \includegraphics[width=\textwidth]{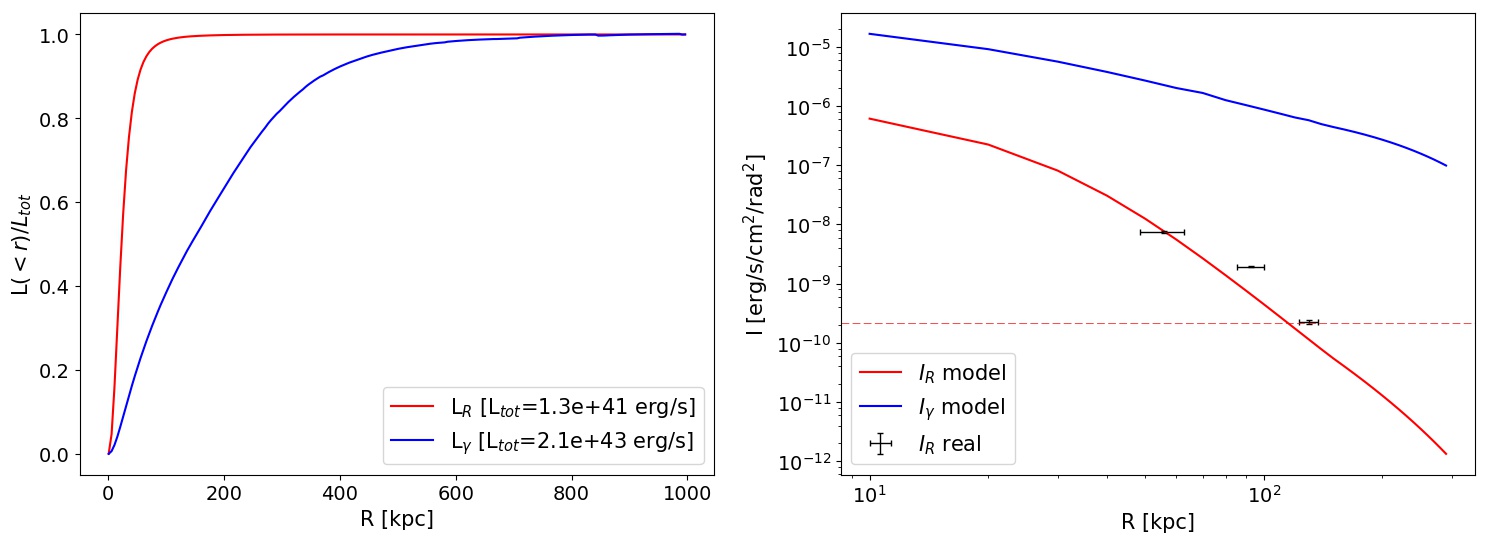}
 \includegraphics[width=\textwidth]{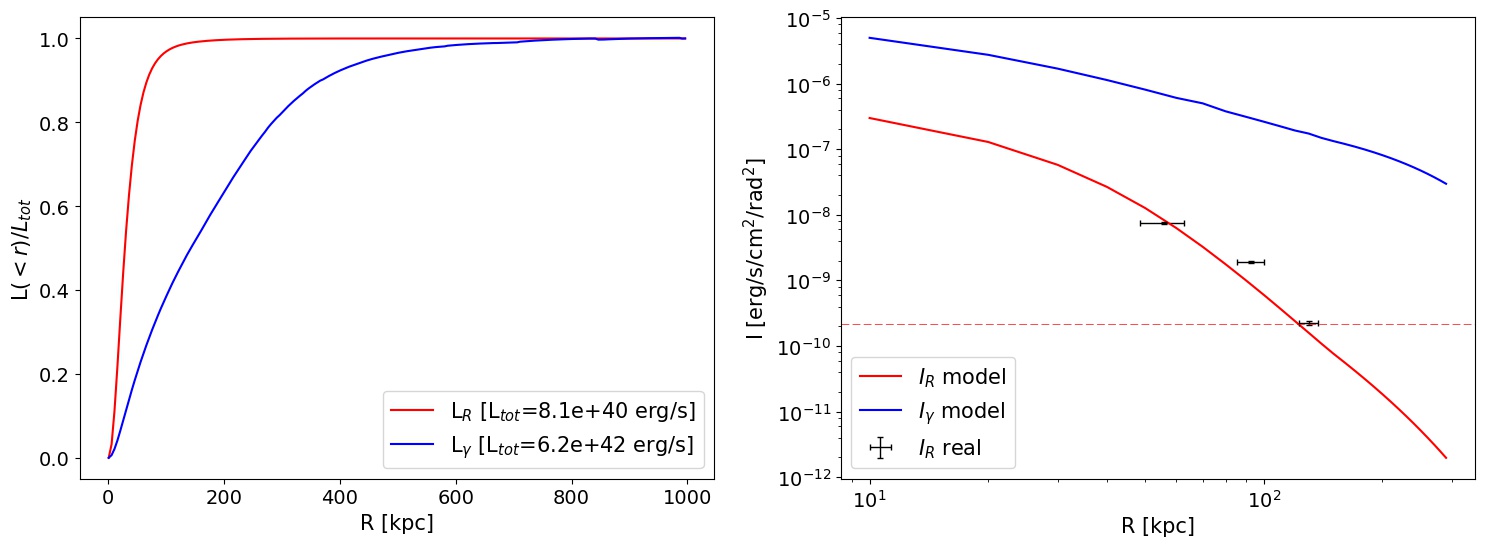}
 \caption{\label{prof_RXC}Results for RXC J1504.1-0248 with $B_0=1$ $\mu$G (top), $B_0=10$ $\mu$G (middle) and $B_0=20$ $\mu$G (bottom). {\it Left:} Integrated radio and $\gamma$-ray luminosity; {\it Right: } radio and $\gamma$-ray surface brightness. We report the observed $I_\text{R}$ profile and the 3$\sigma$ level of the observation (dashed red line). We limited the profile up to $R_\text{MH}$} to avoid the possible contamination by field sources (see B.4). 
\end{figure*}
Finally we compared them with the observed $I_\text{R}$ profiles measured in circular bins with the same resolution of the radio maps. We report in Tab. \ref{res_gamma} the parameters adopted and the results obtained. We estimate that the AGN $L_\text{CRp}$ required in our model to reproduce the observed radio emission is 10$^{44}$-10$^{46}$ erg s$^{-1}$ (see Appendix \ref{D0}). In Fig. \ref{prof_RBS}, \ref{prof_altri} and \ref{prof_RXC} we report, for each cluster, the integrated radio luminosity and the surface brightness radial profiles, predicted and observed, at the observed frequency.

\begin{table*}
\begin{center}
  \caption{ \label{res_gamma} Parameters of the hadronic model }
\begin{tabular}{lcccccccc}
\hline
\hline
  Cluster name & $D_0^\text{1 Gyr}$ [cm$^2$ s$^{-1}$]&R$_\text{$\gamma$}$ [kpc]&$\delta$ &$B_0$ [$\mu$G]&$\eta$ &$Q_0/D_0$& $L_\text{CRp}^{>\text{0.2 GeV}}$ [erg s$^{-1}$]&$S_\text{$\gamma$}^{>1\text{Gev}}$ [erg s$^{-1}$ cm$^{-2}$]\\
\hline
~&~&~\\
RBS 797&6.9$\cdot10^{29}$&650&2.2&11.8$\pm$4.8&0.5&1.7&1.5$\cdot10^{44}$&1.3$\cdot10^{-14}$\\
&&&3.0&18.5$\pm$5.5&0.5&1.6$\cdot10^{-9}$&8.1$\cdot10^{45}$&2.1$\cdot10^{-14}$\\
~&~&~\\
Abell 3444&1.4$\cdot10^{30}$&400&2.6&18.5$\pm$5.5&0.5&1.1$\cdot10^{-5}$&4.2$\cdot10^{44}$&7.3$\cdot10^{-15}$\\

RXC J1504.1-0248&1.5$\cdot10^{30}$&320&2.5&20.0$\pm$18.5&1.0&4.4$\cdot10^{-3}$&4.2$\cdot10^{45}$&9.4$\cdot10^{-14}$\\
&&&&10.0&1.0&1.5$\cdot10^{-2}$&1.4$\cdot10^{46}$&3.3$\cdot10^{-13}$\\
&&&&1.0&1.0&2.2&$2.0\cdot10^{48}$&$4.7\cdot10^{-11}$\\
~&~&~\\
RX J1532.9+3021&1.1$\cdot10^{30}$&300&2.4&$>14.5$&0.5&7.1$\cdot10^{-3}$&$<1.2\cdot10^{45}$&$<5.2\cdot10^{-15}$\\
\hline
\end{tabular}
\tablefoot{From left to right: Cluster name; Diffusion coefficient for which CRp reach $R_\text{MH}$ in 1 Gyr; Radius in which $85\%$ of $\gamma$-rays are emitted; Index of the CRp injecion spectrum (Eq. \ref{inj}); Central magnetic field (Eq. \ref{mag}); Slope between the ICM magnetic field and the thermal plasma (Eq. \ref{mag}); Normalization of non-thermal emissivity, in units of $\left[\left(\frac{\text{g cm}}{\text{s}}\right)^{\delta}\text{cm}^{-2}\right]$ (Eq. \ref{inj}); AGN CRp luminosity ; $\gamma$-ray flux expected from the total emitting region inside $R_{\gamma}$.}
\end{center}
\vspace{-0.2in}

\end{table*}

\subsection{Resulting $\gamma$-rays emission and comparison with current and future observations}
\label{gamma}
In the previous Sections we used the $I_\text{R}$-$I_\text{X}$ scaling to derive constraints on the model parameters.
In this Section we check if the $\gamma$-ray fluxes are consistent with current observational limits. We calculated numerically the $\gamma$-ray emission produced by the $\pi_0$ decay described in Sec. \ref{eq_model} within the same physical boundaries adopted in Sec. \ref{dc} by implementing numerically Eq. \ref{jgamma}.
However, the approximation of thermal density adopted to reproduce the MH volume (Eq. \ref{therm}) could extrapolate incorrectly the thermal density beyond $R_\text{MH}$. Therefore, to calculate the total $\gamma$-ray luminosity we used a double $\beta$-model to better describe the radial decline of the thermal gas density beyond $R_\text{MH}$. \\
 
In Tab. \ref{res_gamma} we report the radius containing the 85$\%$ of the $\gamma$-ray emission and the $\gamma$-ray flux. In Fig. \ref{prof_RBS}, \ref{prof_altri} and \ref{prof_RXC} are reported the integrated $\gamma$-ray luminosity and surface brightness radial profiles at 1 GeV. The radio and $\gamma$-ray halos differ in terms of size, because the $j_\text{R}$ (Eq. \ref{prof}) declines faster than $j_{\gamma}$ (Eq. \ref{jgamma}). According to our results, the $\gamma$-ray halos extend beyond the cooling region, which contains instead almost the totality of the radio emission. In the case of RXC J1504.1-0248, the value of $B_0$ is poorly constrained from the analysis presented in Sec. \ref{magnetic} ($B_0$=20.0$\pm$18.5 $\mu$G). Therefore, for this cluster we compute $L_\text{CRp}$ and the $\gamma$-ray emission assuming three values of the central magnetic field, namely $B_0=$1.0, 10.0, 20.0 $\mu$G. We note that assuming $B_0>20$ $\mu$G would produce results close to the case $B_0=20.0$ $\mu$G (Eq. \ref{gamma-b}).\\

\begin{figure}
\centering
 \includegraphics[width=.51\textwidth]{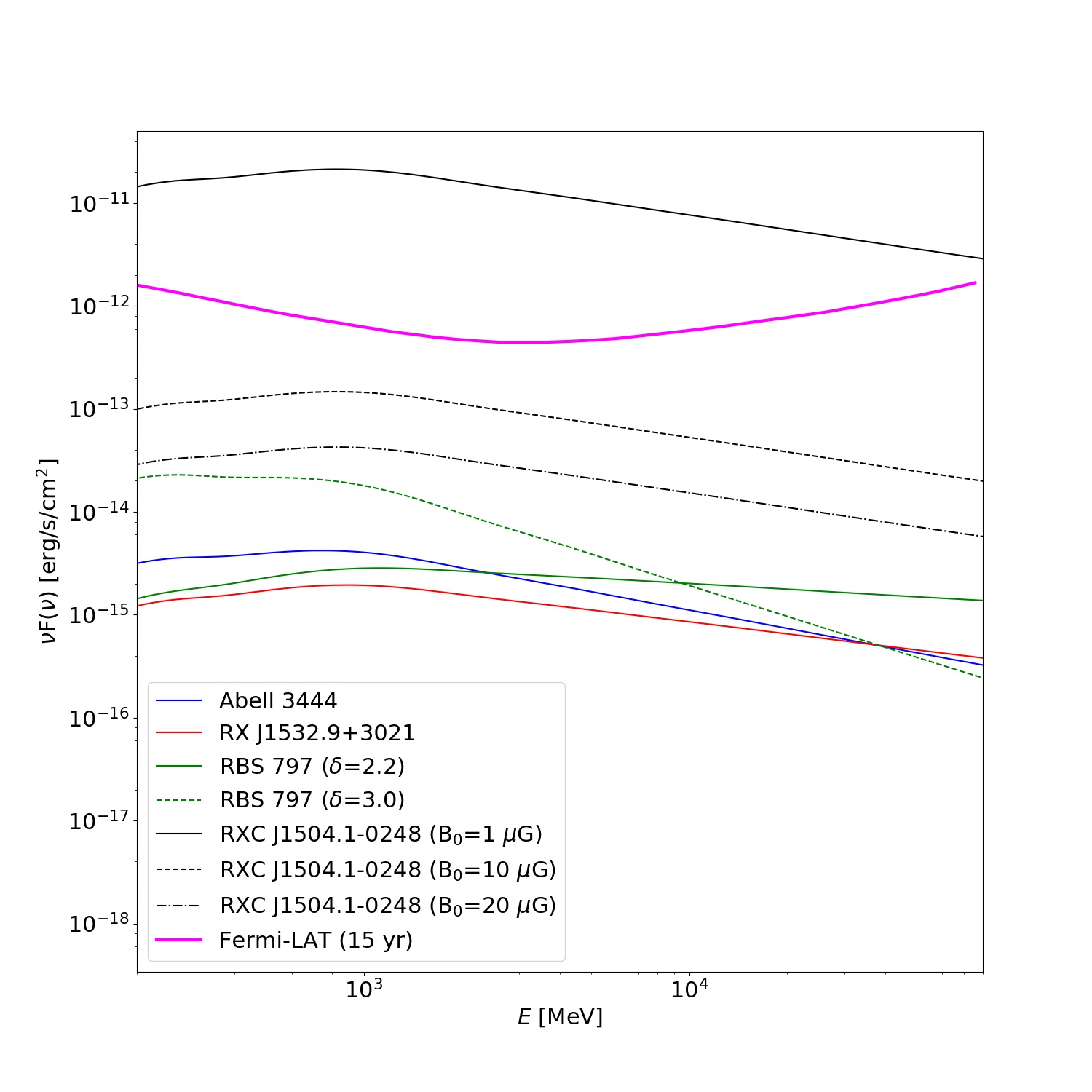}
 \caption{\label{spettri_gamma} $\gamma$-rays spectrum for the MHs for parameters reported in Tab. \ref{res_gamma} compared with the Fermi-LAT 15 yrs detection limit.}
\end{figure}

Diffuse $\gamma$-ray emission from galaxy clusters has never been firmly detected, so we tested the consistency of our model with the observational constraints.
This is shown in Fig. \ref{spettri_gamma}, where we compare the expected $\gamma$-ray emission, computed with the parameters reported in Tab. \ref{res_gamma}, with the Fermi-LAT detection limit after 15 years. 
In general, we find that the $\gamma$-ray fluxes predicted for the four MHs are below the Fermi-LAT detection limit, hence our model constrained from the $I_\text{R}$-$I_\text{X}$ scaling does not violate the current non-detection of diffuse $\gamma$-ray emission. The Fermi-LAT detection limits can also be used to infer complementary limits to the central magnetic field $B_0$ in our model, because in our model fainter magnetic field will result in stronger $\gamma$-ray emission (Eq. \ref{gamma-b}). For RBS 797, Abell 3444 and RX J1532.9+3021 we used the results of the Fermi-LAT 15yrs observations as constraints, whereas for RXC J1504.1-0248 we used the limit obtained by \citet[][]{Dutson_2013} with the Fermi-LAT. We report in Tab. \ref{B0} the limits inferred for the configurations of ICM magnetic field and $L_\text{CRp}$ constrained by our results. Fermi-LAT detection limit provides lower limits below 1 $\mu$G for RBS 797, Abell 3444 and RX J1532.9+3021. On the contrary, for RXC J1504.1-0248 the lower limit is 5.9 $\mu$G, due to the higher $\gamma$-ray emissivity predicted by our model.

\begin{table}
\centering
  \caption{Fermi lower-limits for the central magnetic field $B_0$}
\begin{tabular}{lcc}
\hline
\hline
  Cluster name & $\eta$ & $B_\text{0,min}$  \\
  &&[$\mu$G]\\
\hline
~&~&~\\

RBS 797&0.3&0.5\\
$''$&0.5&0.7\\
Abell 3444 &0.3&0.8 \\
$''$ &0.5&0.8 \\
RXC J1504.1-0248&1.0&5.9 \\
RX J1532.9+3021&0.3&0.5\\
$''$& 0.5&0.8\\
\hline
\label{B0}
\end{tabular}
\tablefoot{From left to right: Cluster name; Magnetic field configuration (see Eq. \ref{mag}); Lowest central magnetic field $B_0$ allowed by Fermi detection limit.}
\end{table}

\section{Discussion and Summary}
\label{discussion}
In this work, for the first time, we have carried out a systematic study of the spatial connection between thermal and non-thermal ICM components in relaxed clusters.
\subsection{New scaling relation for MH}
 The most important result of our paper comes from the study of the spatial correlation between non-thermal radio and thermal X-ray brightness for a sample of seven MHs. We extended the strategy based on a single grid, which has been applied to giant radio halos, by including a Monte Carlo approach in the generation of the grid. This approach allows us to avoid the biases generated by the relatively small ($\sim$20-30) number of independent radio beams that cover the emission of MH (after excluding the regions contaminated by discrete sources). We found evidence of a spatial correlation between $I_\text{R}$ and $I_\text{X}$, where the radio emission is generally more peaked than the thermal emission, thus indicating that the ICM non-thermal components are more concentrated around the central AGN. Our result further confirms the connection between thermal and non-thermal ICM components in MHs, that has already been claimed by previous works that studied the correlations between radio and X-ray luminosity \citep[][]{Gitti_2015,Bravi_2016,Gitti_2018,Giacintucci_2019} and the morphological connection between cold fronts and MHs \citep[][]{Mazzotta-Giacintucci_2008}. Furthermore, the values of $k$ that we measure for MHs differ from the case of radio halos reported in the literature, where a sub-linear or linear scaling is generally found \citep[][]{Govoni_2001, Feretti_2001,Giacintucci_2005, Vacca_2010,Hoang_2019}. This may suggest an intrinsic difference in the nature of these radio sources. The steep decline of the radio emission in MH also suggests that secondary particles injected at the center by the AGN play a role, both directly or as seed particles re-accelerated by other mechanisms. 
\subsection{Comparison with hadronic models}

We consider a simple, reference, hadronic model based on the injection of CRp by the central AGN. The model is a pure hadronic scenario without including the effect of re-acceleration. Furthermore, we assume that the combination of CRp injection and diffusion generates stationary conditions in the ICM. We note that this simple scenario connecting the AGN activity and the MH has been already proposed to explain the origin of the Perseus MH \citep[e.g.,][]{Boehringer_1988,Pfrommer-Ensslin_2004} and, in general, to evaluate the effect of CRp-driven streaming instability on the heating of the cool cores and the connection with the formation of MH \citep[e.g.,][]{Fujita_2013,Jacob-Pfrommer_2017a, Ehlert_2018}.
We compare the observed scalings between $I_\text{R}$ and $I_\text{X}$ with our simple model to infer combined constraints on the AGN CRp luminosity and the ICM magnetic field. We selected the four MHs of our sample with a roundish shape for which it is possible to assume spherical symmetry in 3D. In this case, constraints deriving from point-to-point correlations are similar to those deriving from the azimuthally averaged brightness profile.   
Specifically, we derive $B_0$ in the range 10-40 $\mu$G assuming $\eta=0.5$, where smaller values of $B_0$ would require smaller values of $\eta$, and a $L_\text{CRp}=10^{44}$-$10^{46}(D_0/D_0^\text{1 Gyr})$ erg s$^{-1}$, where we assumed a $D_0^\text{1 Gyr}$ that allows the diffusion of CRp in the MH volume in 1 Gyr. 
We stress that, although these results are based on the assumption of stationary conditions, they are obtained by sampling spatial scales of a few 100 kpc and consequently they do not depend significantly on local variations. As a consequence, we expect that only a strong violation of stationary conditions can affect our conclusions.\\

The values of $B_0$ that we found are consistent, although slightly larger than the values reported by \citet[][]{Carilli_2002} for relaxed clusters. An independent observational test of pure hadronic models, where the AGN plays the main role for the injection of the primary CRp, would thus result from detailed studies of Faraday rotation measure and depolarization of the discrete radio sources embedded in the cluster core or in background \citep[e.g.,][]{Bonafede_2011}.

\subsubsection{$\gamma$-ray emission}
The unavoidable consequence of a hadronic scenario is the production of $\gamma$-rays with a luminosity that is close to the CRp luminosity of the AGN, where the $\gamma$-ray luminosity depends on the model parameters. In Sec. \ref{gamma} we calculate the $\gamma$-ray emission under the assumption of the parameters reported in Tab. \ref{res_gamma}. The expectations do not violate Fermi upper limits (Fig. \ref{spettri_gamma}). We found that smaller $B_0$ produce a larger $\gamma$-ray luminosity, whereas larger $D_0$ entail larger $L_\text{CRp}$ and fulfill the stationary conditions for the CRp distribution in shorter time-scales. The Fermi-LAT detection limit allowed us to provide a lower limit for the central magnetic field that we report in Tab. \ref{B0}. The  $I_\text{$\gamma$}$ profile is broader than the $I_\text{R}$ profile (Fig. \ref{prof_RBS}, \ref{prof_altri} \ref{prof_RXC}), although for the parameters used in Sec. \ref{application} we find that the radius where 85$\%$ of the emission is produced is larger than the core radius of the cluster. 

\subsubsection{Observation tests and limitations}
The size of MH predicted by our model depends on our assumptions on the diffusion coefficient (see Sec. \ref{D0}).
Large diffusion coefficients produce MHs that extend beyond the radius measured by current observations (Fig. \ref{prof_RBS}, \ref{prof_altri} and \ref{prof_RXC}). Deeper observations of the MHs of our sample will allow testing if the emission can extend on larger scales or is more confined, for example within the region defined by cold fronts \citep[][]{Mazzotta-Giacintucci_2008}. These tests will allow understanding whether additional mechanisms, e.g. turbulent re-acceleration, are necessary to explain observations \citep[][]{ZuHone_2013}.\\

One of the main caveats in our analysis is the assumption of stationary conditions. On the one hand, these are justified by the fact that in a cooling time the CRp can diffuse on scales similar or larger than that of MHs. On the other hand, the duty cycles of AGN last for $10^{7}-10^{8}$ years \citep[][]{Morganti_2017} and, consequently, the MHs would be powered by numerous bursts of injection of CRp, whose diffusion scale is $r_\text{max}=\sqrt{4D_0t}$, with $t$ being the look-back time for the single burst. Consequently, stationary conditions also imply that the phase of the interplay between the AGN and MH is much longer than a single burst of activity of the AGN and, thus, that the MH results from the integrated effect of many bursts/AGN active phases. However, if the system has recently experienced an unusually powerful AGN activity (injection of CRp), the resulting spatial distribution of CR would be steeper than in our approximation. This would also result from a scenario where the $L_\text{CRp}$ of the AGN active phases decreases with look-back time. The case of RX1504 where, indeed, a very steep trend between radio and X-ray brightness is found, might suggest that the system had a very strong CRp activity in the last 100 Myr or so.

\subsection{Future prospects}

The Monte Carlo point-to-point analysis and BCES fitting procedure presented in this work can be extended to a larger sample of targets, including both mini and giant radio halos, to confirm the different behavior that we observe. Moreover, the estimates provided by our model could work as constraints for future theoretical work aimed to address the connection between AGN feedback and cooling flow quenching. Our results suggest that simple hadronic models can still match the main observations of MH. Further studies are now required to address the implications of secondary production of electrons in presence of ICM turbulence, and the implications of interplay with the leptonic models in general in the origin of the diffuse radio emission.
The incoming Athena X-ray observatory will play a crucial role in these studies by providing an unprecedented spectral resolution. In particular, by combining radio images and Athena X-ray Integral Field Unit \citep[X-IFU,][]{Barcons_2017} observations in a point-to-point analysis, we will be able to explore the spatial connection between the energy of CRe and the ICM turbulence.\\

The study of radio emission in galaxy clusters will greatly benefit from the present and new radio observatories, as LOFAR and SKA, that could potentially observe thousands of new sources. Interestingly, LOFAR observations are already showing that relaxed clusters can host diffuse, ultra-steep spectrum emission extended far beyond the sloshing region, suggesting a more complex scenario involving "gentle" CRe re-acceleration due to ICM turbulence on large scales \citep[][]{Savini_2018,Savini_2019}. In this case, follow-up studies of point-to-point brightness distribution based on our approach open to the possibility to discriminate the contribution of hadronic collisions (pure hadronic or re-accelerated secondaries) from that of turbulent re-acceleration of primary seeds, because the two regimes should produce different scalings. By probing the steep spectrum emission on larger scales at lower frequencies, we may expect to observe a flattening of the radio and X-ray scaling, similar to what is observed in giant radio halos. 

\section*{Acknowledgments}
We thank the Referee for their comments that improved the presentation of the work. 
AI thanks M. Sereno for the useful discussion. Basic research in radio astronomy at the Naval Research Laboratory is supported by 6.1 Base funding. This research made use of APLpy, an open-source plotting package for Python \citep[][]{Robitaille_2012}

\vspace{-0.15in}
\bibliographystyle{aa}
\bibliography{bibliography}

\begin{thebibliography}{59}
\expandafter\ifx\csname natexlab\endcsname\relax\def\natexlab#1{#1}\fi

\bibitem[{{Akritas} \& {Bershady}(1996)}]{Akritas_1996}
{Akritas}, M.~G. \& {Bershady}, M.~A. 1996, \apj, 470, 706

\bibitem[{{Barcons} {et~al.}(2017){Barcons}, {Barret}, {Decourchelle}, {den
  Herder}, {Fabian}, {Matsumoto}, {Lumb}, {Nandra}, {Piro}, {Smith}, \&
  {Willingale}}]{Barcons_2017}
{Barcons}, X., {Barret}, D., {Decourchelle}, A., {et~al.} 2017, Astronomische
  Nachrichten, 338, 153

\bibitem[{{Blasi} \& {Colafrancesco}(1999)}]{Blasi-Colafrancesco_1999}
{Blasi}, P. \& {Colafrancesco}, S. 1999, Astroparticle Physics, 12, 169

\bibitem[{{Boehringer} \& {Morfill}(1988)}]{Boehringer_1988}
{Boehringer}, H. \& {Morfill}, G.~E. 1988, The Astrophysical Journal, 330, 609

\bibitem[{{Bonafede} {et~al.}(2011){Bonafede}, {Govoni}, {Feretti}, {Murgia},
  {Giovannini}, \& {Br{\"u}ggen}}]{Bonafede_2011}
{Bonafede}, A., {Govoni}, F., {Feretti}, L., {et~al.} 2011, \aap, 530, A24

\bibitem[{{Bravi} {et~al.}(2016){Bravi}, {Gitti}, \& {Brunetti}}]{Bravi_2016}
{Bravi}, L., {Gitti}, M., \& {Brunetti}, G. 2016, \mnras, 455, L41

\bibitem[{{Brunetti}(2004)}]{Brunetti_2004}
{Brunetti}, G. 2004, in IAU Colloq. 195: Outskirts of Galaxy Clusters: Intense
  Life in the Suburbs, ed. A.~{Diaferio}, 148--154

\bibitem[{{Brunetti} \& {Blasi}(2005)}]{Brunetti-Blasi_2005}
{Brunetti}, G. \& {Blasi}, P. 2005, \mnras, 363, 1173

\bibitem[{{Brunetti} \& {Jones}(2014)}]{Brunetti-Jones_2014}
{Brunetti}, G. \& {Jones}, T.~W. 2014, International Journal of Modern Physics
  D, 23, 30007

\bibitem[{{Brunetti} {et~al.}(2017){Brunetti}, {Zimmer}, \&
  {Zandanel}}]{Brunetti_2017}
{Brunetti}, G., {Zimmer}, S., \& {Zandanel}, F. 2017, \mnras, 472, 1506

\bibitem[{{Carilli} \& {Taylor}(2002)}]{Carilli_2002}
{Carilli}, C.~L. \& {Taylor}, G.~B. 2002, \araa, 40, 319

\bibitem[{{Cassano} {et~al.}(2008){Cassano}, {Gitti}, \&
  {Brunetti}}]{Cassano-Gitti_2008}
{Cassano}, R., {Gitti}, M., \& {Brunetti}, G. 2008, \aap, 486, L31

\bibitem[{{Cavaliere} \& {Fusco-Femiano}(1976)}]{Cavaliere-Fusco_1976}
{Cavaliere}, A. \& {Fusco-Femiano}, R. 1976, \aap, 49, 137

\bibitem[{{Chandra} {et~al.}(2004){Chandra}, {Ray}, \&
  {Bhatnagar}}]{Chandra_2004}
{Chandra}, P., {Ray}, A., \& {Bhatnagar}, S. 2004, \apj, 612, 974

\bibitem[{{Donnert} {et~al.}(2010){Donnert}, {Dolag}, {Brunetti}, {Cassano}, \&
  {Bonafede}}]{Donnert_2010}
{Donnert}, J., {Dolag}, K., {Brunetti}, G., {Cassano}, R., \& {Bonafede}, A.
  2010, \mnras, 401, 47

\bibitem[{{Doria} {et~al.}(2012){Doria}, {Gitti}, {Ettori}, {Brighenti},
  {Nulsen}, \& {McNamara}}]{Doria_2012}
{Doria}, A., {Gitti}, M., {Ettori}, S., {et~al.} 2012, \apj, 753, 47

\bibitem[{{Dutson} {et~al.}(2013){Dutson}, {White}, {Edge}, {Hinton}, \&
  {Hogan}}]{Dutson_2013}
{Dutson}, K.~L., {White}, R.~J., {Edge}, A.~C., {Hinton}, J.~A., \& {Hogan},
  M.~T. 2013, \mnras, 429, 2069

\bibitem[{{Ehlert} {et~al.}(2018){Ehlert}, {Weinberger}, {Pfrommer}, {Pakmor},
  \& {Springel}}]{Ehlert_2018}
{Ehlert}, K., {Weinberger}, R., {Pfrommer}, C., {Pakmor}, R., \& {Springel}, V.
  2018, \mnras, 481, 2878

\bibitem[{{Feretti} {et~al.}(2001){Feretti}, {Fusco-Femiano}, {Giovannini}, \&
  {Govoni}}]{Feretti_2001}
{Feretti}, L., {Fusco-Femiano}, R., {Giovannini}, G., \& {Govoni}, F. 2001,
  \aap, 373, 106

\bibitem[{{Fujita} \& {Ohira}(2013)}]{Fujita_2013}
{Fujita}, Y. \& {Ohira}, Y. 2013, Monthly Notices of the Royal Astronomical
  Society, 428, 599

\bibitem[{{Giacintucci} {et~al.}(2011){Giacintucci}, {Markevitch}, {Brunetti},
  {Cassano}, \& {Venturi}}]{Giacintucci_2011a}
{Giacintucci}, S., {Markevitch}, M., {Brunetti}, G., {Cassano}, R., \&
  {Venturi}, T. 2011, \aap, 525, L10+

\bibitem[{{Giacintucci} {et~al.}(2014{\natexlab{a}}){Giacintucci},
  {Markevitch}, {Brunetti}, {ZuHone}, {Venturi}, {Mazzotta}, \&
  {Bourdin}}]{Giacintucci_2014b}
{Giacintucci}, S., {Markevitch}, M., {Brunetti}, G., {et~al.}
  2014{\natexlab{a}}, \apj, 795, 73

\bibitem[{{Giacintucci} {et~al.}(2017){Giacintucci}, {Markevitch}, {Cassano},
  {Venturi}, {Clarke}, \& {Brunetti}}]{Giacintucci_2017}
{Giacintucci}, S., {Markevitch}, M., {Cassano}, R., {et~al.} 2017, \apj, 841,
  71

\bibitem[{{Giacintucci} {et~al.}(2019){Giacintucci}, {Markevitch}, {Cassano},
  {Venturi}, {Clarke}, {Kale}, \& {Cuciti}}]{Giacintucci_2019}
{Giacintucci}, S., {Markevitch}, M., {Cassano}, R., {et~al.} 2019, \apj, 880,
  70

\bibitem[{{Giacintucci} {et~al.}(2014{\natexlab{b}}){Giacintucci},
  {Markevitch}, {Venturi}, {Clarke}, {Cassano}, \&
  {Mazzotta}}]{Giacintucci_2014a}
{Giacintucci}, S., {Markevitch}, M., {Venturi}, T., {et~al.}
  2014{\natexlab{b}}, \apj, 781, 9

\bibitem[{{Giacintucci} {et~al.}(2005){Giacintucci}, {Venturi}, {Brunetti},
  {Bardelli}, {Dallacasa}, {Ettori}, {Finoguenov}, {Rao}, \&
  {Zucca}}]{Giacintucci_2005}
{Giacintucci}, S., {Venturi}, T., {Brunetti}, G., {et~al.} 2005, \aap, 440, 867

\bibitem[{{Gitti} {et~al.}(2012){Gitti}, {Brighenti}, \&
  {McNamara}}]{Gitti_2012}
{Gitti}, M., {Brighenti}, F., \& {McNamara}, B.~R. 2012, Advances in Astronomy,
  2012

\bibitem[{{Gitti} {et~al.}(2018){Gitti}, {Brunetti}, {Cassano}, \&
  {Ettori}}]{Gitti_2018}
{Gitti}, M., {Brunetti}, G., {Cassano}, R., \& {Ettori}, S. 2018, \aap, 617,
  A11

\bibitem[{{Gitti} {et~al.}(2002){Gitti}, {Brunetti}, \& {Setti}}]{Gitti_2002}
{Gitti}, M., {Brunetti}, G., \& {Setti}, G. 2002, \aap, 386, 456

\bibitem[{{Gitti} {et~al.}(2006){Gitti}, {Feretti}, \&
  {Schindler}}]{Gitti_2006}
{Gitti}, M., {Feretti}, L., \& {Schindler}, S. 2006, \aap, 448, 853

\bibitem[{{Gitti} {et~al.}(2013){Gitti}, {Giroletti}, {Giovannini}, {Feretti},
  \& {Liuzzo}}]{Gitti_2013a}
{Gitti}, M., {Giroletti}, M., {Giovannini}, G., {Feretti}, L., \& {Liuzzo}, E.
  2013, \aap, 557, L14

\bibitem[{{Gitti} {et~al.}(2015){Gitti}, {Tozzi}, {Brunetti}, {Cassano},
  {Dallacasa}, {Edge}, {Ettori}, {Feretti}, {Ferrari}, {Giacintucci},
  {Giovannini}, {Hogan}, \& {Venturi}}]{Gitti_2015}
{Gitti}, M., {Tozzi}, P., {Brunetti}, G., {et~al.} 2015, in proceedings of
  {"}Advancing Astrophysics with the Square Kilometre Array{"},
  PoS(AASKA14)076, 76

\bibitem[{{Govoni} {et~al.}(2001){Govoni}, {En{\ss}lin}, {Feretti}, \&
  {Giovannini}}]{Govoni_2001}
{Govoni}, F., {En{\ss}lin}, T.~A., {Feretti}, L., \& {Giovannini}, G. 2001,
  \aap, 369, 441

\bibitem[{{Hlavacek-Larrondo} {et~al.}(2013){Hlavacek-Larrondo}, {Allen},
  {Taylor}, {Fabian}, {Canning}, {Werner}, {Sanders}, {Grimes}, {Ehlert}, \&
  {von der Linden}}]{Hlavacek_2013}
{Hlavacek-Larrondo}, J., {Allen}, S.~W., {Taylor}, G.~B., {et~al.} 2013, \apj,
  777

\bibitem[{{Hoang} {et~al.}(2019){Hoang}, {Shimwell}, {van Weeren}, {Brunetti},
  {R{\"o}ttgering}, {Andrade-Santos}, {Botteon}, {Br{\"u}ggen}, {Cassano},
  {Drabent}, {de Gasperin}, {Hoeft}, {Intema}, {Rafferty}, {Shweta}, \&
  {Stroe}}]{Hoang_2019}
{Hoang}, D.~N., {Shimwell}, T.~W., {van Weeren}, R.~J., {et~al.} 2019, \aap,
  622, A20

\bibitem[{{Jacob} \& {Pfrommer}(2017)}]{Jacob-Pfrommer_2017a}
{Jacob}, S. \& {Pfrommer}, C. 2017, \mnras, 467, 1449

\bibitem[{{Mazzotta} {et~al.}(2003){Mazzotta}, {Edge}, \&
  {Markevitch}}]{Mazzotta_2003}
{Mazzotta}, P., {Edge}, A.~C., \& {Markevitch}, M. 2003, \apj, 596, 190

\bibitem[{{Mazzotta} \& {Giacintucci}(2008)}]{Mazzotta-Giacintucci_2008}
{Mazzotta}, P. \& {Giacintucci}, S. 2008, \apjl, 675, L9

\bibitem[{{Mazzotta} {et~al.}(2001){Mazzotta}, {Markevitch}, {Vikhlinin},
  {Forman}, {David}, \& {van Speybroeck}}]{Mazzotta_2001a}
{Mazzotta}, P., {Markevitch}, M., {Vikhlinin}, A., {et~al.} 2001, \apj, 555,
  205

\bibitem[{{McNamara} \& {Nulsen}(2012)}]{McNamara-Nulsen_2012}
{McNamara}, B.~R. \& {Nulsen}, P.~E.~J. 2012, New Journal of Physics, 14,
  055023

\bibitem[{{Morganti}(2017)}]{Morganti_2017}
{Morganti}, R. 2017, Nature Astronomy, 1, 596

\bibitem[{{Murgia} {et~al.}(2009){Murgia}, {Govoni}, {Markevitch}, {Feretti},
  {Giovannini}, {Taylor}, \& {Carretti}}]{Murgia_2009}
{Murgia}, M., {Govoni}, F., {Markevitch}, M., {et~al.} 2009, \aap, 499, 679

\bibitem[{{Pfrommer}(2008)}]{Pfrommer_2008}
{Pfrommer}, C. 2008, \mnras, 385, 1242

\bibitem[{{Pfrommer} \& {En{\ss}lin}(2004)}]{Pfrommer-Ensslin_2004}
{Pfrommer}, C. \& {En{\ss}lin}, T.~A. 2004, \aap, 413, 17

\bibitem[{{Rajpurohit} {et~al.}(2018){Rajpurohit}, {Hoeft}, {van Weeren},
  {Rudnick}, {R{\"o}ttgering}, {Forman}, {Br{\"u}ggen}, {Croston},
  {Andrade-Santos}, \& {Dawson}}]{Rajpurohit_2018}
{Rajpurohit}, K., {Hoeft}, M., {van Weeren}, R.~J., {et~al.} 2018, \apj, 852,
  65

\bibitem[{{Robitaille} \& {Bressert}(2012)}]{Robitaille_2012}
{Robitaille}, T. \& {Bressert}, E. 2012, {APLpy: Astronomical Plotting Library
  in Python}

\bibitem[{{Rybicki} \& {Lightman}(1979)}]{Rybicki-Lightman_1979}
{Rybicki}, G.~B. \& {Lightman}, A.~P. 1979, {Radiative processes in
  astrophysics}, ed. {Rybicki, G.~B.~\& Lightman, A.~P.}

\bibitem[{{Sanders} {et~al.}(2009){Sanders}, {Fabian}, \&
  {Taylor}}]{Sanders_2009b}
{Sanders}, J.~S., {Fabian}, A.~C., \& {Taylor}, G.~B. 2009, \mnras, 396, 1449

\bibitem[{{Sarazin}(1986)}]{Sarazin_1986}
{Sarazin}, C.~L. 1986, Reviews of Modern Physics, 58, 1

\bibitem[{{Sarazin} {et~al.}(1995){Sarazin}, {Baum}, \& {O'Dea}}]{Sarazin_1995}
{Sarazin}, C.~L., {Baum}, S.~A., \& {O'Dea}, C.~P. 1995, \apj, 451, 125

\bibitem[{{Savini} {et~al.}(2019){Savini}, {Bonafede}, {Br{\"u}ggen},
  {Rafferty}, {Shimwell}, {Botteon}, {Brunetti}, {Intema}, {Wilber}, \&
  {Cassano}}]{Savini_2019}
{Savini}, F., {Bonafede}, A., {Br{\"u}ggen}, M., {et~al.} 2019, \aap, 622, A24

\bibitem[{{Savini} {et~al.}(2018){Savini}, {Bonafede}, {Br{\"u}ggen}, {van
  Weeren}, {Brunetti}, {Intema}, {Botteon}, {Shimwell}, {Wilber}, {Rafferty},
  {Giacintucci}, {Cassano}, {Cuciti}, {de Gasperin}, {R{\"o}ttgering}, {Hoeft},
  \& {White}}]{Savini_2018}
{Savini}, F., {Bonafede}, A., {Br{\"u}ggen}, M., {et~al.} 2018, \mnras, 478,
  2234

\bibitem[{{Sutherland} \& {Dopita}(1993)}]{Sutherland-Dopita_1993}
{Sutherland}, R.~S. \& {Dopita}, M.~A. 1993, \apjs, 88, 253

\bibitem[{{Vacca} {et~al.}(2010){Vacca}, {Murgia}, {Govoni}, {Feretti},
  {Giovannini}, {Orr{\`u}}, \& {Bonafede}}]{Vacca_2010}
{Vacca}, V., {Murgia}, M., {Govoni}, F., {et~al.} 2010, \aap, 514, A71

\bibitem[{{van Weeren} {et~al.}(2019){van Weeren}, {de Gasperin}, {Akamatsu},
  {Br{\"u}ggen}, {Feretti}, {Kang}, {Stroe}, \& {Zandanel}}]{VanWeeren_2019}
{van Weeren}, R.~J., {de Gasperin}, F., {Akamatsu}, H., {et~al.} 2019, \ssr,
  215, 16

\bibitem[{{Venturi} {et~al.}(2007){Venturi}, {Giacintucci}, {Brunetti},
  {Cassano}, {Bardelli}, {Dallacasa}, \& {Setti}}]{Venturi_2007}
{Venturi}, T., {Giacintucci}, S., {Brunetti}, G., {et~al.} 2007, \aap, 463, 937

\bibitem[{{Xie} {et~al.}(2020){Xie}, {van Weeren}, {Lovisari},
  {Andrade-Santos}, {Botteon}, {Br{\"u}ggen}, {Bulbul}, {Churazov}, {Clarke},
  {Forman}, {Intema}, {Jones}, {Kraft}, {Lal}, {Mroczkowski}, \&
  {Zitrin}}]{Xie_2020}
{Xie}, C., {van Weeren}, R.~J., {Lovisari}, L., {et~al.} 2020, arXiv e-prints,
  arXiv:2001.04725

\bibitem[{{ZuHone} {et~al.}(2015){ZuHone}, {Brunetti}, {Giacintucci}, \&
  {Markevitch}}]{ZuHone_2015b}
{ZuHone}, J.~A., {Brunetti}, G., {Giacintucci}, S., \& {Markevitch}, M. 2015,
  \apj, 801, 146

\bibitem[{{ZuHone} {et~al.}(2013){ZuHone}, {Markevitch}, {Brunetti}, \&
  {Giacintucci}}]{ZuHone_2013}
{ZuHone}, J.~A., {Markevitch}, M., {Brunetti}, G., \& {Giacintucci}, S. 2013,
  \apj, 762, 78

\end{thebibliography}

\appendix
\section{Cluster sample}
\label{descr}
 Here we present a brief morphological description of each cluster of the sample.
\begin{itemize}
 \item{\it 2A0335+096:} The MH was first imaged at 1.4 GHz and 5.5 GHz by \citet[][]{Sarazin_1995}. The MH morphology in our images, obtained from the same radio data \citep[see][for details]{Giacintucci_2019}, is consistent with the structure previously mapped. The central radio galaxy is a core-dominated, double-lobe source and another patch of extended emission, which is interpreted as a fossil lobe from an older AGN outburst, is detected at $\sim 25 ''$($\sim$18 kpc) from the cluster center. The MH surrounds this structure extending for $\sim100"$ ($\sim$ 70 kpc). 
 In the X-ray band we observe two cavities, which coincide with the radio lobes, and a cold front located at $\sim$40 kpc from the center. The region inside the cold front shows a number of small, dense gas blobs that may be the shred of a cooling core disturbed by either Kelvin-Helmotz instabilities or intermittent AGN activity. All of these properties relate to processes that may act to disrupt or destroy any cooling flow \citep[][]{Mazzotta_2003, Sanders_2009b}. The cluster hosts a head-tail radio galaxy whose radio tail is close to the MH with a projected distance of $\sim90$ kpc. This suggests the possibility that the close-by passage of the galaxy may have played a role in the injection of both CRe and turbulence in the ICM;

 \item{\it RBS 797:} This cluster shows radio emission on three different scales. VLA observations at 4.8 GHz at high resolutions ($\sim0.4''$) revealed the presence of a pair of jets connected to the BCG, oriented to the north-south direction and extended for $\sim$15 kpc. On the larger scale, the radio emission observed at 1.4 GHz coincides with a striking system of cavities extended for $\sim$26 kpc in the east-west direction observed by Chandra in the X-ray band. The misalignment of the cavities with respect to the inner jet system suggests that the central AGN had different cycle of activities with the jets oriented in different directions \citep[][]{Gitti_2006}. Finally, the cluster shows diffuse radio emission with a rough-spherical morphology and a radius of $\sim$100 kpc \citep[][]{Gitti_2006,Doria_2012}. We excluded the region of the cavities from the analysis of the MH;
  \item{\it Abell 3444:} The MH was reported first by \citet[][]{Venturi_2007} and then confirmed in \citet[][]{Giacintucci_2019}. The BCG at the center of the low-entropy cool core does not show jets (Giacintucci et al., in prep.). The morphology of the radio emission seems orthogonal to the X-ray emission, with the $I_\text{R}$ decreasing rapidly toward east;

 \item{\it MS 1455.0+2232:} The MH is composed by a central region and a tail located at south-east. The northern part of the MH is delimited by a cold front. The cluster, along with RX J1720.1+2637, has been reported by \citet[][]{Mazzotta-Giacintucci_2008} as a first evidence of the connection between cold fronts and MHs;
 
 \item{\it RXC J1504.1-0248:}  This cluster is characterized by an extreme X-ray luminosity ($L_\text{bol}=4.3\cdot 10^{45}$ $h_{70}^{-1}$ erg s$^{-1}$), of which more of the $70\%$ is radiated inside the cool core region.
 
 The exeptional X-ray luminosity suggests that the we are observing the AGN-ICM interactions taking place in extreme conditions. The MH surrounds the BCG, extending for $\sim$140 kpc and it has a spectral index $\alpha=1.2$. The cluster shows also a pair of cold fronts located inside the radio emitting region, that highlight the presence of ongoing sloshing processes \citep[][]{Giacintucci_2011a};

 \item{\it RX J1532.9+3021:} The Chandra observation shows a pair of cavities associated with the BCG and a cold front located at $\sim$65 kpc from the center, partially associated with one of the cavities \citep[][]{Hlavacek_2013}. The MH appears more extended toward the northeast, with a radius of $\sim$ 180 kpc, following the morphology of the X-rays surface brightness. \citet[][]{Giacintucci_2014a} estimated the total spectral index of the diffuse radio emission $\alpha=1.2$ by combining observations at 325 MHz, 610 MHz, 1.4 GHz and 4.9 GHz; 
 
 \item{\it RX J1720.1+2637}: This cluster was the first relaxed system in which sloshing cold fronts have been revealed by Chandra \citep[][]{Mazzotta_2001a} as well as one of the first two clusters in which a connection between MH and  cold  fronts  has  been  reported \citep[][]{Mazzotta-Giacintucci_2008}. The MH consists of a bright central region that contains most of its flux density, and a fainter, arc-shaped tail elongated for $\sim$230 kpc and it is delimited by the cold front. \citet[][]{Giacintucci_2014b} combined several radio observation spanning from 0.317 to 8.44 GHz to obtain a detailed spectral index map of the MH. They observed that the spectral index varies within the MH. The central region shows $\alpha\simeq$1, whereas the tail shows $\alpha\simeq$2-2.5. \citet[][]{ZuHone_2015b} demonstrated via numerical simulations that the CRe in the tail could be efficiently re-accelerated by the turbulence injected at the edge of the cold front.
\end{itemize}
\section{Sampling of the diffuse radio emission}
\label{images}
We report here, for each cluster of the sample, the contours of the radio emission, the mask used for the analysis overlapped on the X-ray image. We present also a random mesh generated during the MCptp analysis, the corresponding SMptp analysis and the distribution of $k$ produced by the MCptp analysis. The resolution and the noise of the each map are reported in Tab. \ref{obs.tab}. For each object we report:
\begin{itemize}
 \item {\it Left:} X-ray surface brightness map smoothed with a 1.5$''$ gaussian, with the contours of the radio map at the -3, 3, 24, 96$\sigma$ levels (white), the mask used in the analysis (grey) and a random sampling mash (green). The cell size matches the angular resolution of the radio image;
 \item{\it Center:} $I_\text{R}$ vs $I_\text{X}$ obtained from the presented mesh. The red, green and blue lines are, respectively, the best-fit slopes obtained with the BCES for ($I_\text{X}\mid I_\text{R}$), ($I_\text{R}\mid I_\text{X}$) and the bisector. The estimated value of $k_\text{SM}$ is reported in the label;
 \item{\it Right:} Distribution of indexes $k$ produced after 1000 iterations of MCptp analysis.
\end{itemize}
\begin{figure*}
\begin{multicols}{3}
\includegraphics[width=\linewidth]{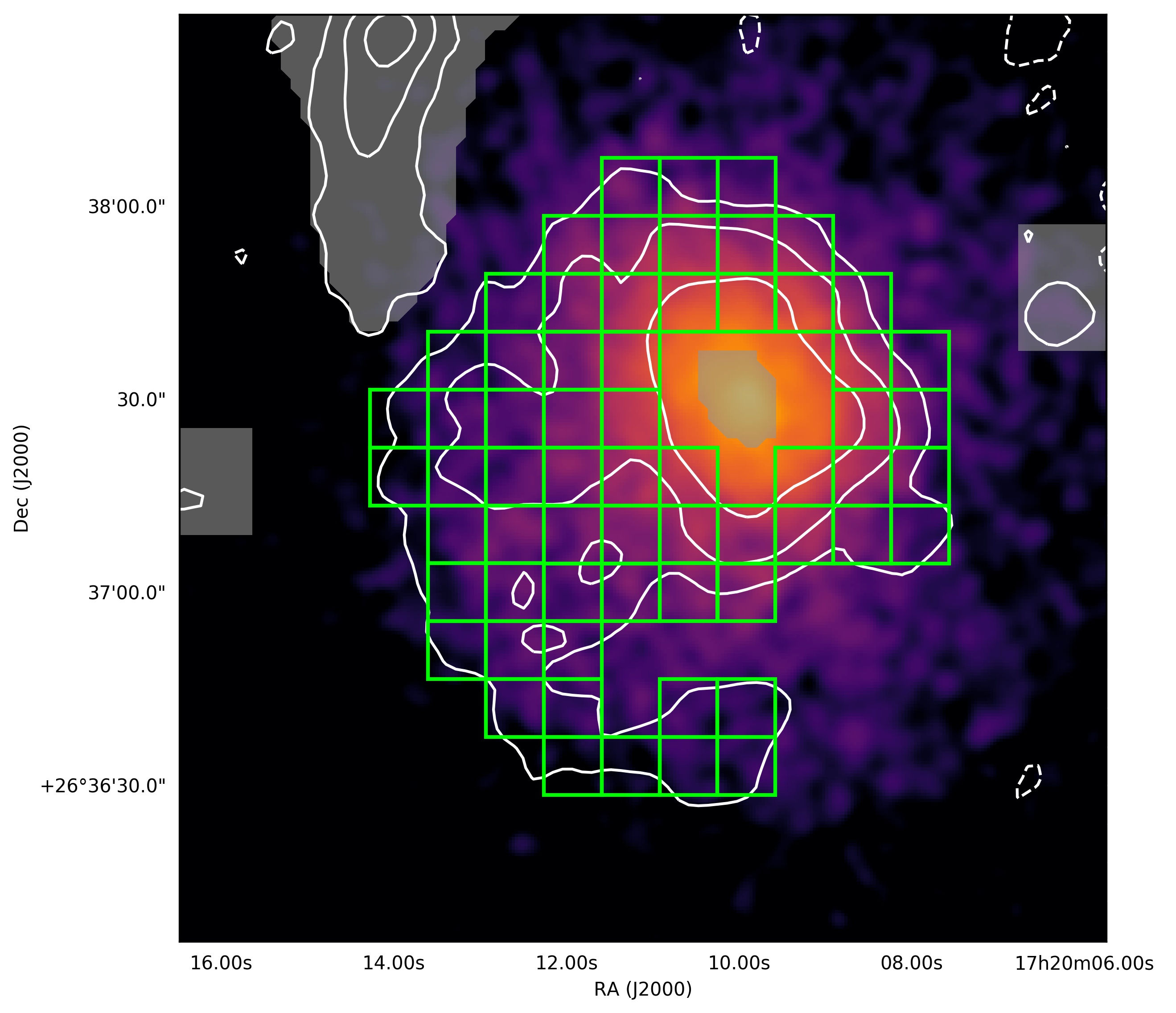}\par
\includegraphics[width=\linewidth]{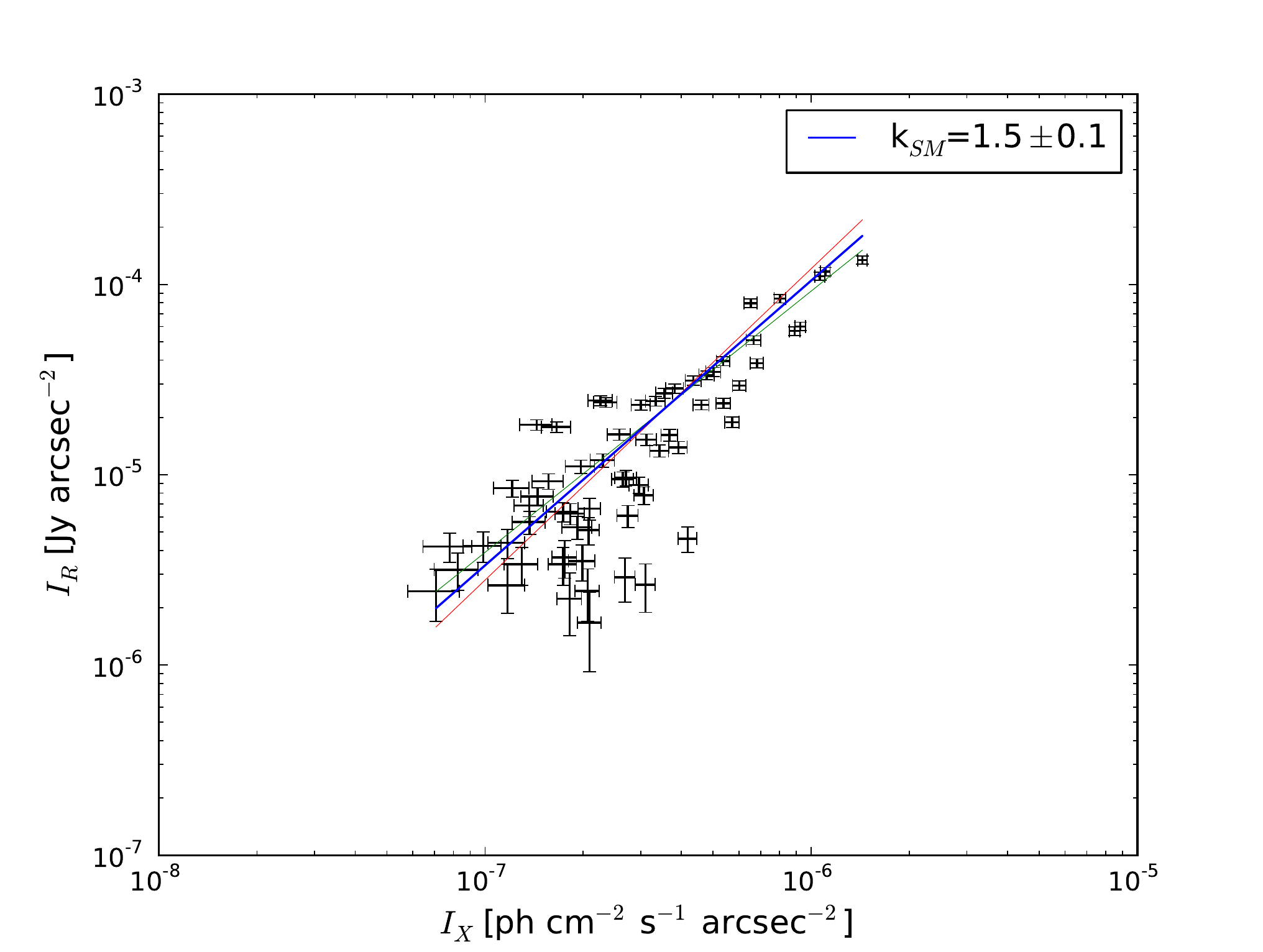}\par
\includegraphics[width=\linewidth]{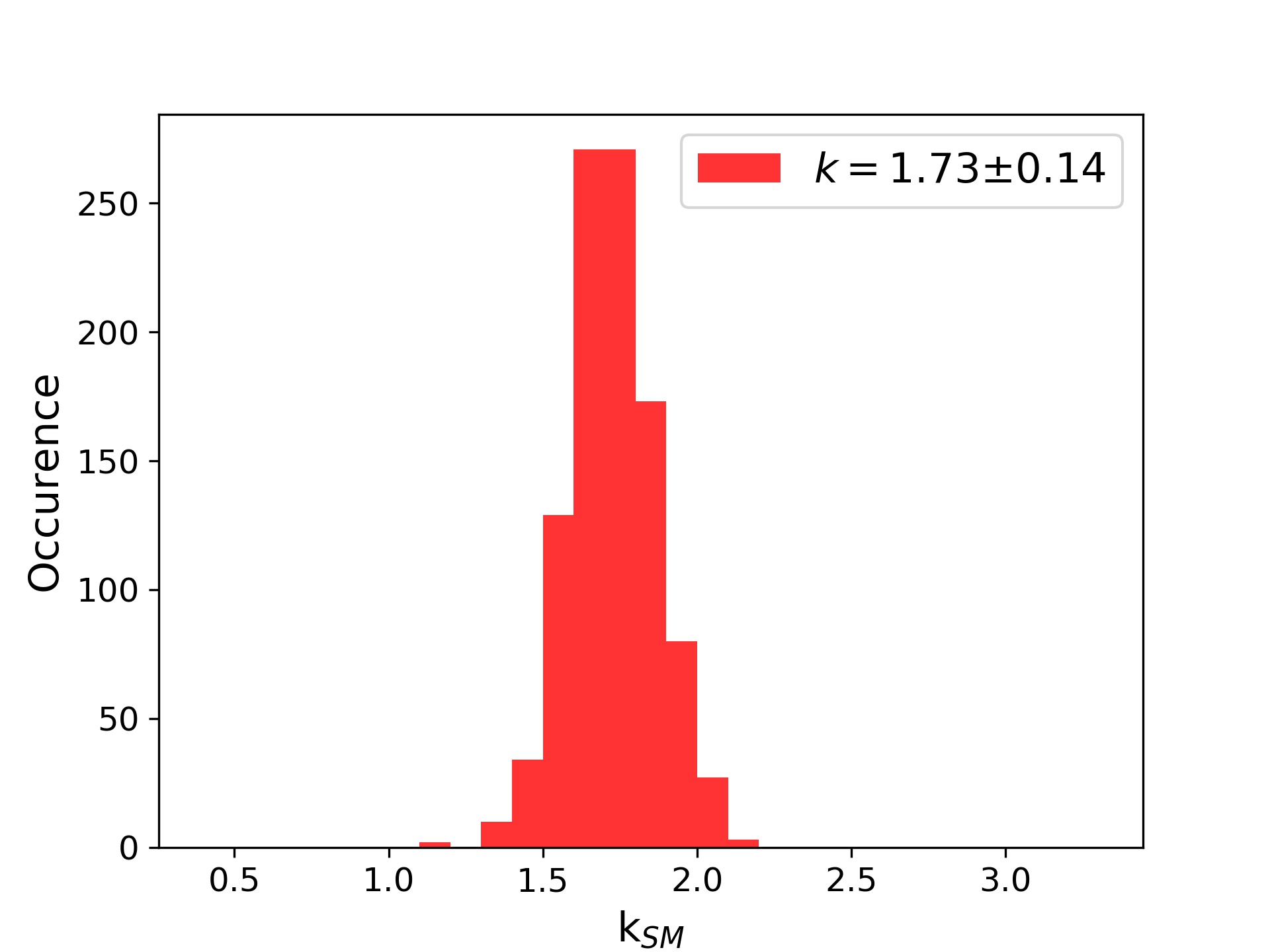}\par
\end{multicols}
\caption{\label{rxj1720.fig} RXJ1720.1+2637 .}
\end{figure*}

\begin{figure*}
\begin{multicols}{3}
\includegraphics[width=\linewidth]{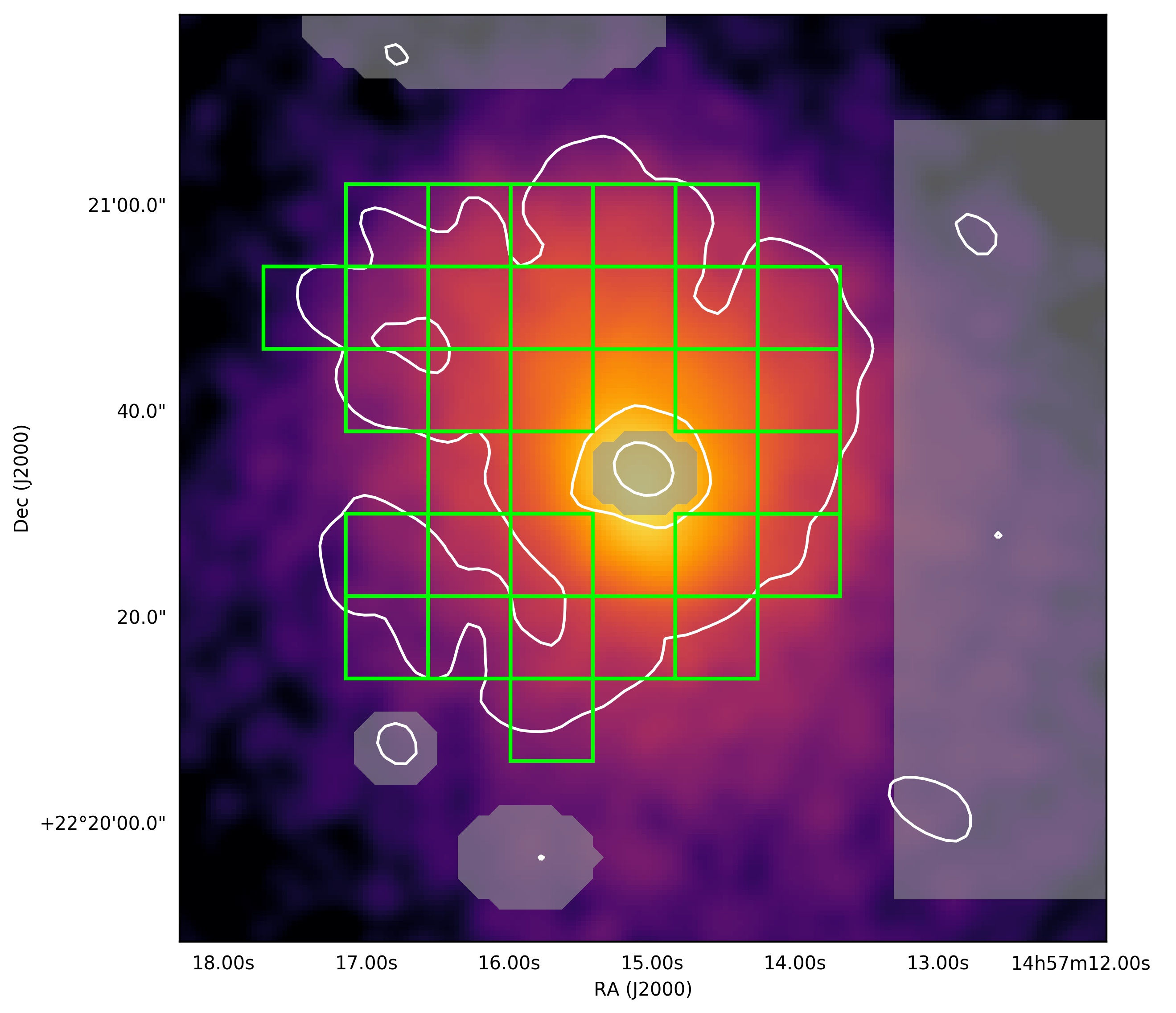}\par
\includegraphics[width=\linewidth]{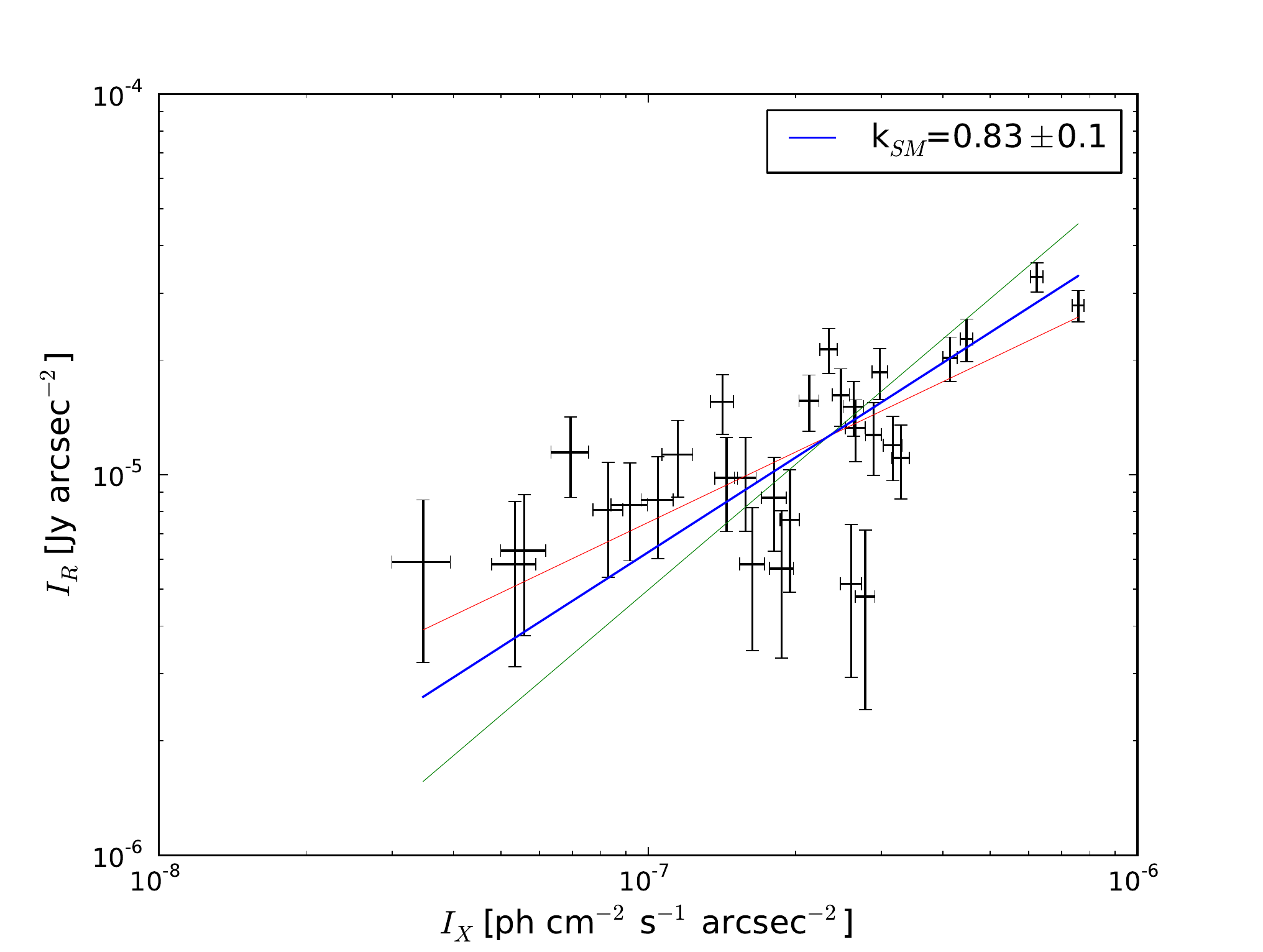}\par
\includegraphics[width=\linewidth]{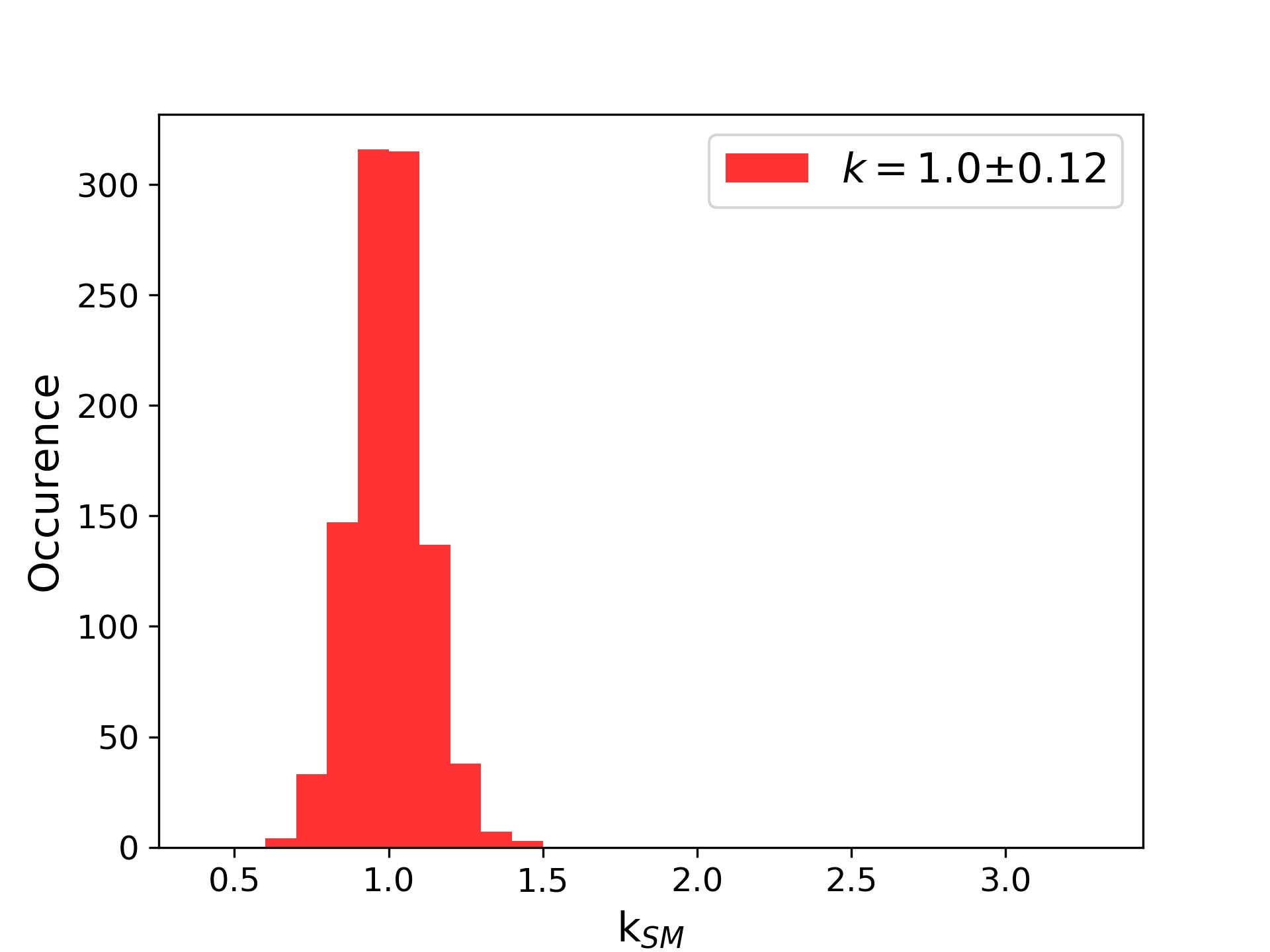}\par
\end{multicols}
\caption{\label{ms1455.fig}  MS 1455.0+2232 .}
\end{figure*} 

\begin{figure*}
\begin{multicols}{3}
\includegraphics[width=\linewidth]{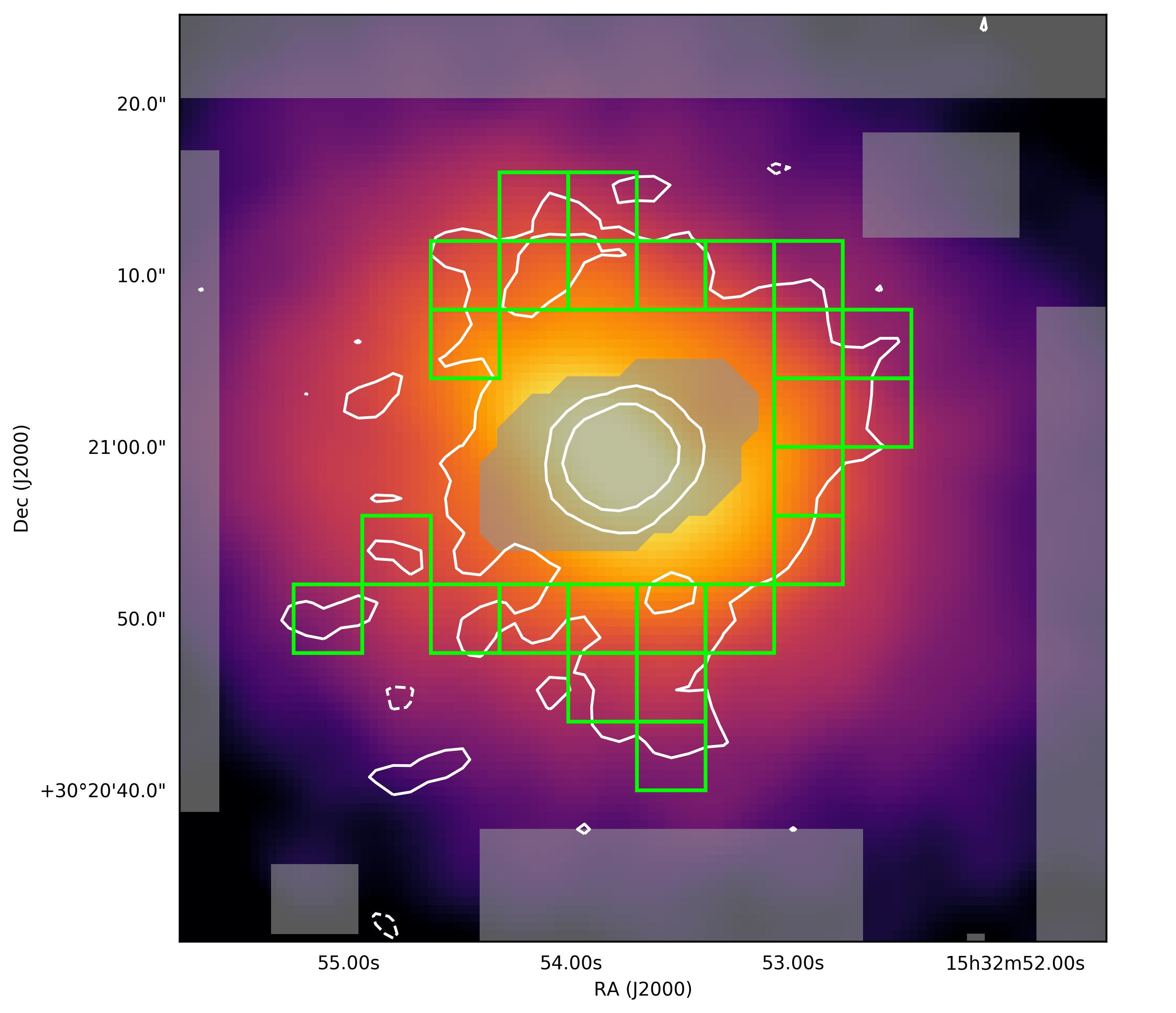}\par
\includegraphics[width=\linewidth]{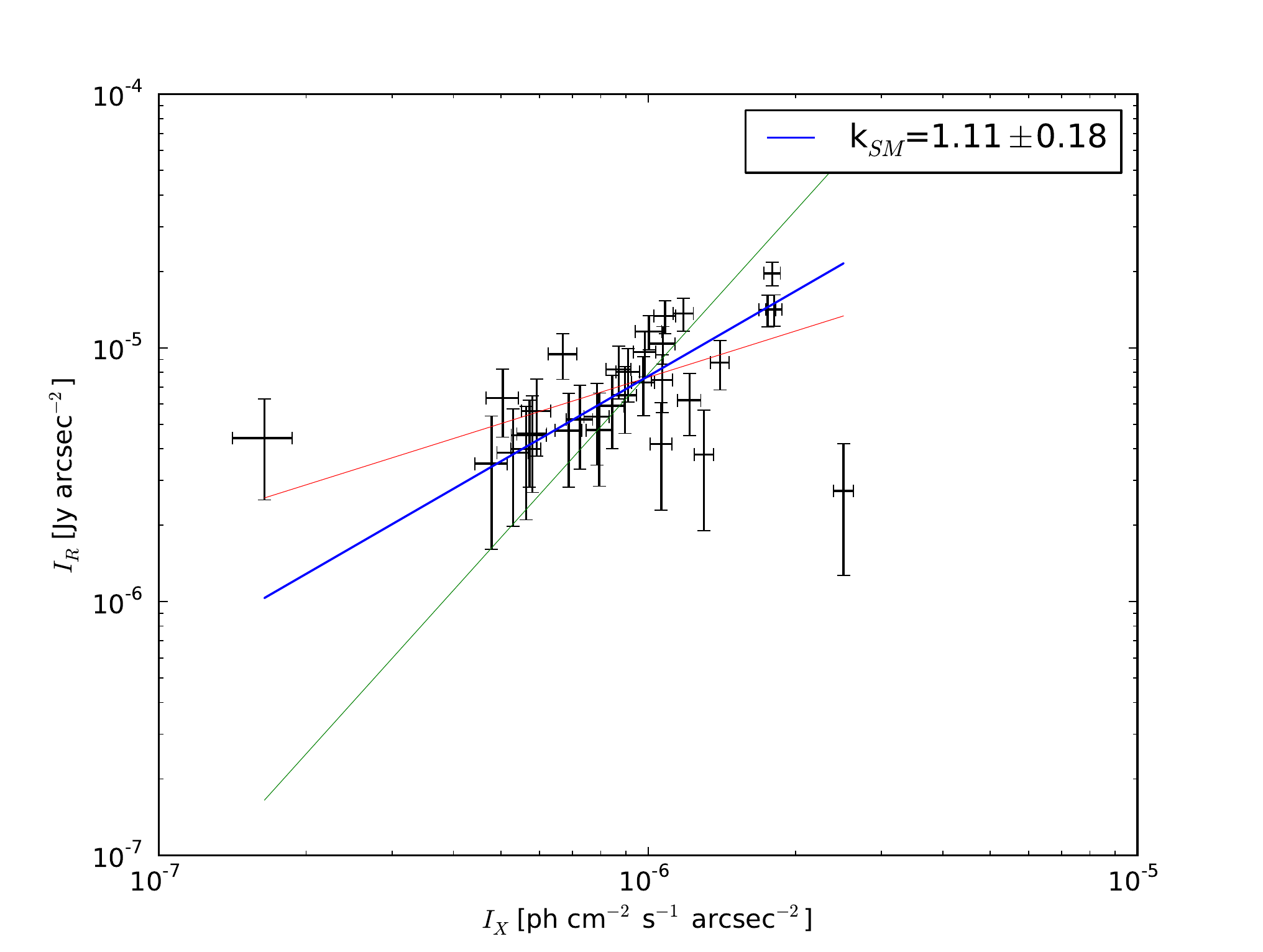}\par
\includegraphics[width=\linewidth]{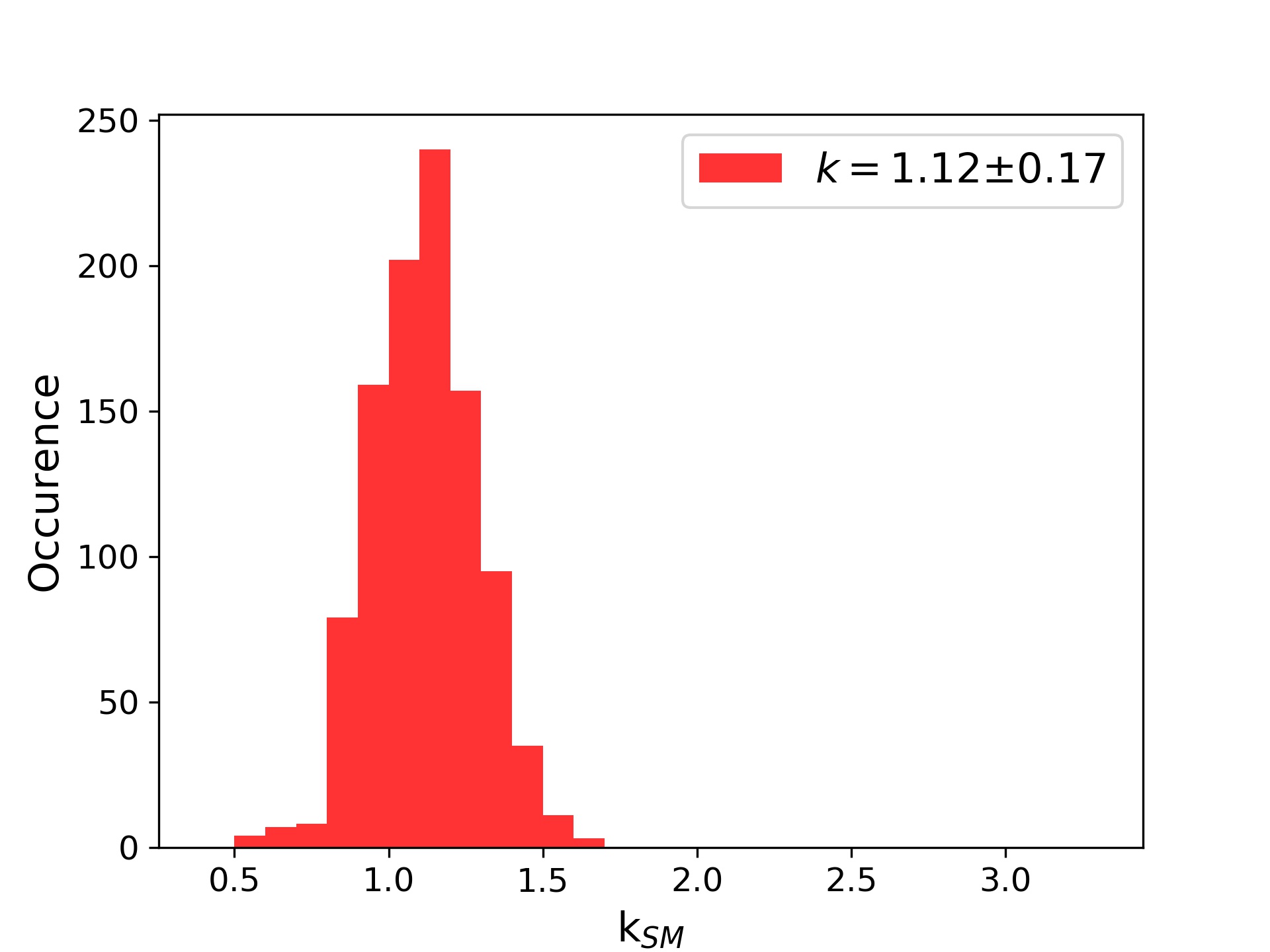}\par
\end{multicols}
   \caption{\label{rxj1532.fig} RX J1532.9+3021 .}
\end{figure*} 

\begin{figure*}
\begin{multicols}{3}
   \includegraphics[width=\linewidth]{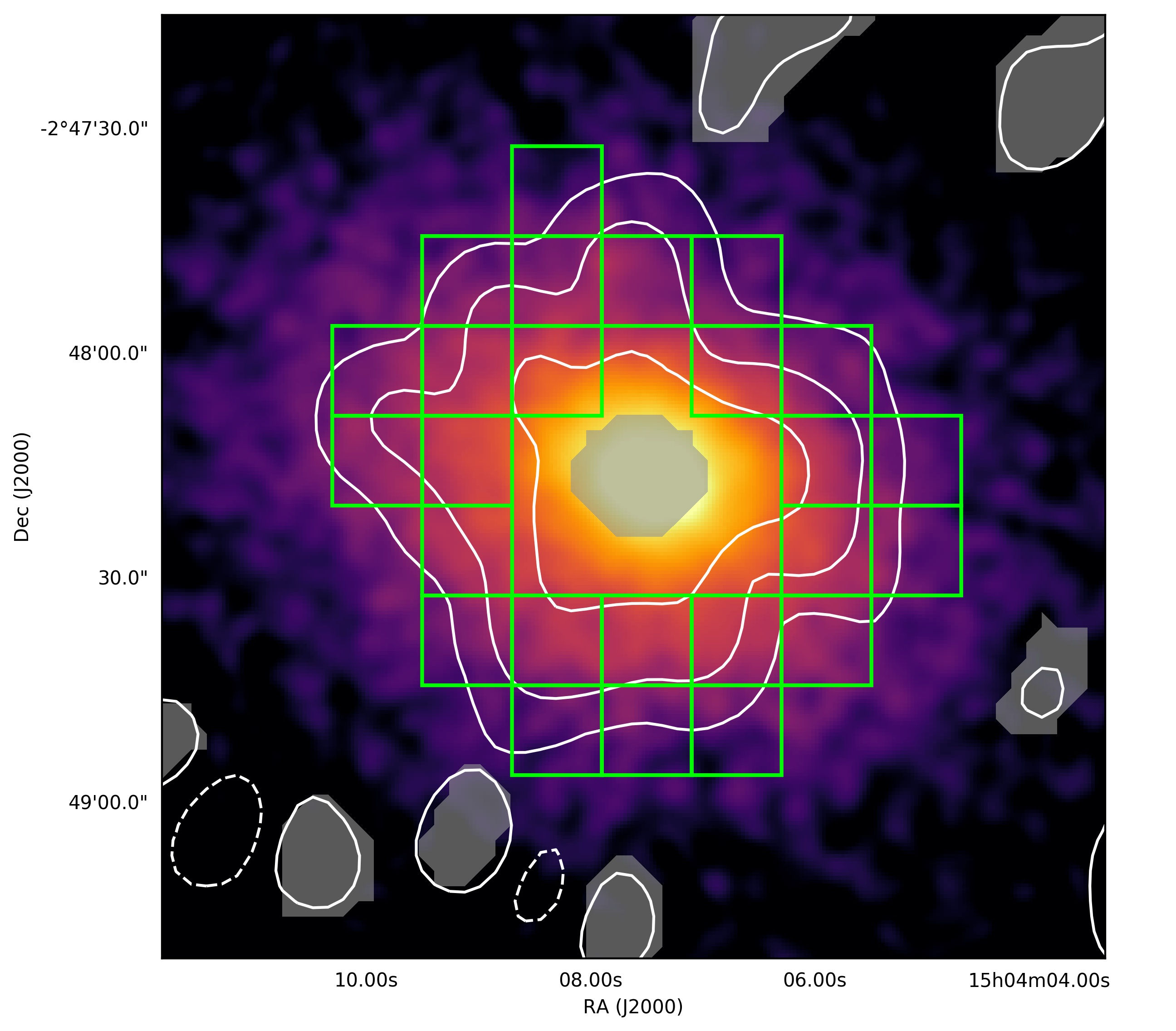}\par
   \includegraphics[width=\linewidth]{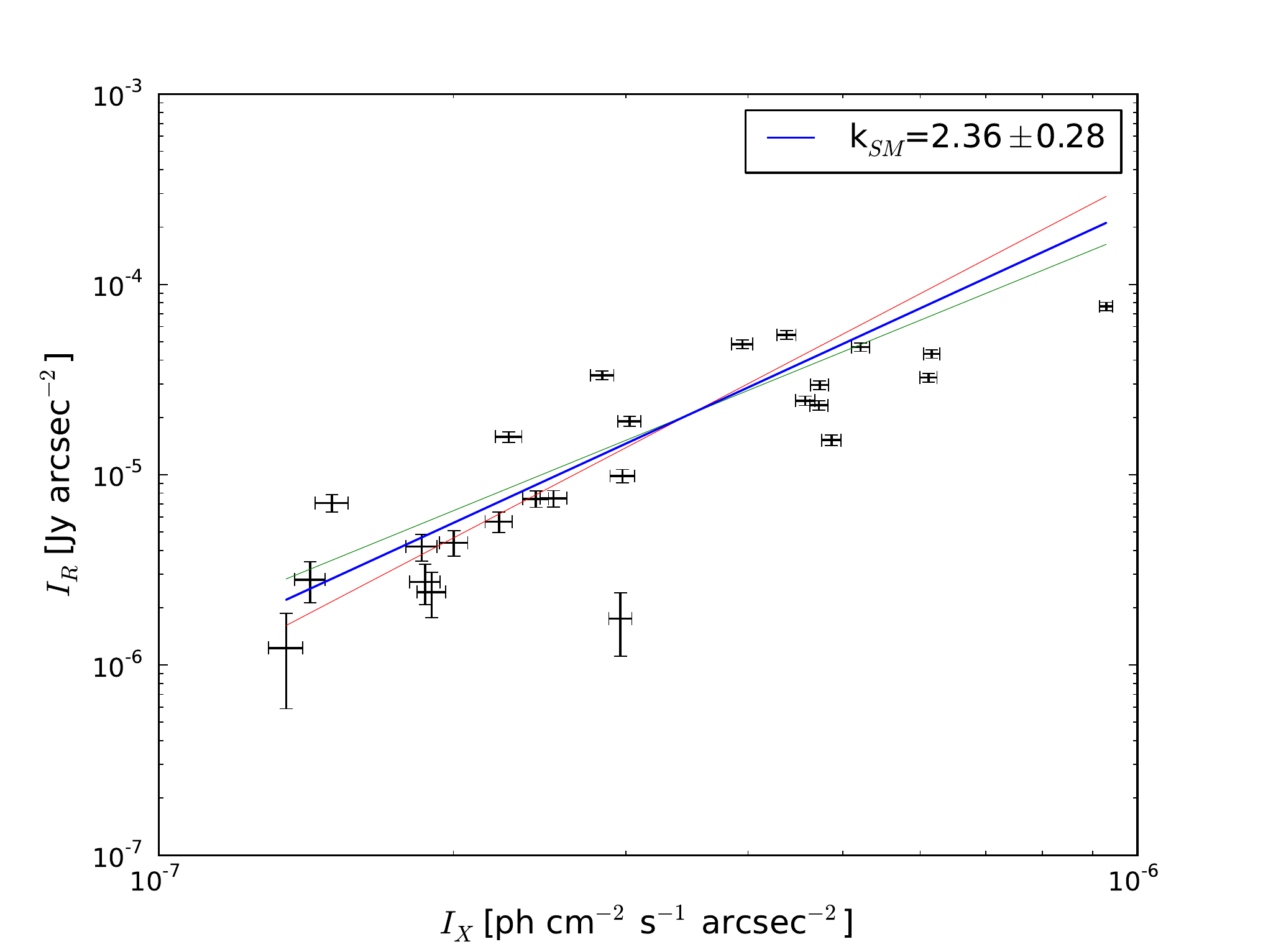}\par
   \includegraphics[width=\linewidth]{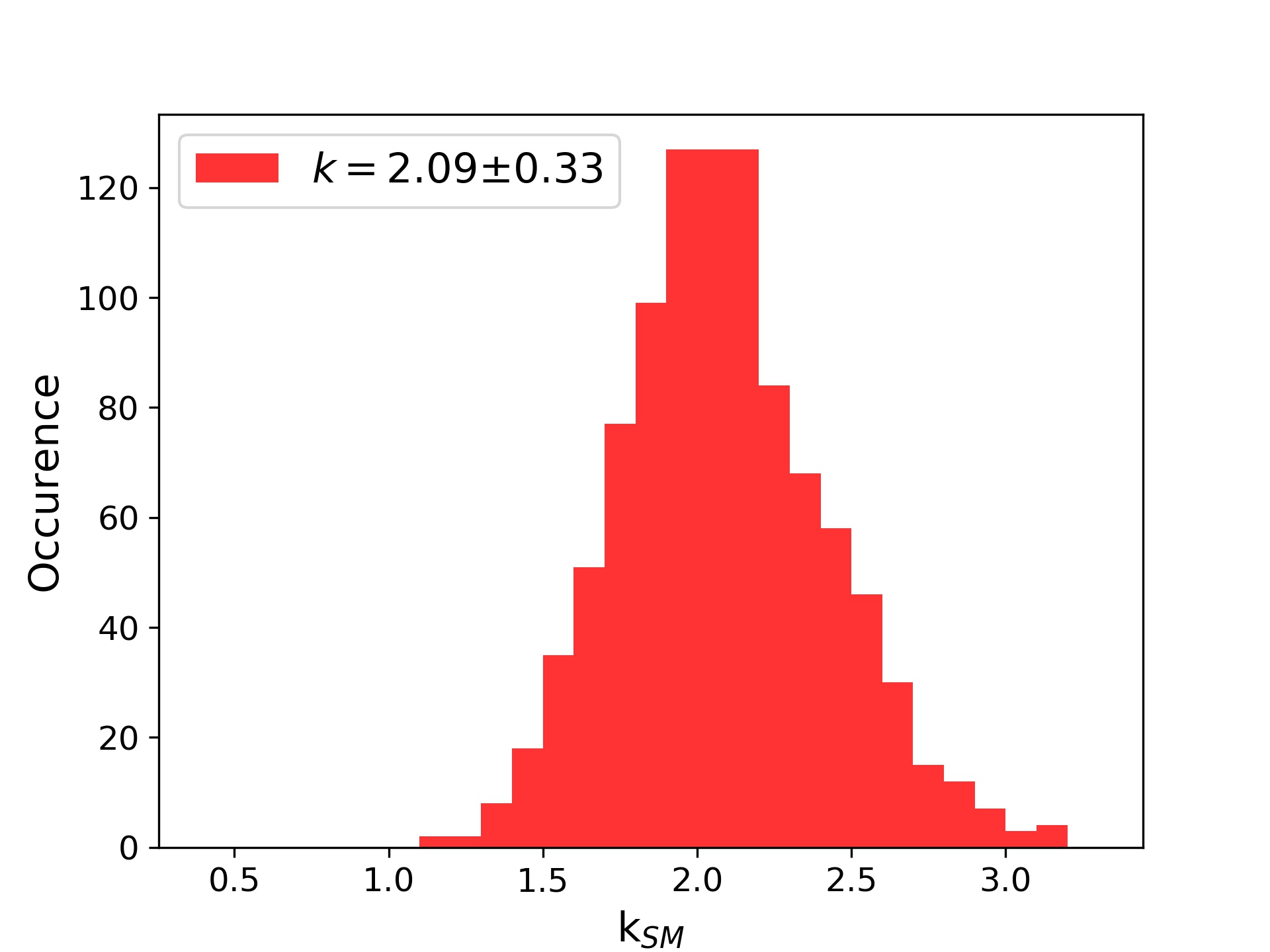}\par
\end{multicols}
   \caption{\label{rxcj1504.fig}RXC J1504.1-0248 .}
\end{figure*}

\begin{figure*}
   \begin{multicols}{3}
 \includegraphics[width=\linewidth]{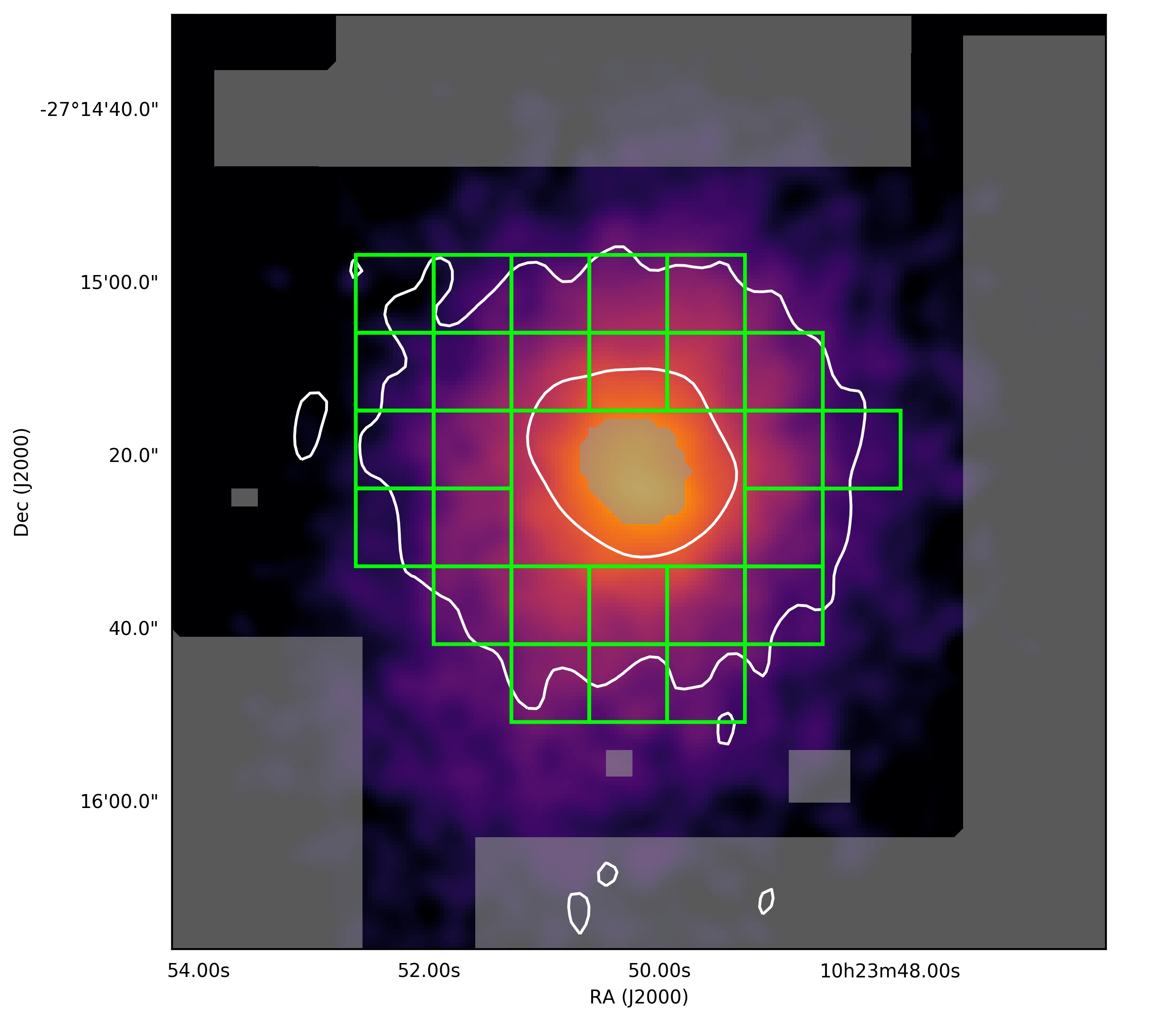}\par
\includegraphics[width=\linewidth]{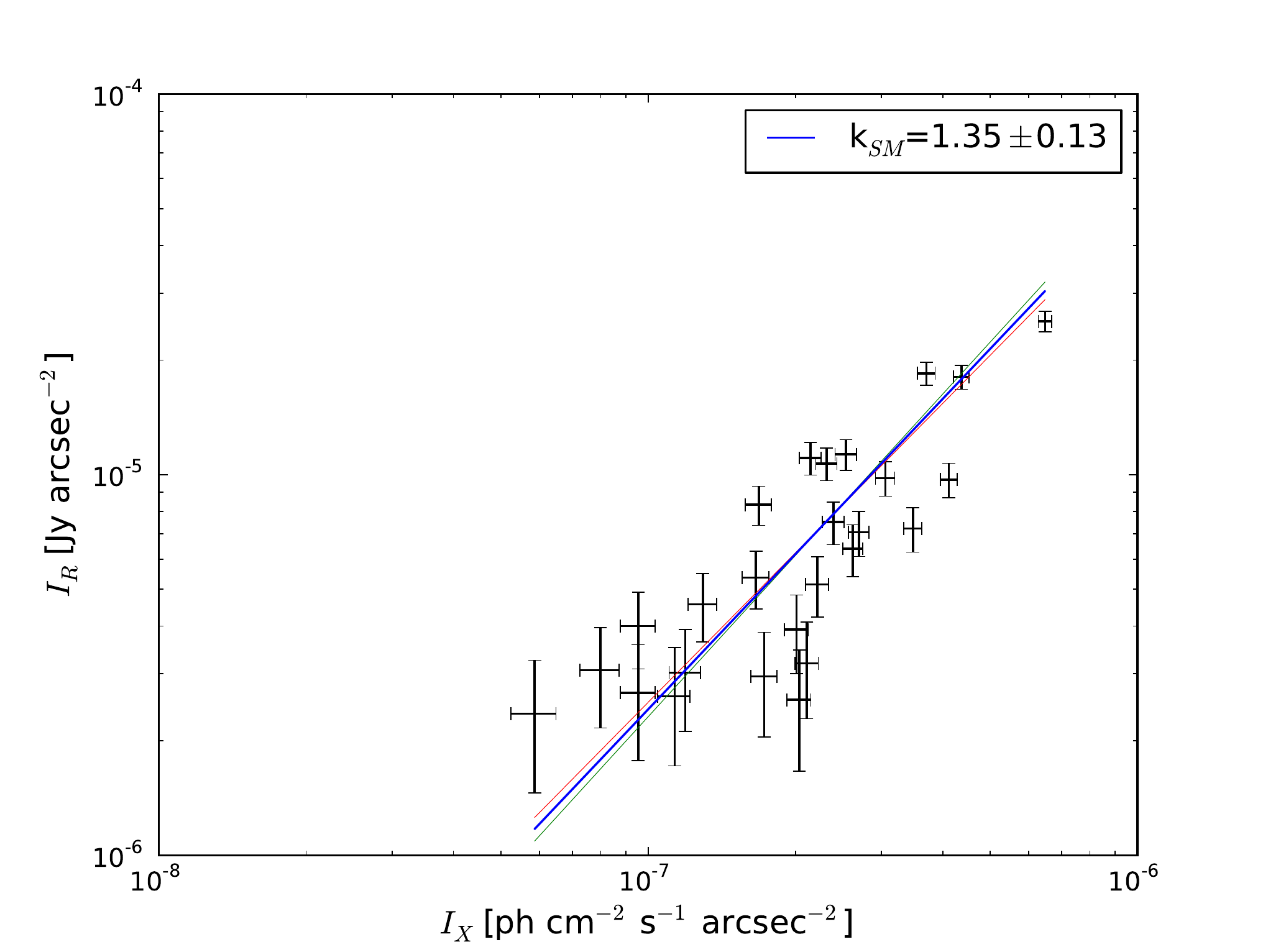}\par
\includegraphics[width=\linewidth]{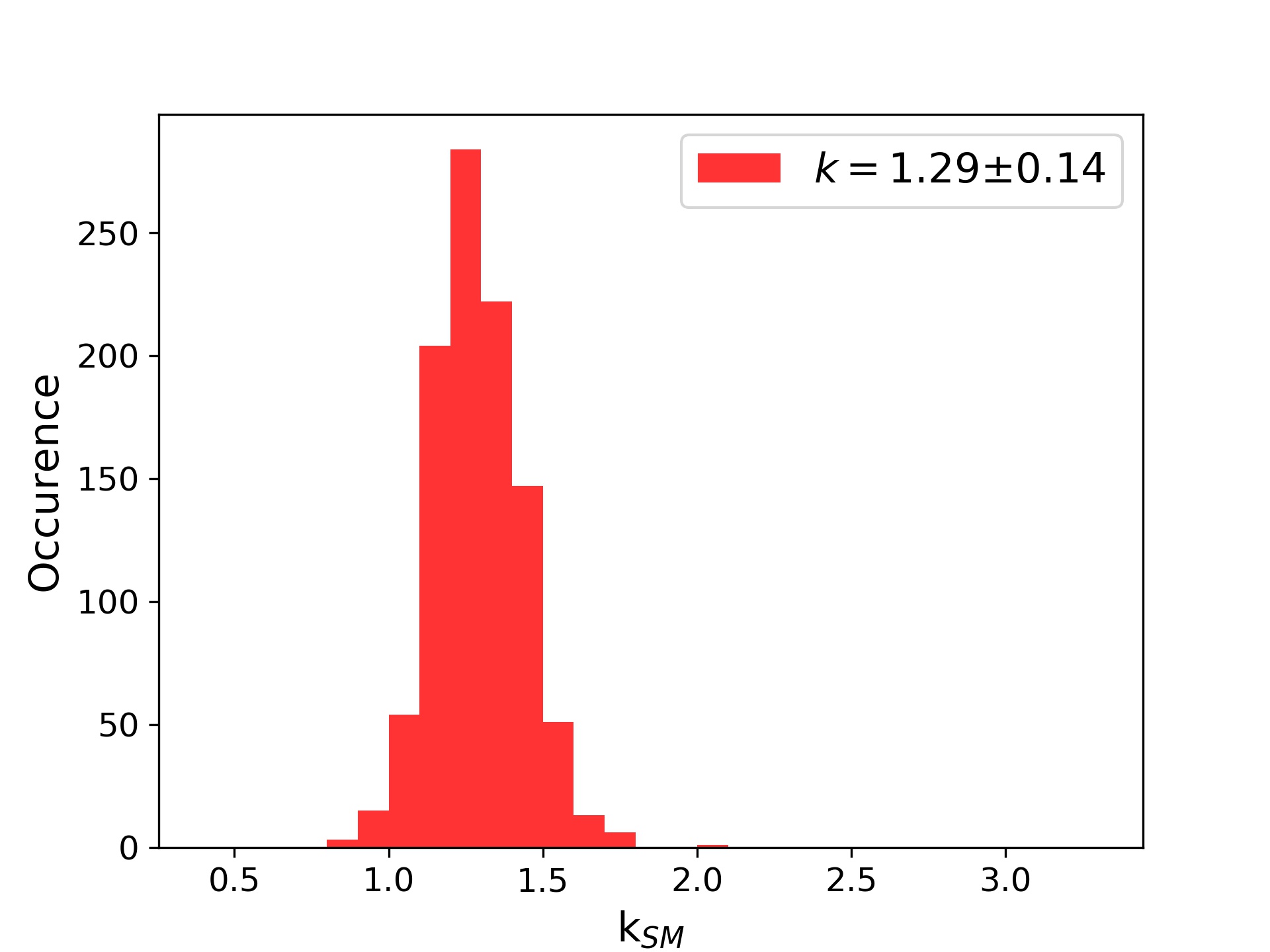}\par
\end{multicols}
  \begin{multicols}{3}
  \includegraphics[width=\linewidth]{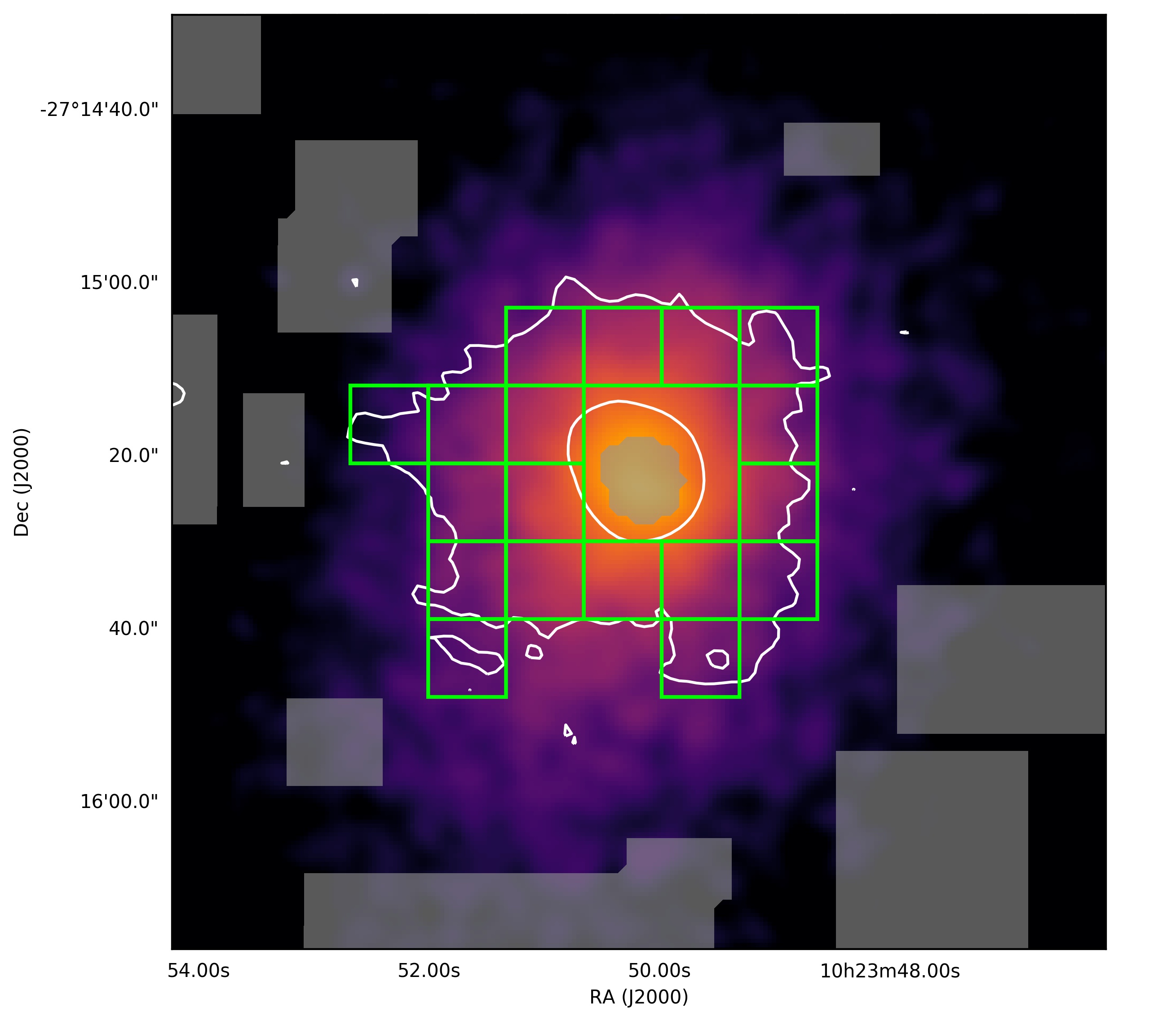}\par
   \includegraphics[width=\linewidth]{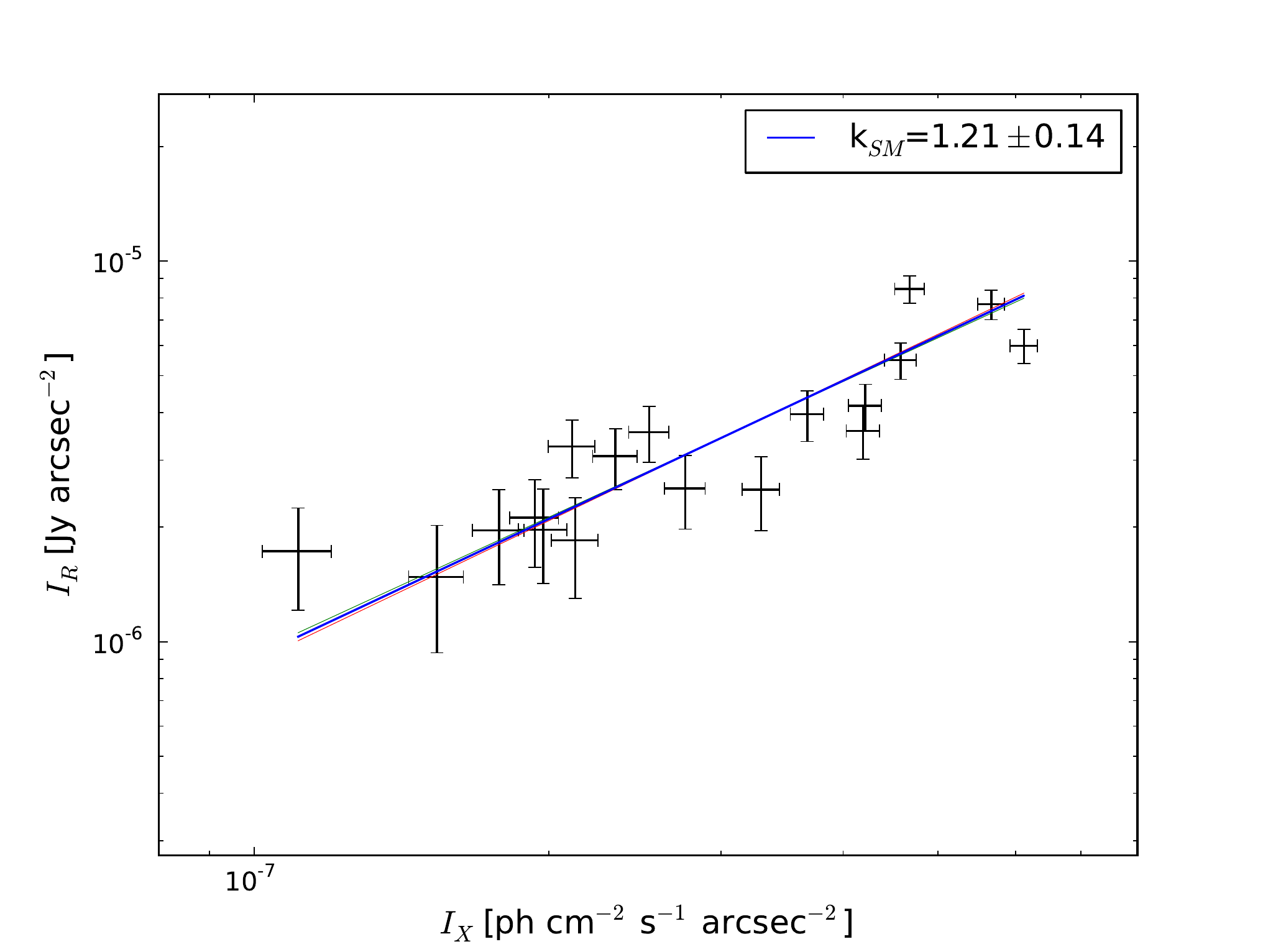}\par
\includegraphics[width=\linewidth]{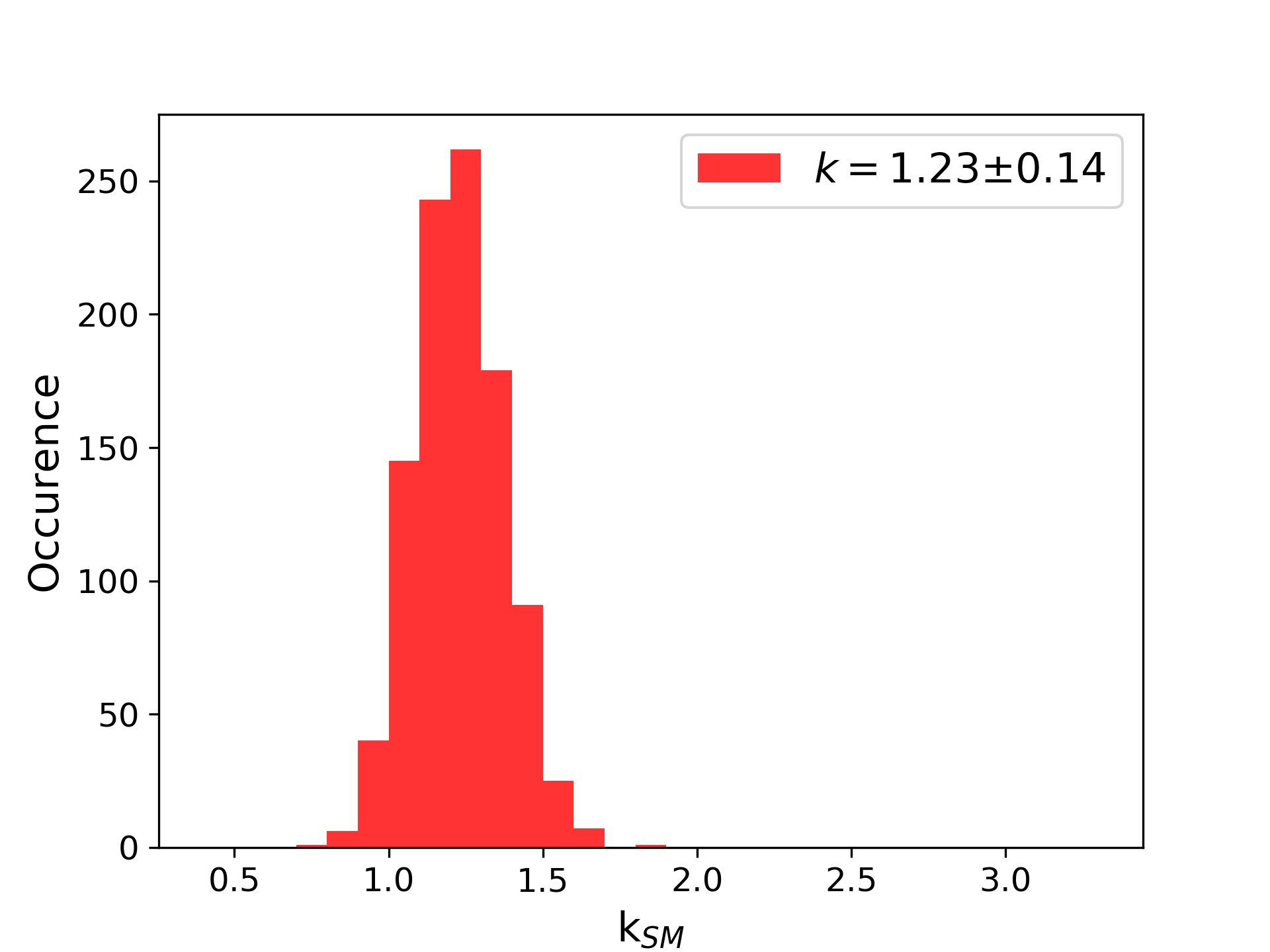}\par
   \end{multicols}
 
   \caption{\label{a3444_610.fig} Abell 3444 at 610 MHz (top) and 1.4 GHz (bottom) .}
\end{figure*} 

\begin{figure*}
   \begin{multicols}{3}
 \includegraphics[width=\linewidth]{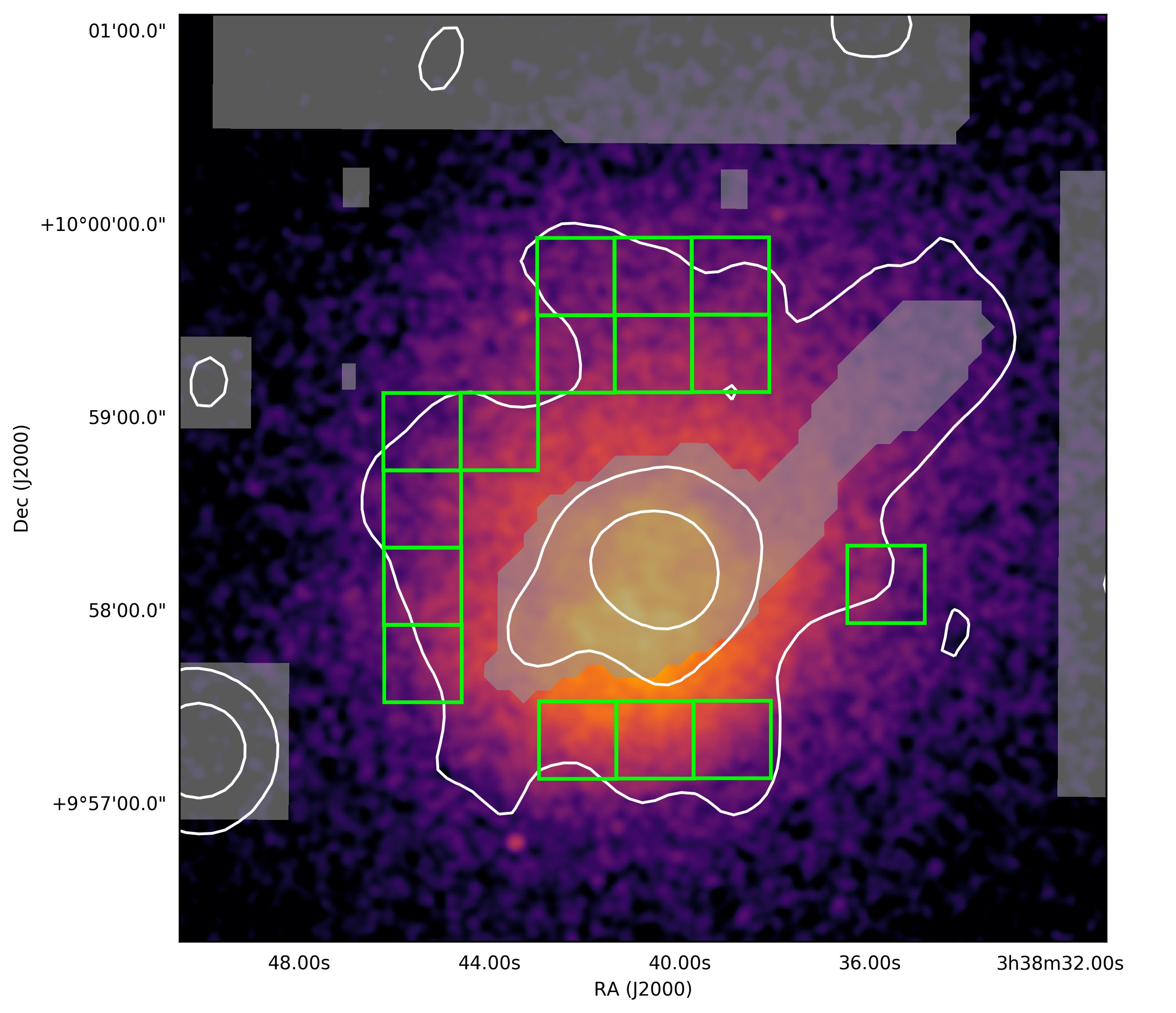}\par
   \includegraphics[width=\linewidth]{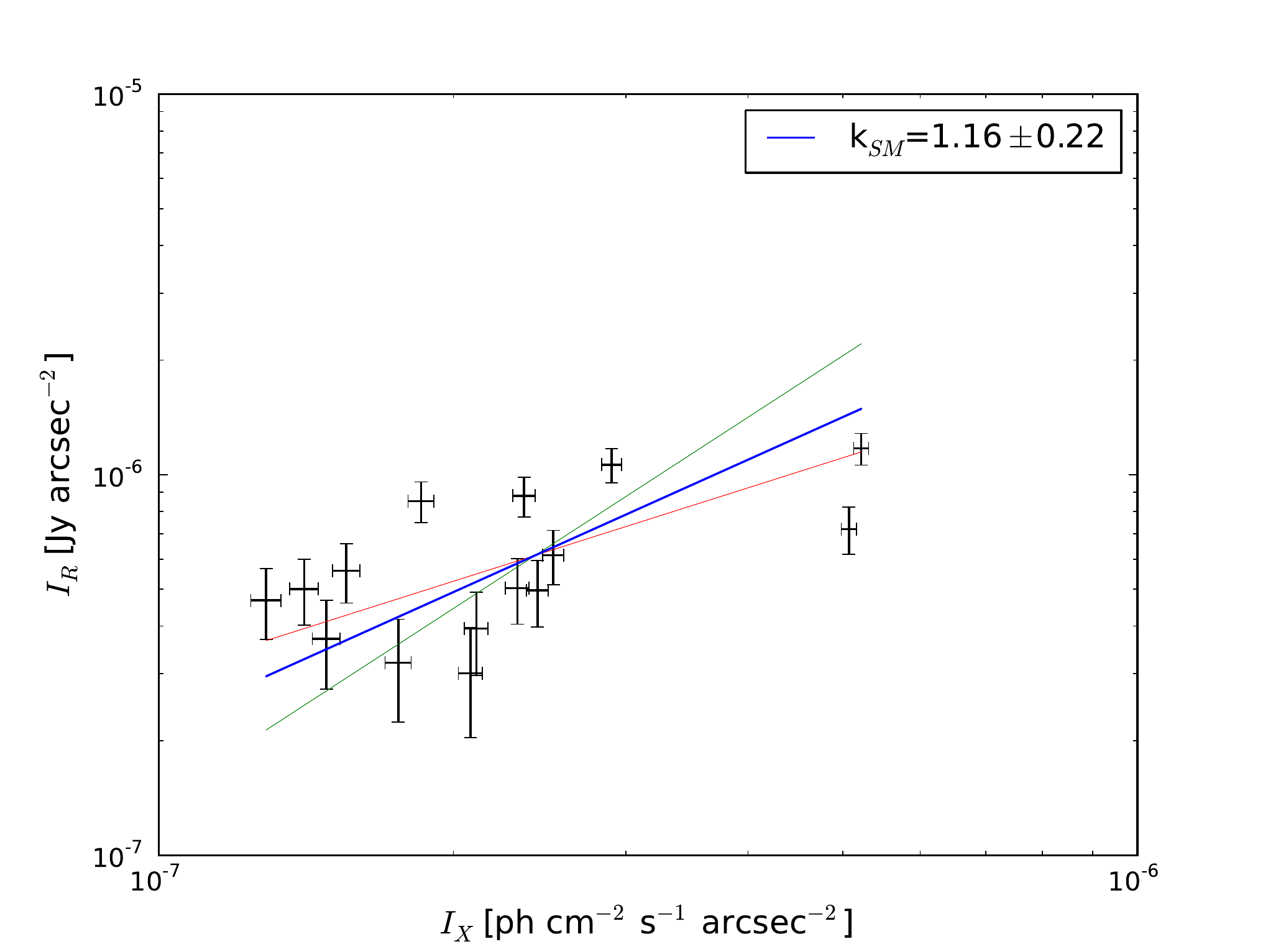}\par
\includegraphics[width=\linewidth]{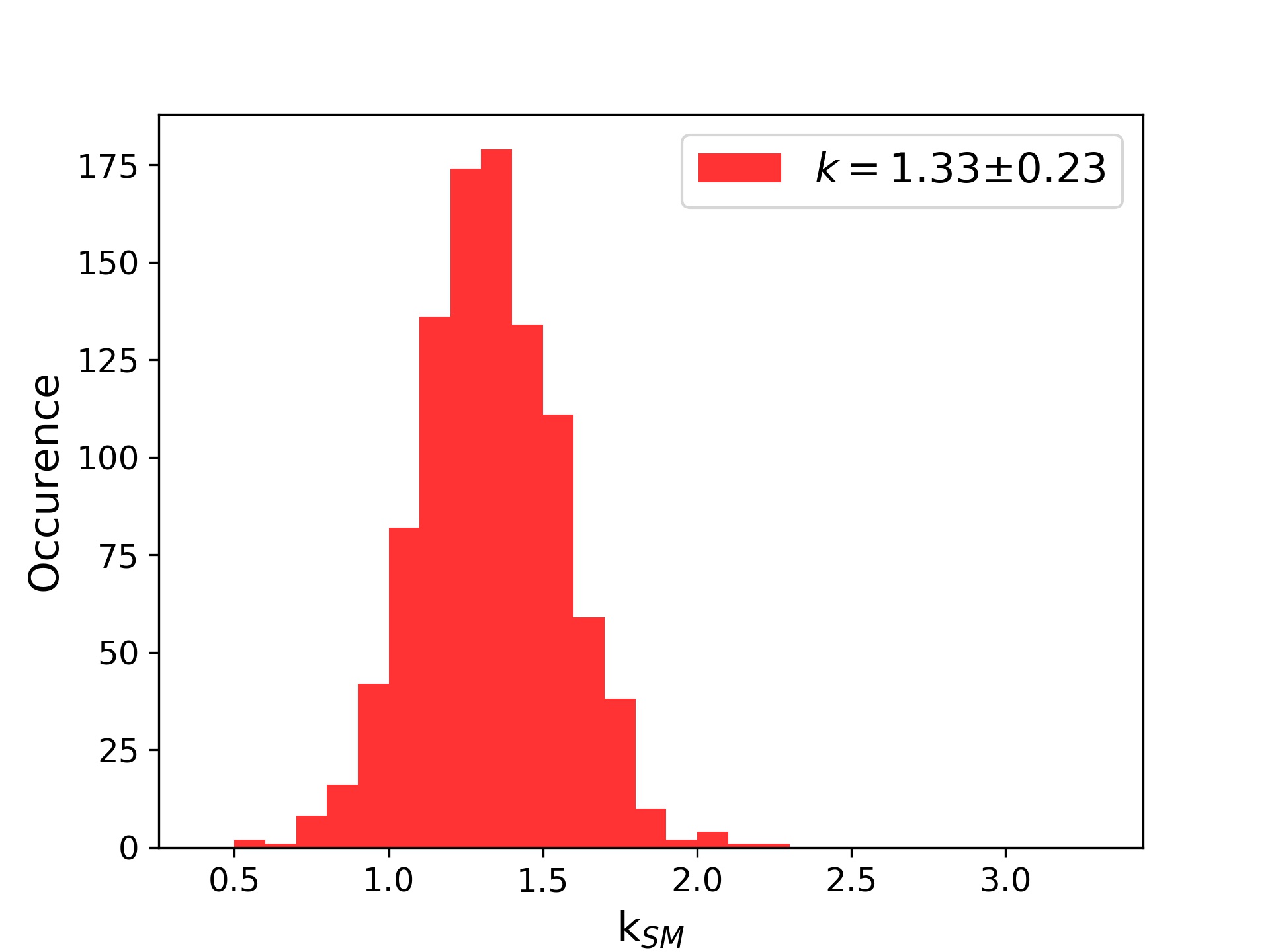}\par
   \end{multicols}
  \begin{multicols}{3}
  \includegraphics[width=\linewidth]{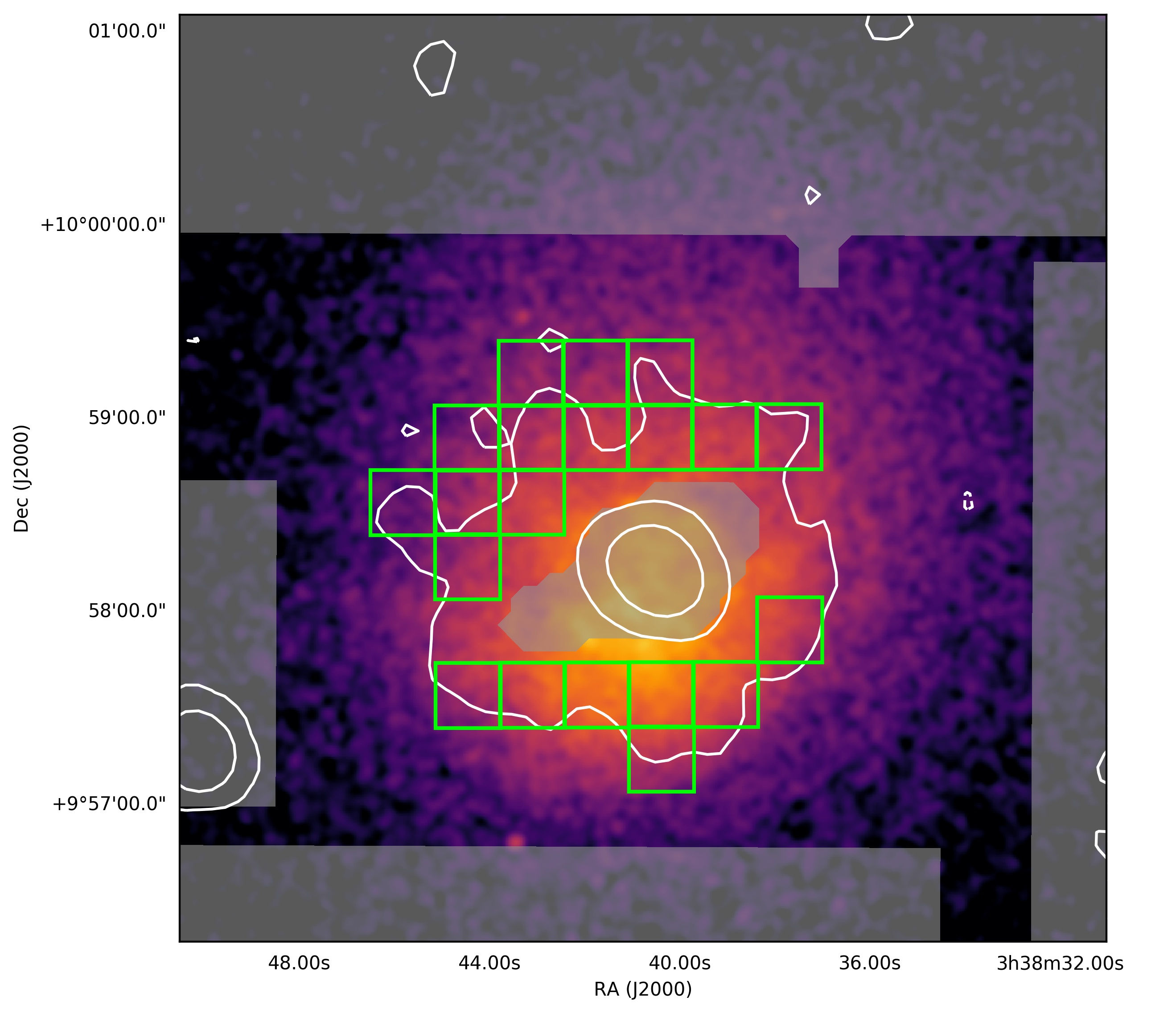}\par
   \includegraphics[width=\linewidth]{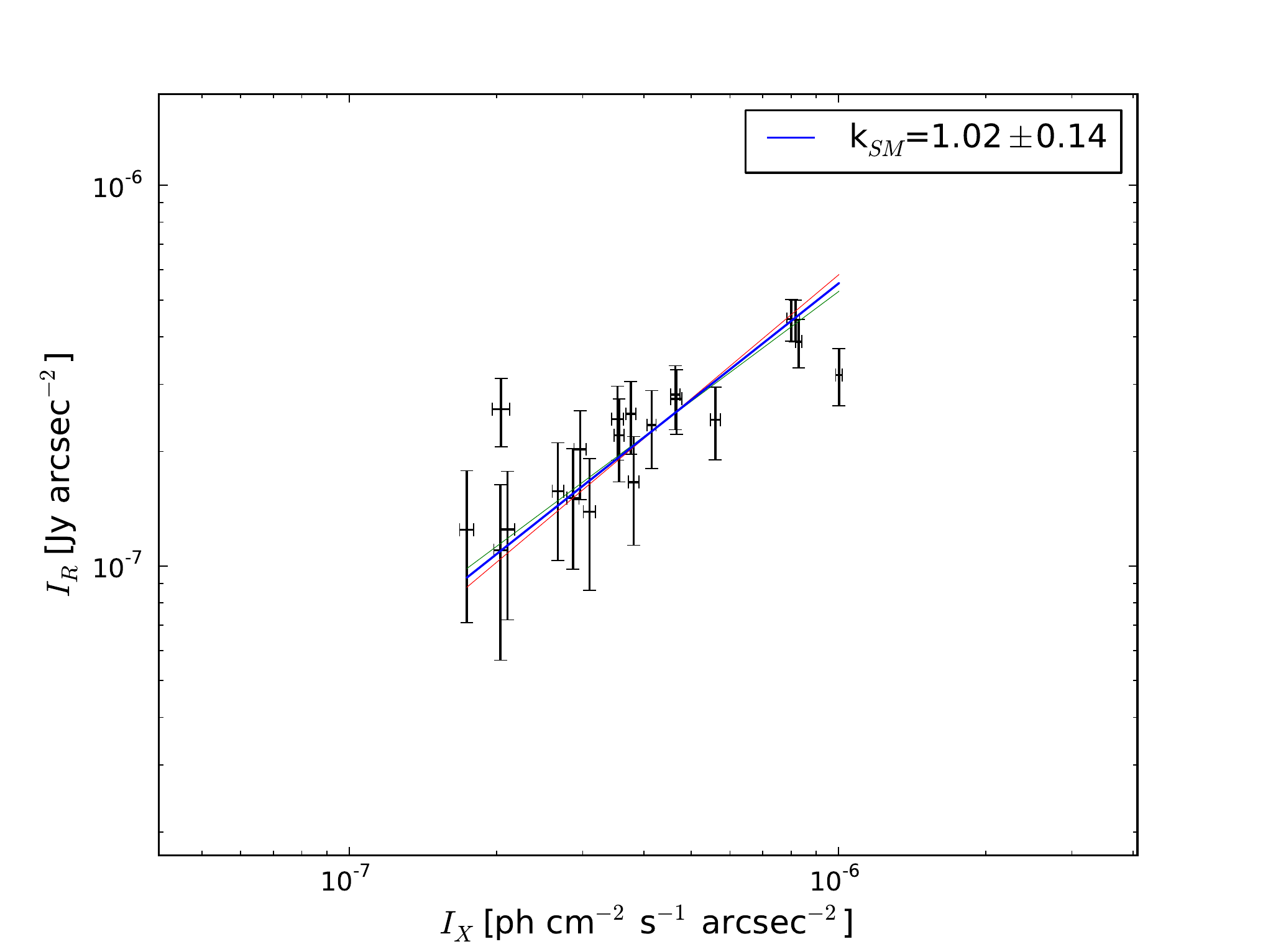}\par
 \includegraphics[width=\linewidth]{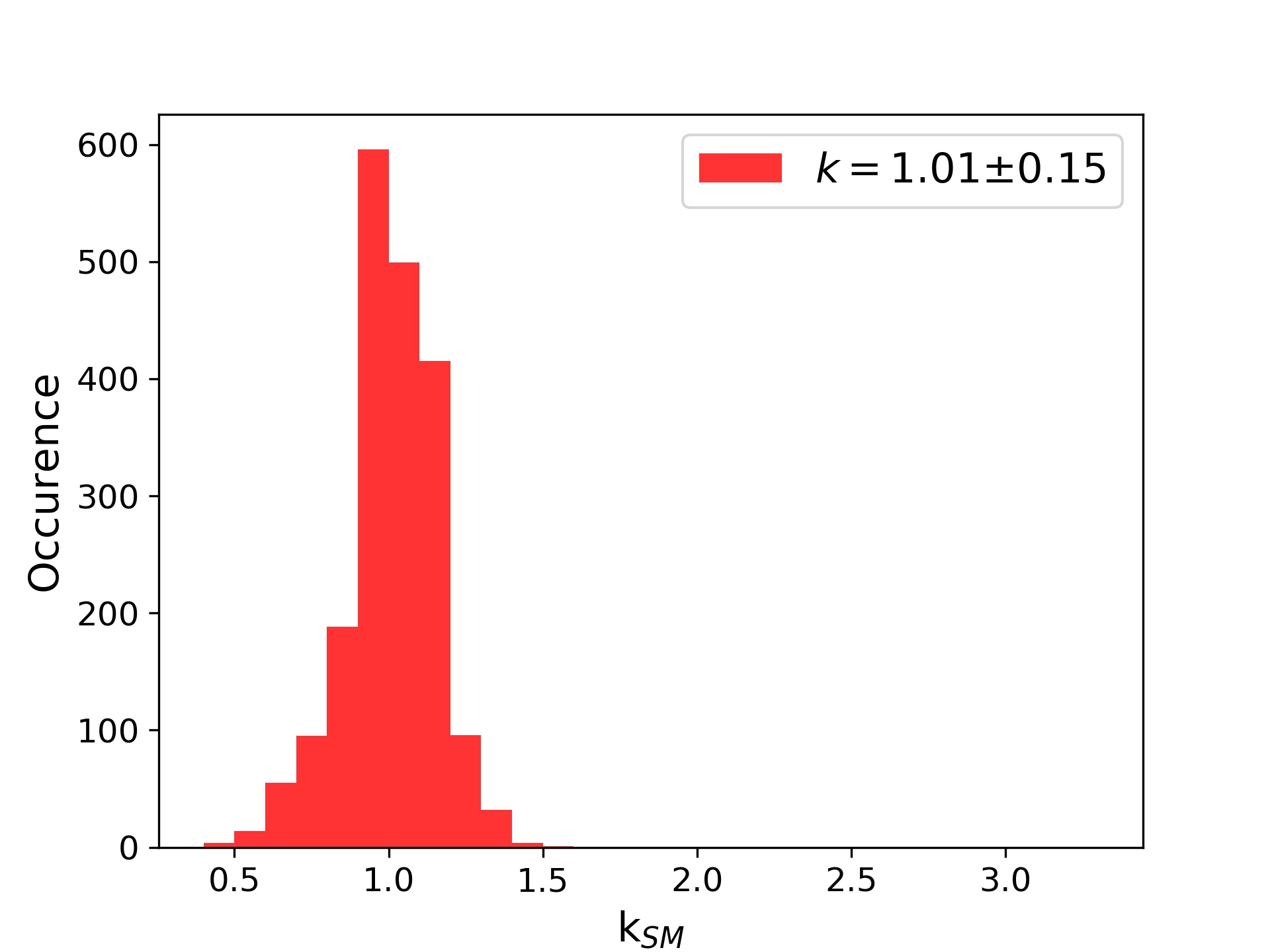}\par
   \end{multicols}
 
   \caption{\label{a3444_610.fig} 2A03335+096 at 1.4 GHz (top) and 5.5 GHz (bottom) .}
\end{figure*} 
\section{Considerations on the role of central source in SMptp analysis}
\label{subtr}
\begin{figure*}
\begin{multicols}{3}
\includegraphics[width=\linewidth]{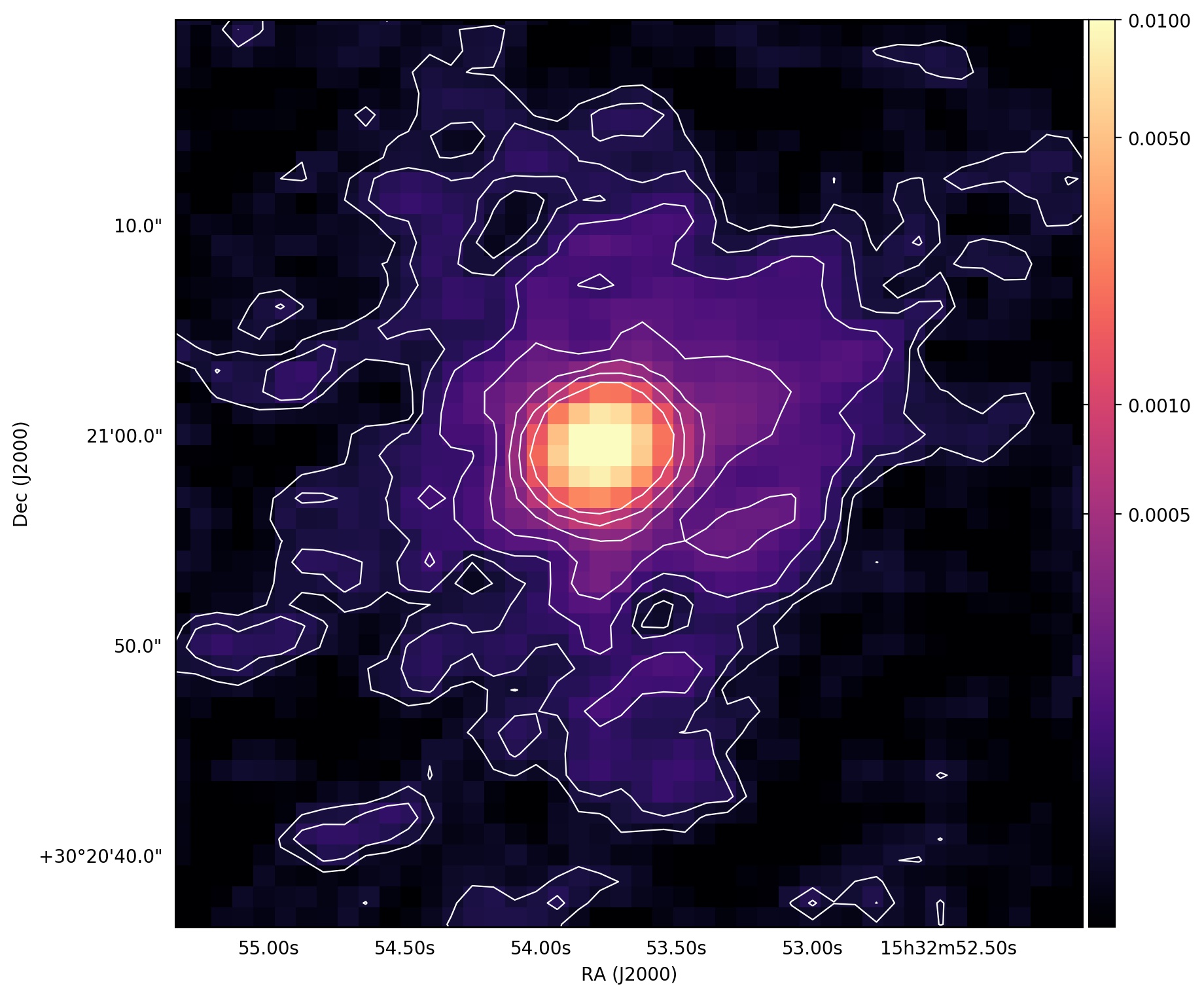}\par
\includegraphics[width=\linewidth]{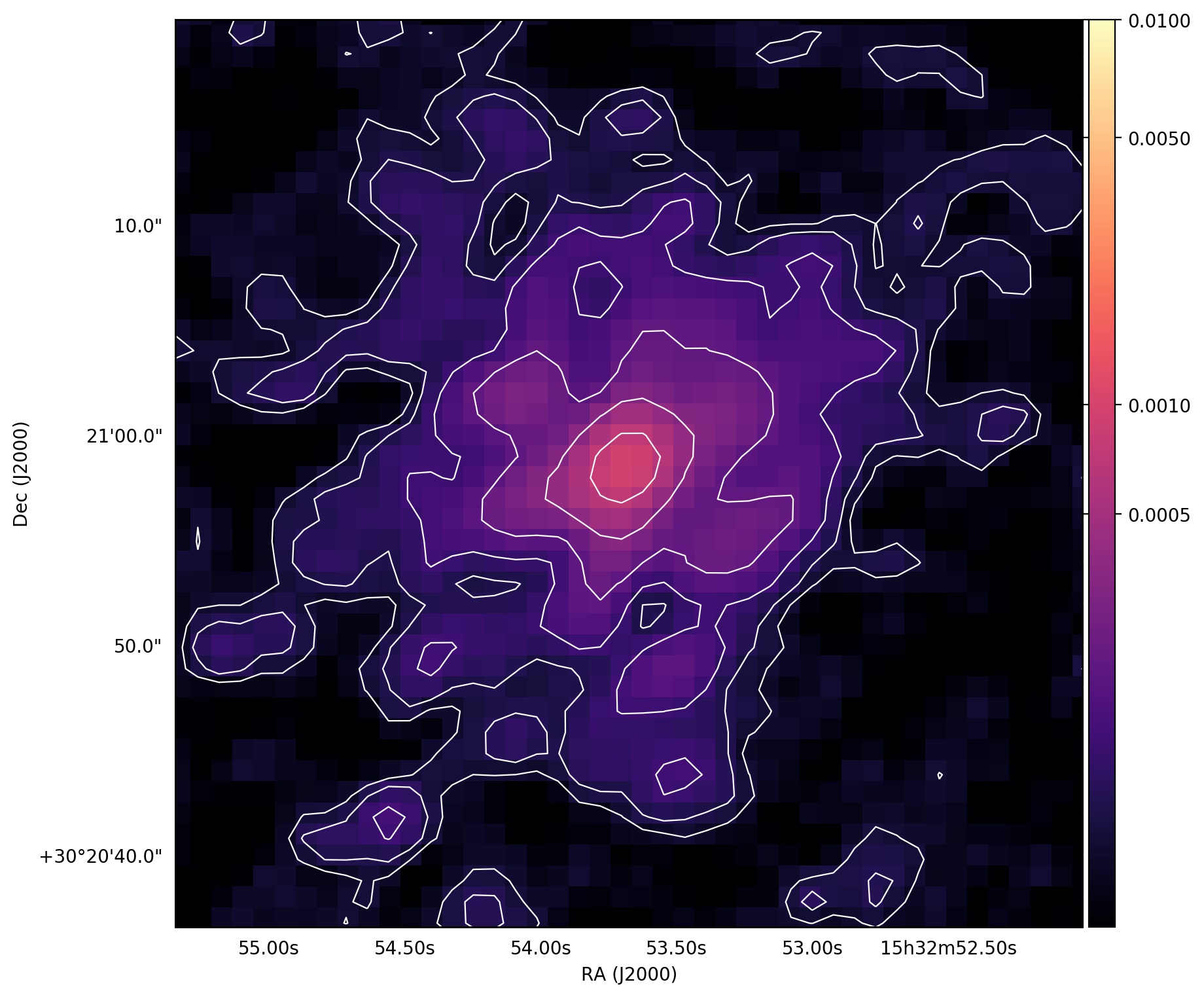}\par
\includegraphics[width=\linewidth]{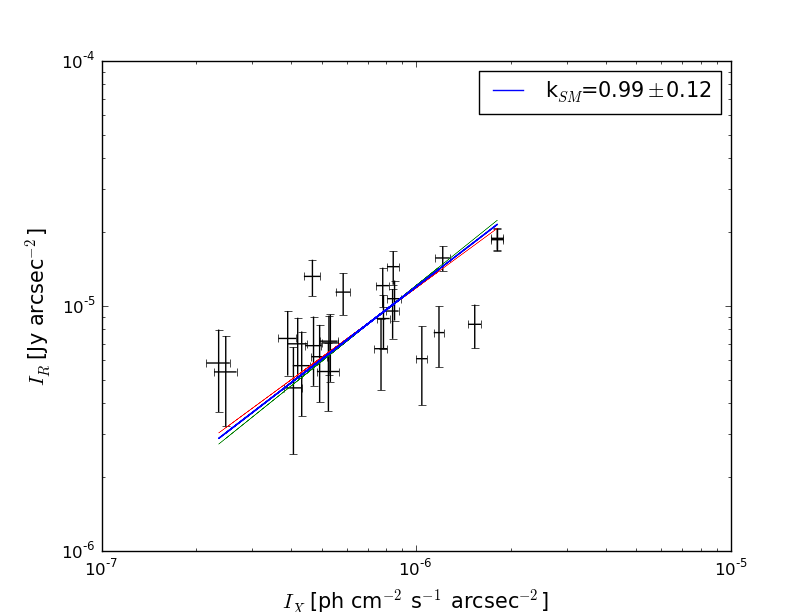}\par
\end{multicols}
\begin{multicols}{3}
\includegraphics[width=\linewidth]{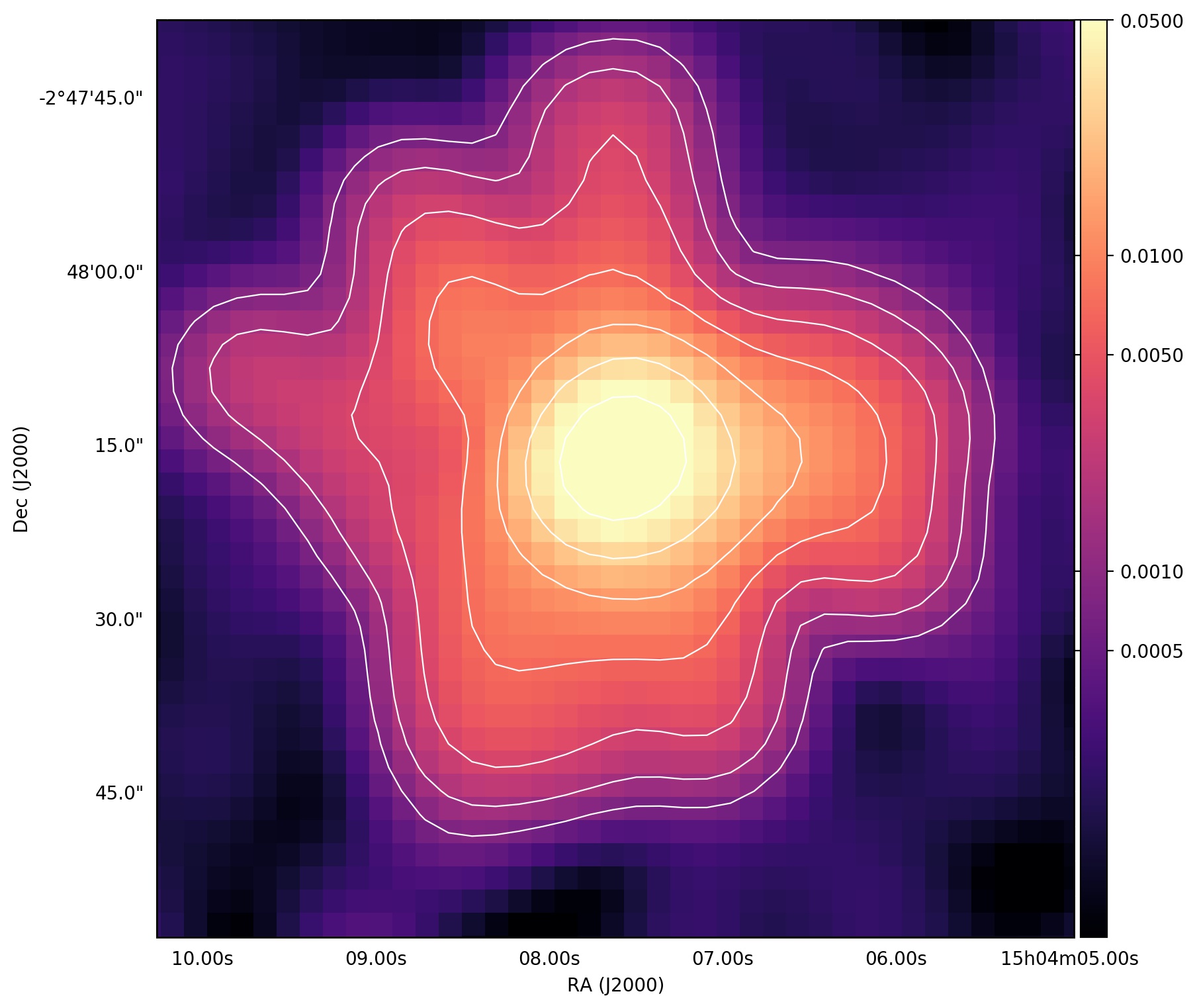}\par
\includegraphics[width=\linewidth]{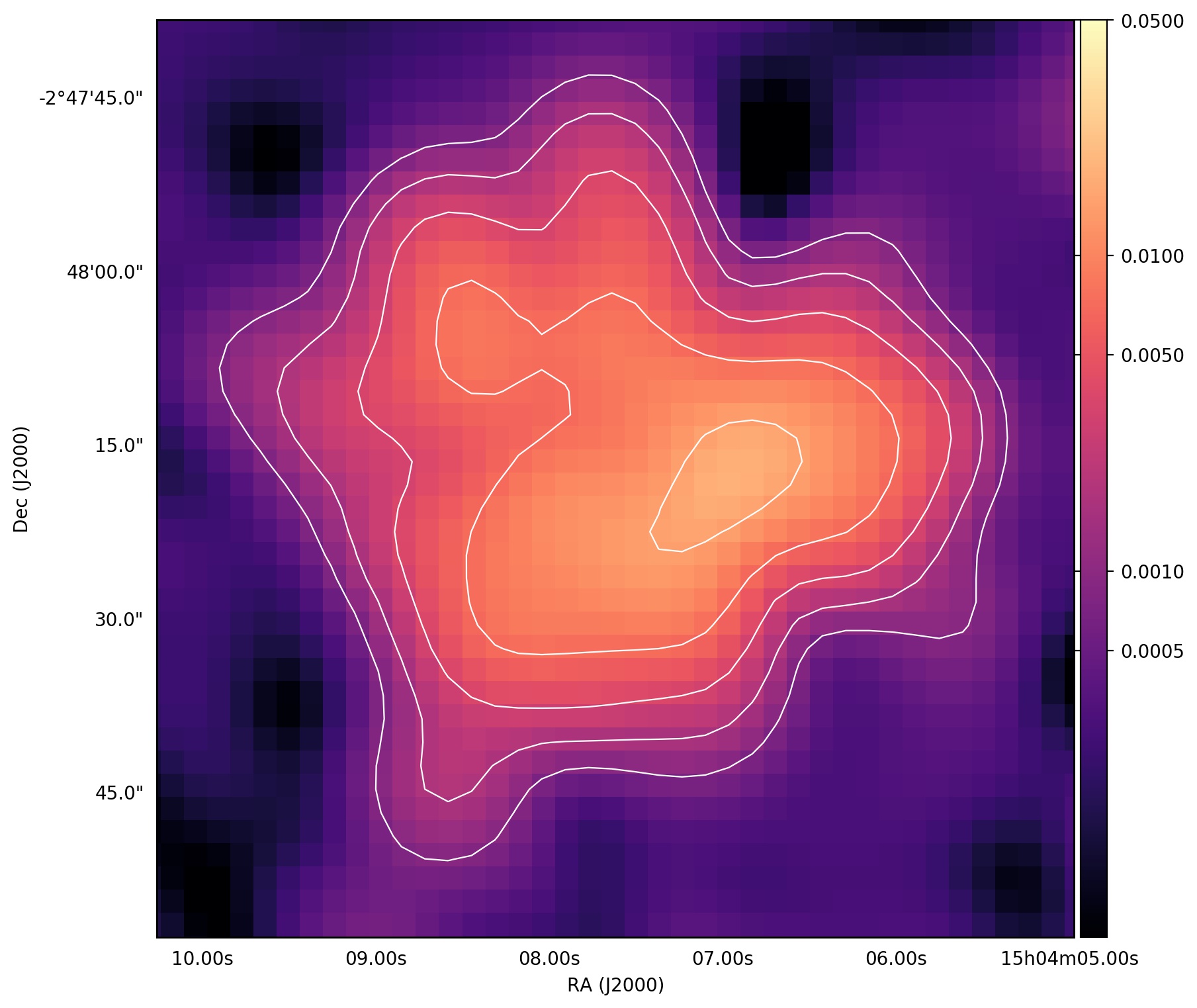}\par
\includegraphics[width=\linewidth]{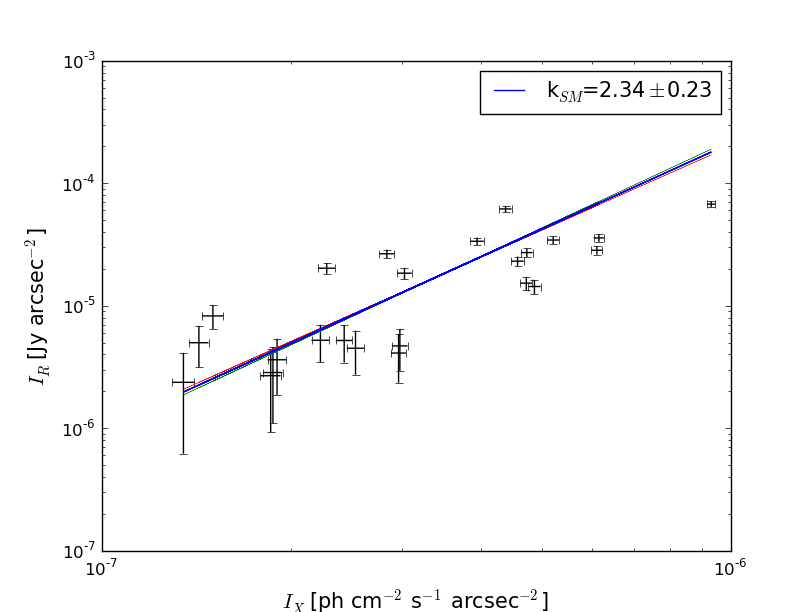}\par
\end{multicols}
\caption{\label{subsub}  {\it Top:} RX J1532.9+3021, before (left) and after the subtractions (center). The contours are at 2, 4, 8, 16, 32, 64 $\times$ 22 $\mu$Jy beam$^{-1}$. The plot (right) is the result of the SMptp analysis performed on the subtracted image with the same grid presented in B.3; {\it Bottom:} RXC J1504.1-0248, before (left) and after the subtractions (center). The contours are at 2, 4, 8, 16, 32, 64 $\times$ 0.9 mJy beam$^{-1}$. The plot (right) is the result of the SMptp analysis performed on the subtracted image with the same grid presented in B.4.}
\end{figure*}

In order to test that the adopted masking is sufficient to contain the contamination of the central radio source in our analysis, we present here the comparison of SMptp analysis for two of our objects after the subtraction of the central source, namely clusters RX J1532.9+3021 and RXC J1504.1-0248. We selected RXC J1504.1-0248 because it hosts the most luminous radio sources of our sample, therefore it should be the more sensitive to possible contaminations. In the two observations, the central sources were first imaged by selecting only baselines longer than, respectively, 10 and 15 k$\lambda$. The clean components were then subtracted from the uv-data to obtain images of the diffuse emission alone. We present in Fig. \ref{subsub} the images before and after the subtraction with the same color-scale and surface brightness levels and the same resolution reported in Tab. \ref{obs.tab}, and the corresponding SMptp analysis performed on the subtracted images by using the same grids presented in Appendix \ref{images} (B.3 and B.4). We found that, for each cluster, the two estimates of $k$ obtained with the two different approaches are consistent within the errors and, thus, that for the aims of this work the central source can be masked instead of subtracted.
\section{Considerations on the diffusion coefficient $D_0$}
\label{D0}
In our model the luminosity of the radio emission depends on the value of the ratio $Q_0/D_0$, hence, assuming the value of $D_0$ has consequences on the estimate of CRp injection amplitude, $Q_0$, and, ultimately, on the AGN CRp luminosity necessary to reproduce the observed radio emission. Specifically, higher values of $D_0$ result in a higher $Q_0$ and $L_\text{CRp}$.
We assumed that CRp can diffuse on the MH scale on time-scales that are shorter than the CRp cooling time. This cooling time is dominated by CRp-p collisions and is of the order of several Gyr \citep[][]{Brunetti-Jones_2014}. 
More quantitatively, the condition is that the optical depth due to CRp-thermal proton collision calculated on a MH scale is $\tau\simeq\sigma_\text{pp}n_\text{th}L$, where $\sigma_\text{pp}=32$ mBarn is the cross-section of the collision and $L$ is the spatial scale. As the CRp diffuse in the ICM, they move within different thermal densities, which contribute to the total optical depth as $d\tau=\sigma_\text{pp}n_\text{th}(r(t))cdt$, where $r(t)=\sqrt{4D_0t}$. Therefore, the time $t_\text{max}$ required to dissipate all the injected CRp in the thermal plasma within $R_\text{MH}$ can be derived by imposing that the total optical depth is: 
\begin{equation}
 \tau=c\sigma_\text{pp}n_0\int_0^{t_\text{max}}\left[1+ \left( \frac{R_\text{MH}}{r_c}\sqrt{\frac{t}{t_\text{max}}}\right)^2\right]^{-\frac{3}{2}\beta }dt=1
\end{equation}
where $n_0$, $r_c$ and $\beta$ are the parameters that describe the $\beta$-model for each cluster. For a given $t_\text{max}$, the associated diffusion coefficient is $D_0^\text{min}=R_\text{MH}^2/4t_\text{max}$. We report in Tab. \ref{D0_tab} the diffusion coefficients that we estimated for each MH.\\
This gives a lower limit to the CRp luminosity and an upper limit to the timescale for diffusion, that results longer than time-scale of cosmological cluster evolution. Assuming a larger value of the diffusion coefficient allows the diffusion of CRp on the MH scale on shorter time-scales, thus establishing the stationary CRp distribution faster, and entails that a higher $L_\text{CRp}$ is requested to reproduce the observed radio emission. Therefore, as a reference value, we assumed a coefficient that allows the diffusion of CRp over the MH radius in 1 Gyr ($D_0^\text{1 Gyr}$), that we report in Tab. \ref{D0_tab} and in Tab. \ref{res_gamma} with the corresponding $L_\text{CRp}$. We note that adopting $D_0^\text{min}$, instead of $D_0^\text{1 Gyr}$, results in values of $L_\text{CRp}$ that are a factor $D_0^\text{min}$/$D_0^\text{1 Gyr}\simeq 0.1$ lower that the values that we report. The $\gamma$-ray luminosity does not change, because it depends, instead, on the ratio $Q_0$/$D_0$, that is constrained by the observed radio luminosity.

\begin{table}
  \caption{Diffusion coefficients}
\begin{tabular}{lccc}
\hline
\hline
  Cluster name & $D_0^\text{1 Gyr}$ &$D_0^\text{min}$& $t_\text{max}$  \\
  &[$10^{29}$ cm$^2$ s$^{-1}$]&[$10^{29}$ cm$^2$ s$^{-1}$]&[Gyr]\\
\hline
~&~&~\\

RBS 797&6.9&1.1&6.4\\
Abell 3444 &13.8&1.1&12.3 \\
RXC J1504.1-0248&15.0&1.3&11.4 \\
RX J1532.9+3021& 10.5&0.9&12.1\\
\hline
\label{D0_tab}
\end{tabular}
\tablefoot{From left to right: Cluster name; Diffusion coefficient that allows the diffusione of CRp within $R_\text{MH}$ in 1 Gyr; Diffusion coefficent that assures the complete dissipation of CRp within $R_\text{MH}$; Time required to dissipate all the injected CRp within $R_\text{MH}$ by adopting $D_0^\text{min}$.}
\vspace{-0.2in}
\end{table}

\end{document}